\documentclass[preprint]{aastex}
\accepted{}

\slugcomment{13 February 2006 Draft}

\author{
Gordon T. Richards,\altaffilmark{1,2}
Michael A. Strauss,\altaffilmark{1}
Xiaohui Fan,\altaffilmark{3}
Patrick B. Hall,\altaffilmark{4}
Sebastian Jester,\altaffilmark{5,6}
Donald P. Schneider,\altaffilmark{7}
Daniel E. Vanden Berk,\altaffilmark{7}
Chris Stoughton,\altaffilmark{5}
Scott F. Anderson,\altaffilmark{8}
Robert J. Brunner,\altaffilmark{9}
Jim Gray,\altaffilmark{10}
James E. Gunn,\altaffilmark{1}
\v{Z}eljko Ivezi\'{c},\altaffilmark{8}
Margaret E. Kirkland,\altaffilmark{1}
G.R. Knapp,\altaffilmark{1}
Jon Loveday,\altaffilmark{11}
Avery Meiksin,\altaffilmark{12}
Adrian Pope,\altaffilmark{2}
Alexander S. Szalay,\altaffilmark{2}
Anirudda R. Thakar,\altaffilmark{2}
Brian Yanny,\altaffilmark{5}
Donald G. York,\altaffilmark{13,14}
J. C. Barentine,\altaffilmark{15}
Howard J. Brewington,\altaffilmark{15}
J. Brinkmann,\altaffilmark{15}
Masataka Fukugita,\altaffilmark{16}
Michael Harvanek,\altaffilmark{15}
Stephen M. Kent,\altaffilmark{5,13}
S. J. Kleinman,\altaffilmark{15}
Jurek Krzesi\'{n}ski,\altaffilmark{15,17}
Daniel C. Long,\altaffilmark{15}
Robert H. Lupton,\altaffilmark{1}
Thomas Nash,\altaffilmark{5}
Eric H. Neilsen, Jr.,\altaffilmark{5}
Atsuko Nitta,\altaffilmark{15}
David J. Schlegel,\altaffilmark{18}
and Stephanie A. Snedden\altaffilmark{15}
}

\altaffiltext{1}{Princeton University Observatory, Peyton Hall, Princeton, NJ 08544.}
\altaffiltext{2}{Department of Physics and Astronomy, The Johns Hopkins University, 3400 North Charles Street, Baltimore, MD 21218-2686.}
\altaffiltext{3}{Steward Observatory, University of Arizona, 933 North Cherry Avenue, Tucson, AZ 85721.}
\altaffiltext{4}{Department of Physics and Astronomy, York University, 4700 Keele Street, Toronto, Ontario, M3J 1P3, Canada.}
\altaffiltext{5}{Fermi National Accelerator Laboratory, P.O. Box 500, Batavia, IL 60510.}
\altaffiltext{6}{School of Physics and Astronomy, Southampton University, Southampton SO17 1BJ, UK.}
\altaffiltext{7}{Department of Astronomy and Astrophysics, The Pennsylvania State University, 525 Davey Laboratory, University Park, PA 16802.}
\altaffiltext{8}{Department of Astronomy, University of Washington, Box 351580, Seattle, WA 98195.}
\altaffiltext{9}{Department of Astronomy, University of Illinois at Urbana-Champaign, 1002 West Green Street, Urbana, IL 61801-3080.}
\altaffiltext{10}{Microsoft Research, 301 Howard Street, No. 830, San Francisco, CA 94105.}
\altaffiltext{11}{Astronomy Centre, University of Sussex, Falmer, Brighton BN1 9QJ, UK.}
\altaffiltext{12}{Institute for Astronomy, Royal Observatory, University of Edinburgh, Blackford Hill, Edinburgh EH9 3HJ, United Kingdom.}
\altaffiltext{13}{Department of Astronomy and Astrophysics, The University of Chicago, 5640 South Ellis Avenue, Chicago, IL 60637.}
\altaffiltext{14}{Enrico Fermi Institute, The University of Chicago, 5640 South Ellis Avenue, Chicago, IL 60637.}
\altaffiltext{15}{Apache Point Observatory, P.O. Box 59, Sunspot, NM 88349.}
\altaffiltext{16}{Institute for Cosmic Ray Research, University of Tokyo, 5-1-5 Kashiwa, Kashiwa City, Chiba 277-8582, Japan.}
\altaffiltext{17}{Obserwatorium Astronomiczne na Suhorze, Akademia Pedagogicazna w Krakowie, ulica Podchor\c{a}\.{z}ych 2, PL-30-084 Krak\'{o}w, Poland.}
\altaffiltext{18}{Lawrence Berkeley National Lab, 1 Cyclotron Road MS 50R-5032, Berkeley, CA 94720-8160.}

\received{2005 September 20}

\begin{document}

\title{The SDSS Quasar Survey: Quasar Luminosity Function from Data Release Three}

\begin{abstract}

We determine the number counts and $z=0$--5 luminosity function for a
well-defined, homogeneous sample of quasars from the Sloan Digital Sky
Survey (SDSS).  We conservatively define the most uniform statistical
sample possible, consisting of 15,343 quasars within an effective area
of 1622 deg$^2$ that was derived from a parent sample of 46,420
spectroscopically confirmed broad-line quasars in the 5282 deg$^2$ of
imaging data from SDSS Data Release Three.  The sample extends from
$i=15$ to $i=19.1$ at $z\lesssim3$ and to $i=20.2$ for $z\gtrsim3$.
The number counts and luminosity function agree well with the results
of the Two-Degree Field QSO Redshift Survey (2QZ) at redshifts and
luminosities where the SDSS and 2QZ quasar samples overlap, but the
SDSS data probe to much higher redshifts than does the 2QZ sample.
The number density of luminous quasars peaks between redshifts 2 and
3, although uncertainties in the selection function in this range do
not allow us to determine the peak redshift more precisely.  Our best
fit model has a flatter bright end slope at high redshift than at low
redshift.  For $z<2.4$ the data are best fit by a redshift-independent
slope of $\beta = -3.1$ ($\Phi(L) \propto L^{\beta}$).  Above $z=2.4$
the slope flattens with redshift to $\beta \gtrsim -2.37$ at $z=5$.  This
slope change, which is significant at the $\gtrsim5$-sigma level, must
be accounted for in models of the evolution of accretion onto
supermassive black holes.

\end{abstract}


\section{Introduction}

The advent of the Two Degree Field (2dF) QSO Redshift Survey (2QZ;
\markcite{bsc+00,csb+04}{Boyle} {et~al.} 2000; {Croom} {et~al.} 2004) and Sloan Digital Sky Survey (SDSS;
\markcite{yaa+00}{York} {et~al.} 2000) has resulted in a more than ten-fold increase in the
number of known quasars over the past decade.  While the evolution of
quasars and active galactic nuclei (AGN) in general has been of
considerable interest since the first identification of quasar
redshifts \markcite{sch63,sch68}({Schmidt} 1963, 1968), there has been a resurgence of interest
in the subject as a result of recent work in understanding the role of
AGN in galaxy evolution.  In particular, the formation of bulges and
supermassive black holes at the centers of galaxies appear to be
intimately related (the so-called ${\rm M_{BH}}-\sigma$ relationship;
\markcite{fm00,gbb+00,tgb+02}{Ferrarese} \& {Merritt} 2000; {Gebhardt} {et~al.} 2000; {Tremaine} {et~al.} 2002), emphasizing the importance of
understanding the role that quasar activity plays in the formation and
evolution of the galaxy population as a whole.  It has also been
argued that feedback mechanisms \markcite{beg04}(e.g., {Begelman} 2004) may play an
important role in determining the ${\rm M_{BH}}-\sigma$ relationship
and the co-evolution of black holes and the spheroid component of
their host galaxies \markcite{dcs+03,wl03,gds+04,so04,dsh05}(e.g., {Di Matteo} {et~al.} 2003; {Wyithe} \& {Loeb} 2003; {Granato} {et~al.} 2004; {Scannapieco} \& {Oh} 2004; {Di Matteo}, {Springel}, \&  {Hernquist} 2005).
Furthermore, an accurate description of the quasar luminosity function
(QLF) is needed to map the black hole accretion history of the
Universe \markcite{yt02}(e.g., {Yu} \& {Tremaine} 2002) and determine how quasars contribute to
the feedback cycle.

Until recently, the quasar population was parameterized by a broken
power law in luminosity with a peak in space density at $z\sim$2--3.
The luminosity at the power-law break has most often been
characterized by ``pure luminosity evolution'', whereby the rarity of
luminous quasars today is a result of a fixed population of quasars
becoming less luminous with time \markcite{who94,csb+04}({Warren}, {Hewett}, \& {Osmer} 1994; {Croom} {et~al.} 2004).  However, pure
luminosity evolution fails beyond the peak (at $z\sim2.5$) of the
luminous quasar space density \markcite{ssg95,fss+01}({Schmidt}, {Schneider}, \& {Gunn} 1995; {Fan} {et~al.} 2001).  Furthermore, hard
X-ray surveys \markcite{Ueda03,bcm+05}(e.g., {Ueda} {et~al.} 2003; {Barger} {et~al.} 2005), which probe both
optically obscured AGN and substantially fainter optical quasars, have
found that AGN evolution is best fit by a model in which less luminous
AGN peak in space density at smaller redshifts.  This behavior has
been termed ``cosmic downsizing,'' whereby the most massive black
holes did most of their accreting in the distant past, while less
massive objects underwent active accretion in the more recent past
\markcite{cbb+03,mer04,hkb+04}(e.g., {Cowie} {et~al.} 2003; {Merloni} 2004; {Heckman} {et~al.} 2004).

While X-ray and infrared surveys
\markcite{Ueda03,hsl+04,tuc+04,bcm+05,hms05}(e.g., {Ueda} {et~al.} 2003; {Haas} {et~al.} 2004; {Treister} {et~al.} 2004; {Barger} {et~al.} 2005; {Hasinger}, {Miyaji}, \& {Schmidt} 2005) provide a more
complete census of non-stellar nuclear activity in galaxies than do
optical surveys, and radio surveys have demonstrated that the decline
of quasars at high redshift is not due to dust obscuration
\markcite{wjs+05}({Wall} {et~al.} 2005), the optical luminosity function remains a powerful
diagnostic tool for our understanding of luminous AGNs.  This is, in
no small part, because of the large areas covered by optical surveys
such as the SDSS ($\sim10,000\,$deg$^2$).  Sensitive hard X-ray and IR
surveys, although sampling much higher AGN densities than optical
surveys, suffer from much smaller survey areas ($\lesssim1\,$deg$^2$)
and have difficulty constraining the AGN population where it is
intrinsically least dense (e.g., the most luminous and highest
redshift objects).

This paper presents a long sought after result: the optical luminosity
function of a large, homogeneous sample of luminous type 1 quasars
covering the entire span of observed quasar redshifts.  This overall
goal has already been roughly met by splicing together different
surveys \markcite{pei95}(e.g., {Pei} 1995): in particular, the combination of
results from the 750 deg$^2$ 2QZ survey to $z\sim2.2$
\markcite{bsc+00,csb+04}({Boyle} {et~al.} 2000; {Croom} {et~al.} 2004) (also earlier work including that of
\markcite{sg83}{Schmidt} \& {Green} 1983, \markcite{kk88}{Koo} \& {Kron} 1988, \markcite{bsp88}{Boyle}, {Shanks}, \& {Peterson} 1988, and \markcite{hfc93}{Hewett}, {Foltz}, \& {Chaffee} 1993,
among others); the photometrically-selected COMBO-17 survey
\markcite{wwb+03}({Wolf} {et~al.} 2003) spanning $1.2<z<4.8$ but only 0.78 deg$^2$; and $z>3$
surveys such as \markcite{who94}{Warren} {et~al.} (1994), \markcite{ssg95}{Schmidt} {et~al.} (1995), \markcite{kdd+95}{Kennefick}, {Djorgovski}, \& {de  Carvalho} (1995), and
\markcite{fss+01}{Fan} {et~al.} (2001).  Herein we accomplish this goal using a single
carefully constructed subset of the SDSS-DR3 data that was designed
for maximal homogeneity; this subsample has $\sim15,000$ quasars
selected over $\sim1600$ deg$^2$.

This paper describes the luminosity function of quasars with
$15.0<i<19.1$ and $0.0 \lesssim z \lesssim 3.0$, and extending to
$i<20.2$ for higher redshifts up to $z\sim5$.  The magnitude limits of
the SDSS survey only just approach the ``break'' magnitude
($b_J\sim19.5$) seen in the 2QZ number counts, and are nearly two
magnitudes brighter than that of the combined SDSS+2dF (2SLAQ) sample
from \markcite{rca+05}{Richards} {et~al.} (2005), both of which are restricted to $z\lesssim2.2$.
The SDSS data complement these smaller but deeper optical surveys both
by extending to $z\sim5$ and having superb multicolor
photometry.\footnote{Work on the QLF at $z\sim6$ involves going beyond
the main selection algorithm of SDSS \markcite{fss+01}({Fan} {et~al.} 2001).}

In \S~\ref{sec:data} we describe the creation of a homogeneous
statistical sample of quasars from the SDSS data.  The selection
function is presented in detail in \S~\ref{sec:selfunct}.  The less
technically-minded reader may choose to skip to
Section~\ref{sec:number_counts} where we present the number counts
relationship for our sample of quasars.  In \S\S~\ref{sec:kcorrect}
and \ref{sec:lumfunct} we discuss the application of $K$-corrections
to the data and the luminosity function (both binned and maximum
likelihood) that we derived after doing so.  Finally some discussion
and conclusions are presented in \S~\ref{sec:conclusions}.  Throughout
this paper we use a $\Lambda$ cosmology with $\Omega_m=0.3$,
$\Omega_\Lambda=0.7$, a Hubble Constant of $H_0=70\,{\rm
km\,s^{-1}\,Mpc^{-1}}$ \markcite{svp+03}(e.g., {Spergel} {et~al.} 2003), and luminosity
distances determined according to \markcite{hog99}{Hogg} (1999) for this cosmology.

\section{Construction of a Uniform Quasar Sample}
\label{sec:data}

\subsection{The Parent Sample}

The SDSS is an imaging and spectroscopic survey of the high Galactic
latitude sky in the Northern Hemisphere \markcite{yaa+00}({York} {et~al.} 2000).  It uses a
dedicated wide-field 2.5m telescope \markcite{gunn+05}({Gunn et al.} 2005) at Apache Point
Observatory, New Mexico with a 140-megapixel imaging camera
\markcite{gcr+98}({Gunn} {et~al.} 1998) and a pair of fiber-fed multi-object double
spectrographs.  The imaging is carried out in five broad bands
($ugriz$; \markcite{fig+96}{Fukugita} {et~al.} 1996, \markcite{slb+02}{Stoughton} {et~al.} 2002) on photometric moonless
nights of good seeing \markcite{hfs+01}({Hogg} {et~al.} 2001).  The imaging data are processed
with a series of pipelines \markcite{lgi+01,pmh+03}({Lupton} {et~al.} 2001; {Pier} {et~al.} 2003), resulting in
astrometric calibration errors of $<0\farcs1$ rms per coordinate, and
photometric calibration to better than 0.03 mag
\markcite{stk+02,ils+04,tucker+05}({Smith} {et~al.} 2002; {Ivezi{\' c}} {et~al.} 2004; {Tucker et al.} 2005).  The photometry we use is corrected
for Galactic extinction using the maps of \markcite{sfd98}{Schlegel}, {Finkbeiner}, \&  {Davis} (1998).
Spectroscopic targets, including quasar candidates \markcite{rfn+02}({Richards} {et~al.} 2002), are
selected from the imaging catalogs, assigned to spectroscopic tiles
\markcite{blm+03}({Blanton} {et~al.} 2003b), and spectra are obtained.  These data have been made
publicly available in a series of data releases (EDR:
\markcite{slb+02}{Stoughton} {et~al.} 2002; DR1: \markcite{aaa+03}{Abazajian} {et~al.} 2003; DR2: \markcite{aaa+04}{Abazajian} {et~al.} 2004; DR3:
\markcite{aaa+05}{Abazajian} {et~al.} 2005; DR4: \markcite{adel+05}{Adelman-McCarthy} {et~al.} 2006).

Main survey quasar candidates are selected for spectroscopic followup
as described in \markcite{rfn+02}{Richards} {et~al.} (2002).  The quasar candidates are
distinguished from the much more numerous stars and normal galaxies in
the SDSS in two ways: either by having distinctive $ugriz$ colors
(subdivided into $ugri$ [$z \lesssim 3.0$, where most sources are
UV-excess] and $griz$ [$z \gtrsim 3.0$] selection criteria), {\em or}
by having FIRST \markcite{bwh95}({Becker}, {White}, \& {Helfand} 1995) 20 cm radio counterparts.  Quasars with
redshifts around 2.7 and 3.5 have colors very close to those of normal
stars \markcite{fan99}({Fan} 1999), which greatly decreases the efficiency and
completeness of the quasar sample in the vicinity of these redshifts.
Radio selection, while adding less than $1$\% to the color-selected
sample (as most radio-selected quasars are also color selected;
\markcite{imk+02}{Ivezi{\' c}} {et~al.} 2002), helps to improve our selection completeness at
these redshifts.  Accounting for this and other sources of
incompleteness is the focus of \S~\ref{sec:selfunct}.

The spectroscopically confirmed quasars (restricted to those objects
that meet a traditional quasar definition [$M_i<-22$ measured in the
rest frame, and a full width at half maximum (FWHM) of lines from the
broad line region greater than $1000\,{\rm km\,s^{-1}}$]) in the SDSS
have been published by \markcite{shr+05}{Schneider} {et~al.} (2005, hereafter DR3Q), and the
current study uses this quasar sample as its basis.  The quasar
identifications and redshifts in DR3Q are identical to those in the
DR3 online database itself in most cases, but visual inspection has
caused the redshifts and line widths of several hundred objects to be
corrected.

\subsection{Statistical Sample Construction}

\begin{figure}
\epsscale{1.0}
\plotone{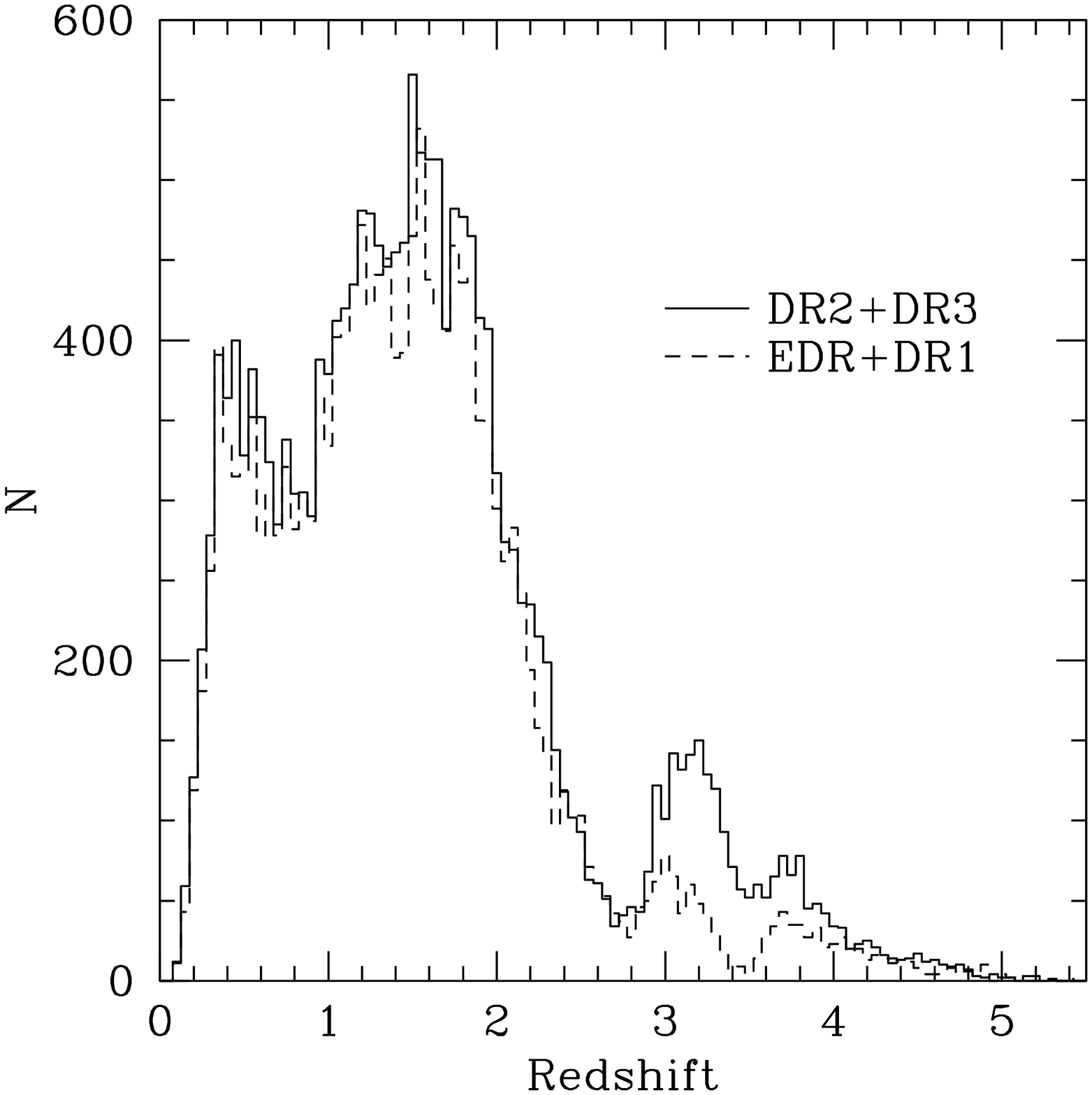}
\caption{Comparison of the SDSS quasar redshift distribution before
(EDR+DR1; {\em dashed line}) and after (DR2+DR3; {\em solid line}) the
\markcite{rfn+02}{Richards} {et~al.} (2002) selection algorithm was put in place; the two
subsamples have similar numbers of objects.  Note the improvement in
completeness at $z=3$--4 for the objects discovered after DR1.  At
lower redshift ($z\lesssim2$), quasars are selected largely by UV
excess, and the EDR and DR1 samples show no evidence of
incompleteness.  The structure in the redshift distribution is due to
selection effects, see \S~\ref{sec:selfunct}.
\label{fig:fig1}}
\end{figure}

The DR3Q sample contains 46,420 quasars, but as the QLF requires the
most homogeneous data set possible, we wish to use only the subset of
objects that were selected uniformly with the quasar target selection
algorithm described by \markcite{rfn+02}{Richards} {et~al.} (2002).  In particular, we reject 
several classes of object:

\begin{enumerate}
\item Objects selected for spectroscopy by algorithms
other than the main quasar target selection code (especially various
``serendipity'' algorithms; \markcite{slb+02,avm+03}{Stoughton} {et~al.} 2002; {Anderson} {et~al.} 2003). Most of these
additional quasars are fainter than the $i=19.1$ magnitude limit of
the UV-excess branch of quasar target selection.  This rejects close
to 28\% of the quasars in the DR3Q sample.

\item The original version of the quasar target selection algorithm
used in the EDR and DR1 (including data taken through June 2001) did a
particularly poor job of selecting quasars with redshifts close to
$z=3.5$.  Explicit color cuts, implemented for targets new to DR2 and
beyond, much improved the situation.  In this paper, we restrict
ourselves solely to quasars selected using this improved algorithm
(v3\_1\_0 of the target selection algorithm), as described by
\markcite{rfn+02}{Richards} {et~al.} (2002), thus rejecting nearly half the DR3Q quasars.
Figure~\ref{fig:fig1} compares the redshift distribution for quasars
from before and after the changes in the algorithm; the $z>3$ region
is much better (but still not perfectly) sampled with the final
algorithm.  The Appendix shows the Structured Query Language (SQL)
query that we used to select the resulting subsample from the SDSS
Catalog Archive Server (CAS).  This query resulted in 53,459 quasar
candidates selected over 2520 deg$^2$ of sky\footnote{The full DR3
spectroscopic area is 3732 deg$^2$}, of which 18,966 are matched to
spectroscopically confirmed quasars from the DR3Q\footnote{ Note that
$>99\%$ of the spectroscopically observed quasars yield an unambiguous
redshift \markcite{vsr+05}({Vanden Berk} {et~al.} 2005).}.

\item The quasar target selection algorithm searches for outliers from
the stellar locus in color space, and as such, is very sensitive to
data with large photometric errors due to 
problems in photometric calibration or in PSF determination.
As we describe in \S~\ref{sec:image_quality}, we identify problematic
stellar loci on a field-by-field basis (a field is an SDSS image in
all five bands, with approximate dimensions of $13^\prime \times
10^\prime$) and reject quasar candidates selected from these fields.
Roughly 21\% of all DR3 quasar candidates are rejected in this stage.
\end{enumerate}

Because of the evolution of the image processing software
\markcite{lig+02,aaa+03}({Lupton} {et~al.} 2002; {Abazajian} {et~al.} 2003), the SDSS includes two versions of
point-spread-function (PSF) photometry for each object: the TARGET
photometry (using the version of the pipeline and calibration current
at the time quasar target selection was carried out) and the BEST
photometry (using the most recent version of the pipeline at the time
of the data release).  The differences between the two are subtle, but
only 94\% of the DR3Q quasars in the computer sample were selected
from both TARGET and BEST photometry (most of the discrepant objects
are near the magnitude or color boundaries of the selection criteria).
The spectroscopic sample is (close to) complete only in TARGET, which
is what we use here to construct a homogeneous, statistical sample.

\subsection{Image Quality Restrictions}
\label{sec:image_quality}

Because of the sensitivity of the quasar target selection algorithm to
data of poor photometric quality, we identify and reject such data
from further consideration in this analysis.

As described by \markcite{ils+04}{Ivezi{\' c}} {et~al.} (2004) and \markcite{aaa+05}{Abazajian} {et~al.} (2005), the quality of the
PSF can be tested by examining the difference between PSF and
large-aperture photometry of bright stars on a field-by-field basis.
Thus, any field in which the median difference between these
quantities was more than a $3\,\sigma$ outlier (with respect to the
entire DR3 sample) in any of the five bands was rejected.

As a further check on the quality of the data, we take advantage of the
fact that the locus of stars in color-color space
\markcite{fif+00}(e.g., {Finlator} {et~al.} 2000) is remarkably narrow (0.03 mag or less in most
projections) and uniform over the sky.  The ridge lines of the stellar
locus allow one to define a series of ``principal colors''
\markcite{ils+04}({Ivezi{\' c}} {et~al.} 2004); deviations of the principal colors from their
canonical values or increases in the width of the stellar locus are
indications of inferior data.  The median principal color in a running
four-field boxcar and the rms principal color around the median are
available in the SDSS runQA outputs for each field.  We rejected
fields in which any of the four principal colors deviated from the
mean (for that run) by more than $3\,\sigma$.

However, the statistics of the rms widths of the principal color
distributions per field show considerable variation from run to run,
with some runs showing substantially narrower principal color
distributions than others.  Thus, if a given field had an rms width
$5\,\sigma$ greater than the mean width over all of DR3, that field
was rejected.  On the other hand, if the rms widths for a field were
$3\,\sigma$ greater than the mean for the given run, but less than
$2\,\sigma$ relative to all of DR3, the field was retained.

Finally, we required that the $r$-band seeing be less than $2\farcs0$
and that the operational database quality flag for that field not be
BAD, MISSING, or HOLE (these latter typically flag missing fields, or
fields in which the photometric pipeline was unable to process the
data; see the discussion in \S~4.6 of \markcite{slb+02}{Stoughton} {et~al.} 2002).

These restrictions reject 16.9\% of the area considered in our initial
query, and remove 11,186 of the 53,459 quasar candidates (20.9\%),
leaving a cleaned statistical sample of $15,343$ quasars (down from
$18,966$).  The removed fields have a higher than average density of
quasar candidates, a reflection of the fact that inferior photometry
tends to push stars out of the stellar locus.

\subsection{Effective Area and Quasar Selection Efficiency}
\label{sec:area}

Calculating the luminosity function requires determining the solid
angle of the survey.  Of the 5282 deg$^2$ of the DR3 imaging, 2520
deg$^2$ was targeted using v3\_1\_0 (or later) of the target selection
algorithm, and is thus covered by the query in the Appendix.  Of this
area, 426 deg$^2$ are rejected as having inferior photometry as
described in the previous section, leaving 2094 deg$^2$.  Only 0.05\%
of the sky is unsearchable for quasars due to nearby bright stars
\markcite{sjd+02}({Scranton} {et~al.} 2002), which is smaller than our uncertainty in the total
area (see below), thus we have not taken this correction into account.

\begin{figure}
\epsscale{1.0}
\plotone{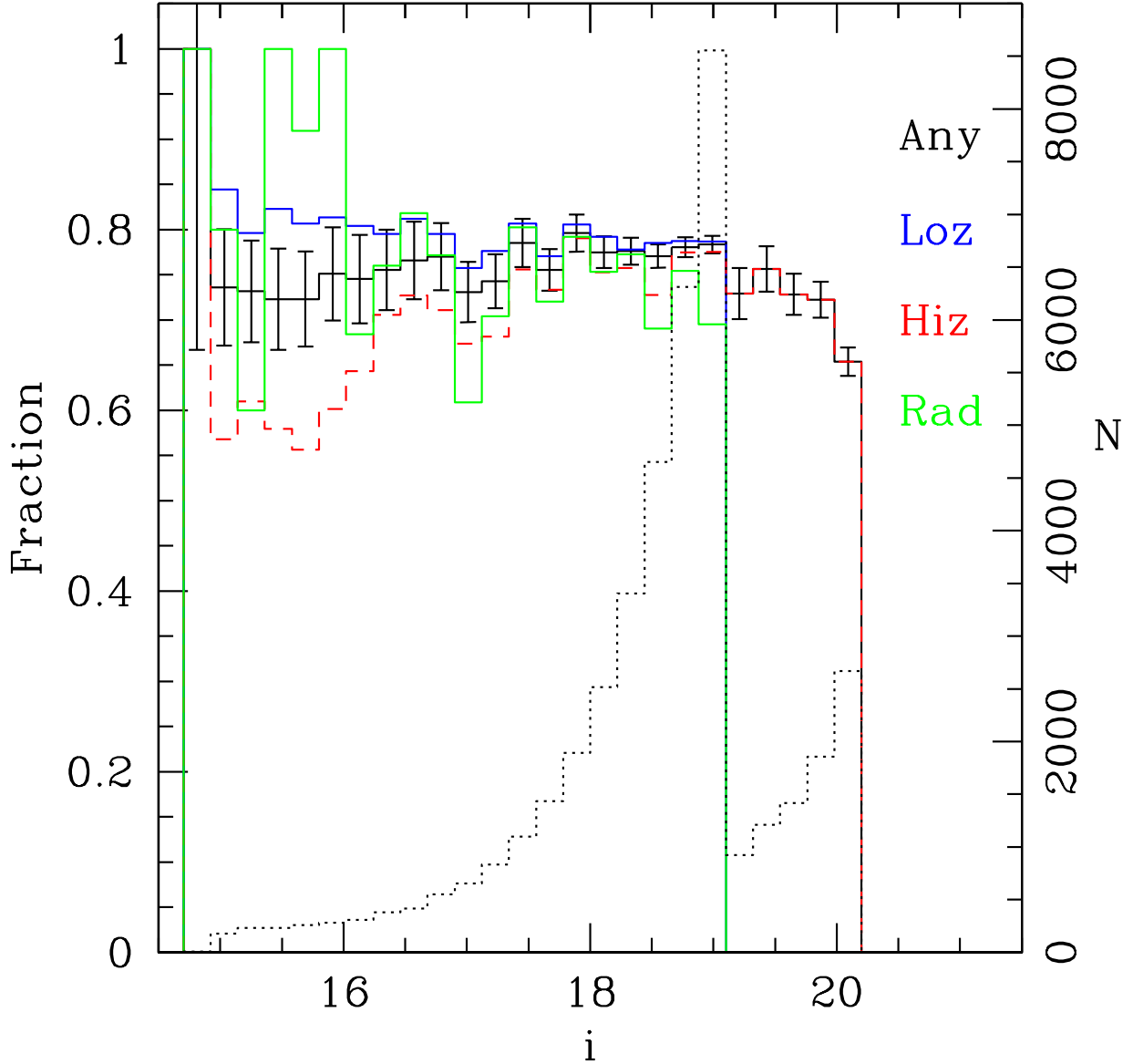}
\caption{Fraction of quasar targets (from TARGET photometry) with
spectroscopic observations.  Different classes of quasars are marked
separately --- black: all (with Poisson errorbars superposed); blue:
$ugri$($z\lesssim3$)-selected (Loz); red: $griz$($z\gtrsim3$)-selected
(Hiz); green: radio-selected (Rad).  The average completeness for all
$i\le19.1$ targets is 77.44\%.  Almost all the incompleteness is due
to the lag of the SDSS spectroscopic survey with respect to the
imaging survey.  The right-hand axis and the dotted line show the
number of DR3Q quasars as a function of magnitude.
\label{fig:fig2}}
\end{figure}

For technical reasons, it is cumbersome to combine the geometrical
information on the imaging and spectroscopic sky coverage, so we use
an empirical technique to determine the fraction of these 2094 deg$^2$
for which we have spectroscopy.  We simply determine the completeness
of the quasar candidate spectroscopy (Fig.~\ref{fig:fig2}) and
multiply the total area by this fraction to find the effective area.
The spectroscopic completeness (fraction of quasars targets with
spectra) for targets with $i<19.1$ is 77.4\% (fainter than 19.1, we
are increasingly sensitive to regions of poor photometry).  The tiling
algorithm is more than 99\% complete \markcite{blm+03}({Blanton} {et~al.} 2003b), so this 22.6\%
incompleteness reflects the area of sky not yet covered by
spectroscopy.  Thus the effective area covered by our sample is 1622
deg$^2$.  The uncertainty on this area is of order 10 deg$^2$, most of
which comes from the empirical correction to the difference between
the imaging and spectroscopic area.

\begin{figure}
\epsscale{1.0}
\plotone{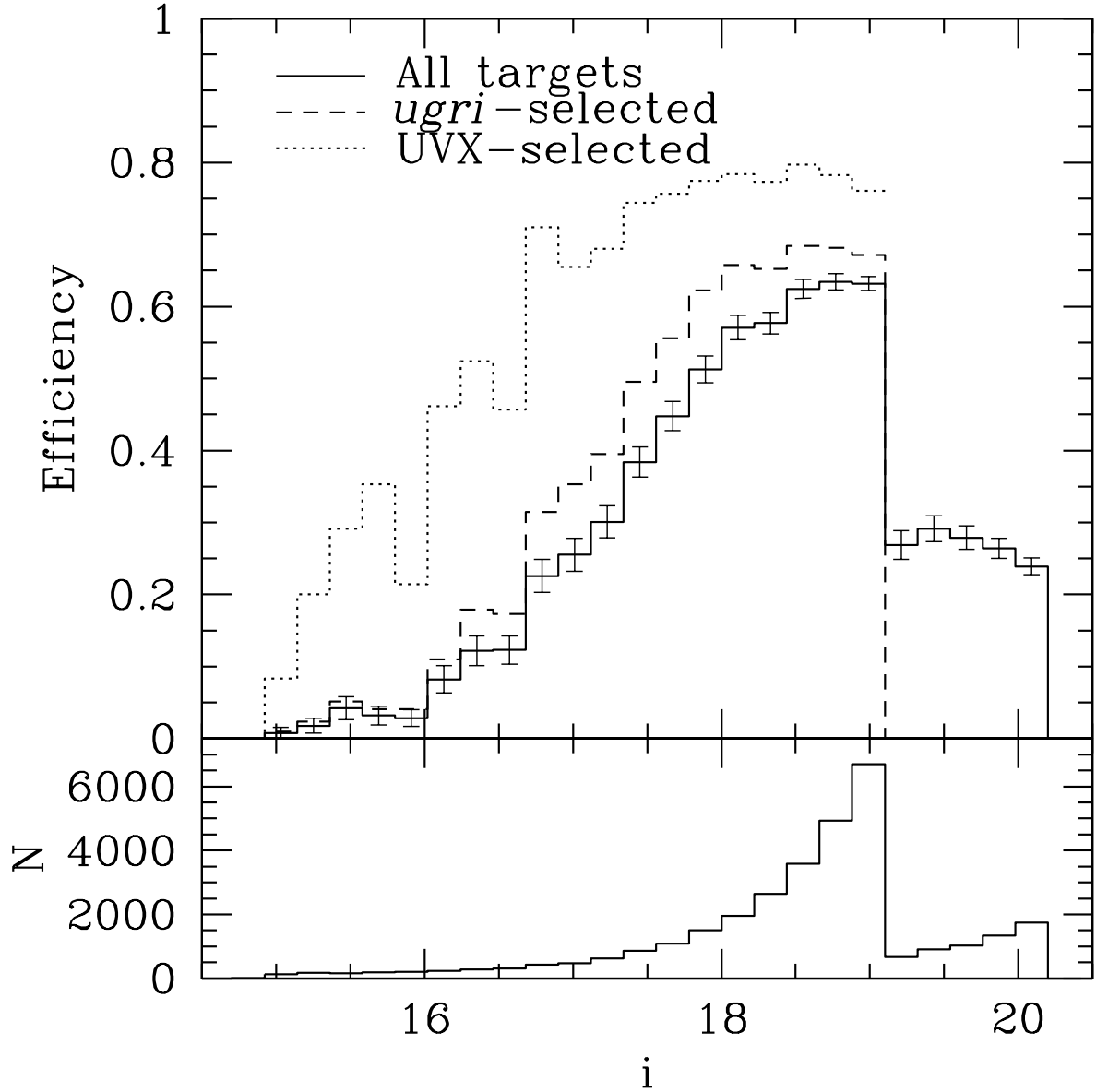}
\caption{Quasar selection efficiency as a function of magnitude in our
statistical sample of SDSS quasar candidates.  The overall efficiency
is given by the thick black line, with Poisson errorbars superposed.
The dashed line is for low-$z$ selection ($ugri$) only, excluding
quasars fainter than $i = 19.1$.  The dotted line shows the efficiency
for UV-excess quasar candidates with $u-g<0.6$ and $g-i>-0.3$, and $i
\le 19.1$.  Note that fainter than $i=19.1$ the efficiency of the
$griz$ branch of the code is substantially smaller.  The overall
efficiency of the algorithm for all quasar candidates, $ugri$-selected
quasars, and UV excess-selected quasars are 49\%, 61\% and 77\%,
respectively.
\label{fig:fig3}}
\end{figure}

\markcite{rfn+02}{Richards} {et~al.} (2002) describe the efficiency (fraction of quasar candidates
that are indeed quasars) of the quasar target selection algorithm;
here we revisit this with the clean data we have described here.
Figure~\ref{fig:fig3} shows the fraction of spectroscopically
confirmed quasars among the quasar targets with spectroscopy, as a
function of magnitude.  The efficiency is worse at brighter
magnitudes, as the ratio of stars to quasars increases dramatically at
the bright end.  The dashed black line gives the efficiency of our
$ugri$ ($z\lesssim3$) selection, which is appreciably higher than for
all objects.  Fainter than $i = 19.1$, candidates are from the $griz$
selection, where the efficiency is known to be significantly lower
than for the $i<19.1$ selection.  The dotted curve shows the
efficiency of a simple selection for UV-excess sources (defined here
as sources with $u-g<0.6$ and $g-i>-0.3$, the latter cut removing most
hot white dwarfs); this efficiency is close to 80\%.  The principal
contaminant of the UV-excess sources is emission-line galaxies. The
overall efficiencies of the algorithm for all quasar candidates,
$ugri$-selected quasars, and UV excess-selected quasars are 49\%, 61\%
and 77\%, respectively.  Correcting for any remaining bad fields
(\S~\ref{sec:image_quality}) and counting lower-luminosity AGN [i.e.,
\markcite{sey43}{Seyfert} (1943) galaxies], narrow-lined AGN
\markcite{hec80,zsk+03}(e.g., {Heckman} 1980; {Zakamska} {et~al.} 2003), and BL Lac objects
\markcite{csh+05}(e.g., {Collinge} {et~al.} 2005) would further increase the selection
efficiency of AGN in general (cf., the 66\% overall efficiency given
in \markcite{rfn+02}{Richards} {et~al.} 2002).

\section{Selection Function and the Completeness of the Quasar Sample}
\label{sec:selfunct}

The quasar target selection algorithm is a function of observed
colors, magnitudes, and radio brightness, and is sensitive to the
completeness and photometric errors of the SDSS imaging data.  In this
section, we determine the selection function of the sample, i.e., the
completeness as a function of apparent magnitude and redshift.

\subsection{Image Quality Completeness}
\label{sec:quality_completeness}

We start with the incompleteness due to the SDSS imaging data and
photometric pipeline.  In particular, as described by \markcite{rfn+02}{Richards} {et~al.} (2002),
objects with the so-called ``fatal'' and ``non-fatal'' flags from the
photometric pipeline, indicating unreliable photometry of the object,
are precluded from being selected on the basis of their colors.
\markcite{vsr+05}{Vanden Berk} {et~al.} (2005) have determined that 3.8\% of all point sources
brighter than $i = 19.1$ are excluded because of these flags; this
fraction is a (weak) function of magnitude.\footnote{This magnitude
dependence, however, may be due to the fact that the \markcite{vsr+05}{Vanden Berk} {et~al.} (2005)
sample is mostly stars, which have a different color and Galactic
latitude dependence than do quasars.}

In the magnitude range $17.5 < i < 18.5$, roughly $94$\% of 2QZ
\markcite{csb+04}({Croom} {et~al.} 2004) quasars, based on completely independent imaging data,
are recovered by SDSS, suggesting an SDSS image quality completeness
of this order, in good agreement with what we found above.  Similarly,
at the bright end, 17/18 (94\%) $z>0.3$ PG \markcite{sg83}({Schmidt} \& {Green} 1983) quasars in the
DR3 sample footprint are free of cosmetic defects in the SDSS imaging
(see the discussion in \markcite{jsr+05}{Jester} {et~al.} 2005).  We thus apply a global 5\%
correction (splitting the difference between the 94\% image quality
completeness from these comparisons and the 96\% value from
\markcite{vsr+05}{Vanden Berk} {et~al.} 2005) to account for image quality incompleteness.

\subsection{Simulated Quasars}
\label{sec:sim}

\markcite{fan99}{Fan} (1999) describes the construction of simulated quasar
photometry in the SDSS photometric system.  These simulations are run
through the quasar target selection code, allowing us to quantify the
fraction of objects selected as a function of magnitude and redshift.
The simulations used herein are similar to those of \markcite{fan99}{Fan} (1999) [see
also \markcite{fss+01}{Fan} {et~al.} (2001) and \markcite{rhv+03}{Richards} {et~al.} (2003)], modulo some changes in the
relative strengths of the emission lines, adoption of a redder
continuum shortward of Ly$\alpha$ emission, and the use of a more
recent characterization of the SDSS filter curves \markcite{slb+02}({Stoughton} {et~al.} 2002).

The colors of each simulated quasar are determined by the power-law
index $\alpha_\nu$ of its continuum ($f_{\nu} \propto
\nu^{\alpha_{\nu}}$), the strength of its emission lines, the
absorption of the Ly$\alpha$ forest and presence of Lyman limit
systems, and its redshift. 
For the power-law continuum, the simulations assume a Gaussian
distribution in $\alpha_\nu$, with mean $-0.5$ and standard deviation
0.3.  This mean spectral index is in reasonable agreement with the
composite SDSS quasar spectrum given by \markcite{vrb+01}{Vanden Berk} {et~al.} (2001), who find
$\alpha_\nu=-0.46$.  At wavelengths shortward of the Ly$\alpha$
emission line we instead use a spectral index derived from a Gaussian
distribution with mean of $-1.57$ and dispersion of 0.17, consistent
with the results of \markcite{tzk+02}{Telfer} {et~al.} (2002)\footnote{A much harder (i.e.,
bluer) UV color was derived from a less luminous, low-redshift sample
of quasars studied by {\em FUSE} \markcite{skb+04}({Scott} {et~al.} 2004).  We found that using
this bluer continuum did a poorer job of matching the observed
redshift histogram; see Figure~\ref{fig:fig9n}.}; this spectral index
is taken to be uncorrelated with that used at longer wavelengths.
Note that the use of this steeper spectral index shortward of
Ly$\alpha$ represents a significant change from our previous use of
these simulations (such as discussed in \markcite{rfn+02}{Richards} {et~al.} 2002) in which we
generally used the same power law for all optical/UV wavelengths; we
discuss below how this affects the $z>2.2$ selection function.

We simulate 200 quasars at each grid point in apparent magnitude and
redshift space, with an additional 1000 quasars on grid points with
$1.8 < z < 3.2$, where the selection efficiency is low.  The magnitude
grid points are separated by 0.1 mag in the range $13 <
AB_{\,1\,\micron} < 22.4$ in asinh magnitudes \markcite{lgs+99,slb+02}({Lupton}, {Gunn}, \& {Szalay} 1999; {Stoughton} {et~al.} 2002)
--- fully spanning the space of the SDSS imaging data.\footnote{The
simulations are normalized at $1\,\micron$, a local minimum in the
spectral energy distribution (SED) of quasars \markcite{ewm+94}({Elvis} {et~al.} 1994), as
working longward of the effective wavelength of the SDSS $i$ band
allows us to match the bluer observed color distribution of the
brightest quasars \markcite{jsr+05}(e.g., {Jester} {et~al.} 2005).} The redshift grid points
span the range $0 < z < 5$, spaced at intervals of 0.05.  We add
errors to the magnitudes consistent with the estimated PSF magnitude
errors in the SDSS photometric pipeline (see Fan 1999); they are
Poisson distributed in flux space so that we properly reproduce the
rollover of the asinh magnitude errors \markcite{lgs+99}({Lupton} {et~al.} 1999).  Errors from
photometric calibration uncertainties
\markcite{ils+04,tucker+05}(e.g., {Ivezi{\' c}} {et~al.} 2004; {Tucker et al.} 2005) of 0.02 mag are added in quadrature
to the $g,r$ and $i$ measurement and 0.03 mag to the $u$ and $z$
measurement.  The output of the simulations are the $ugriz$ photometry
and errors of each object, which can be input directly into the target
selection code.  Any object with errors of greater than 0.2 mag (i.e.,
less than a $5\,\sigma$ detection) in a given band is considered to be
undetected in that band.

\begin{figure}[p]
\epsscale{1.0}
\plotone{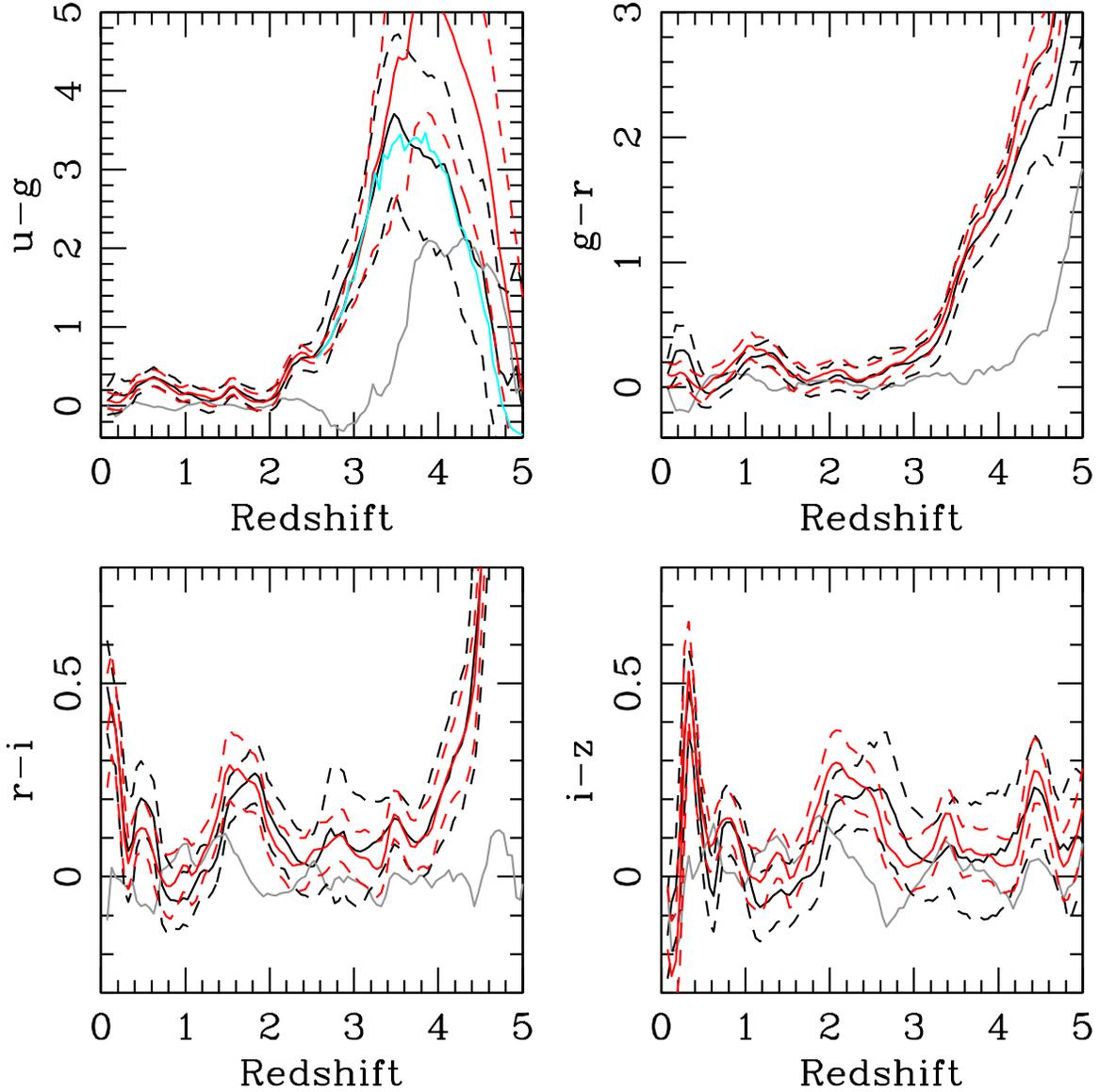}
\caption{Mean DR3Q quasar colors (solid black line) and 68\%
confidence limits (dashed black lines) and mean simulated quasar
colors (solid red line) and 68\% confidence limits (dashed red lines),
all as a function of redshift.  Gray lines show the difference between
the simulated and observed means.  The cyan line at high redshift in
the upper left-hand panel is the mean magnitude-weighted simulated
quasar color; this weighting properly accounts for the effect of asinh
magnitudes at low signal-to-noise ratio, and is a much better match to
the data.
\label{fig:fig4}}
\end{figure}

Figure~\ref{fig:fig4} compares the observed (black) and simulated
(red) SDSS colors of quasars as a function of redshift; both the mean
colors and the contours containing 68\% of the quasars at each
redshift are shown.  The difference between the observed and simulated
mean colors as a function of redshift is given by the gray curve.  In
general the simulated colors trace the observed colors quite well.  At
higher redshifts, where Ly$\alpha$ forest absorption causes asinh and
logarithmic magnitudes to exhibit magnitude-dependent differences, it
is important to determine a properly magnitude-weighted mean for the
simulations for comparison with the data --- as shown by the cyan line
in the upper left-hand panel of Figure~\ref{fig:fig4}.

\begin{figure}[p]
\epsscale{1.0}
\plotone{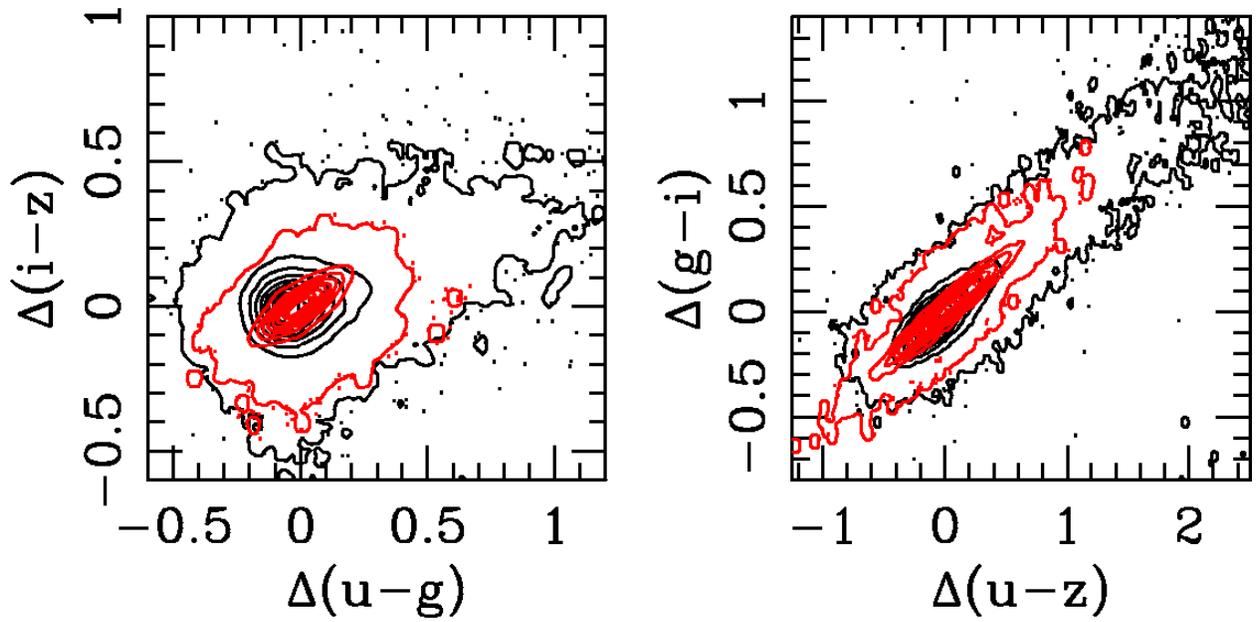}
\caption{Relative colors of DR3 quasars (black contours/dots) and
simulated quasars (red contours/dots).  Only quasars with $0.6<z<2.2$
and $i<19.1$ are considered.  Overall the simulations match the data
quite well, but the data clearly show red outliers and objects with
convex SEDs that are not described by the simulations.  Objects with
pure power-law continua and no photometric errors would show perfect
correlation between the relative colors.
\label{fig:fig5}}
\end{figure}

We next compare the {\em range} of colors between simulations and
observations. Following \markcite{rfs+01,rhv+03}{Richards} {et~al.} (2001, 2003) we subtract the mean
color as a function of redshift from the observed color, removing the
mean emission line contribution from the colors and allowing for a
redshift-independent color comparison.  This process allows for more
robust comparison of quasar colors across a wide range of redshifts
(as compared to using disparate continuum windows at low and high
redshift).  These relative colors ($\Delta (u-g)$, etc.) of quasars
with $0.6<z<2.2$ and $i<19.1$ are plotted against each other for both
the data and simulations in Figure~\ref{fig:fig5}.  The simulations
reproduce the overall trends in relative color; the differences
reflect the sense in which the optical/UV continua of quasars are not
perfectly described by power laws.  The simulations and data agree
well in the $\Delta(u-z), \Delta(g-i)$ plane, but $\Delta(u-g)$ and
$\Delta(i-z)$ are much less correlated in the data than in the
simulations.  Power laws produce an exact correlation between these
relative colors in the absence of errors.  We find that the
$\Delta(u-g)$ color is within $3\sigma$ of the $\Delta(i-z)$ color for
78\% of the observed quasars, while this fraction is 94\% for the
simulated quasars.  Thus roughly 16\% of the quasars in our sample
have continua that deviate significantly from a power law.

Dust reddening \markcite{rhv+03,hsh+04}(e.g., {Richards} {et~al.} 2003; {Hopkins} {et~al.} 2004) can explain at least
some of the (comparatively) red tail of the relative color
distribution and those objects with red [more positive] $\Delta(u-g)$
and blue [more negative] $\Delta(i-z)$, as the simulations do not
include the effects of dust.  However, the fraction of such reddened
objects is relatively small.  \markcite{rhv+03}{Richards} {et~al.} (2003) estimate that 6\% of
broad-line (type 1) quasars in the SDSS are consistent with being
moderately reddened by dust.  Objects with blue $\Delta(u-g)$, red
$\Delta(i-z)$ objects must instead be due either to photometric
errors, atypical emission line ratios, or intrinsic (convex)
curvature.  The latter objects may be interesting for comparison with
accretion disk model SEDs \markcite{hab+00}(e.g., {Hubeny} {et~al.} 2000).

\subsection{Completeness of the Quasar Sample}
\label{sec:completeness} 

We apply the quasar target selection algorithm, using the version
described by \markcite{rfn+02}{Richards} {et~al.} (2002), to the simulated quasar colors.  As we
have discussed, the target selection algorithm includes both color and
radio selection.  For all practical purposes the selection algorithm
considers objects as fitting into one of three classes: point sources
without radio detections, point sources with radio detections, and
extended sources.  To simulate radio sources we simply set the flag
that says that the object is detected in the radio.  For extended
sources we similarly set the extended flag (but note that the
simulations do not include any host galaxy contribution to the
magnitudes or colors).

\begin{figure}
\epsscale{1.0}
\plotone{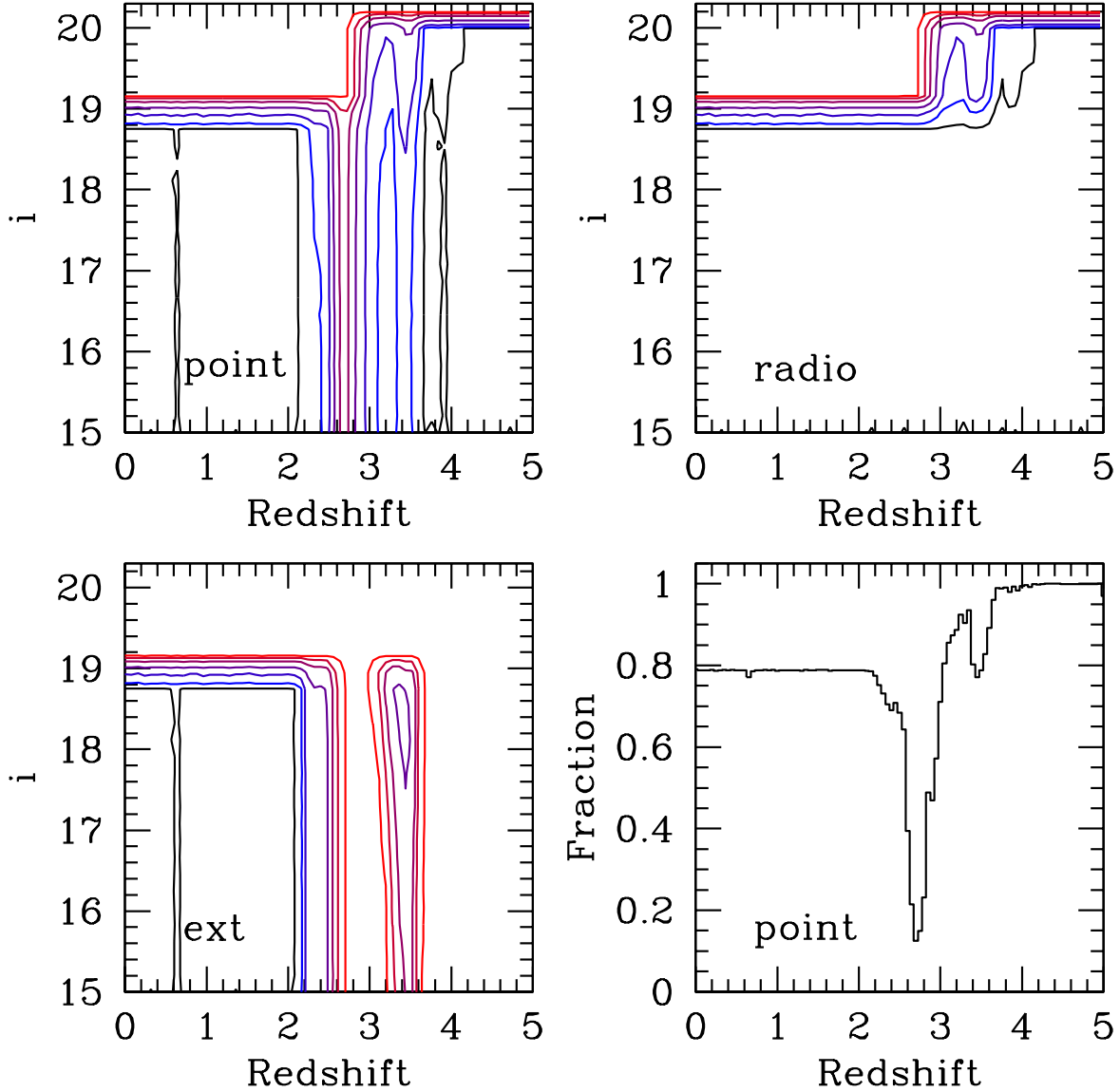}
\caption{Quasar target selection completeness from the simulations as
a function of redshift and $i$ magnitude for point/nonradio (point),
point/radio (radio), and extended (ext) sources.  Contours are at 1,
10, 25, 50, 75, 90, and 99\% completeness.  The 99\% completeness
limit is given in black; the 1\% limit is red.  The bottom right-hand
panel shows the completeness of point non-radio sources from the top
left-hand panel, averaged over $15 < i < 20.2$. Recall that the
limiting magnitude changes from $i=19.1$ to $i=20.2$ at $z\approx3$.
\label{fig:fig6}}
\end{figure}

Figure~\ref{fig:fig6} shows the completeness of the algorithm as a
function of redshift and $i$ magnitude, for radio sources, non-radio
point sources, and extended sources.  Contours are at 1, 10, 25, 50,
75, 90, and 99\% completeness.  The 99\% completeness limit is given
as the black line; 1\% is the red line.  This figure, and the two
following, are similar to those shown in \markcite{rfn+02}{Richards} {et~al.} (2002), but use the
updated quasar simulations as discussed above.  The selection function
shown in Figure~\ref{fig:fig6} is also given in Table~\ref{tab:tab01};
the columns are $i$-band magnitude, redshift, and the selection
function for non-radio point sources, radio-detected point sources,
and extended sources, respectively; image quality completeness is not
included.

Table~\ref{tab:tab01} and the selection function plots give the
fraction of true quasars at a given redshift and apparent magnitude
that would be selected.  Brighter than the magnitude limit, the
selection function is only weakly dependent on magnitude, and we show
the marginalized selection function in the lower right-hand corner of
Figure~\ref{fig:fig6}.  Thus each object is assigned a weight (inverse
of the value of the selection function) depending on its redshift and
apparent magnitude in the determination of the luminosity function
below.  Not surprisingly, the selection function is particularly low
at redshift $2.7$, where quasars have colors very similar to A/F stars.
There is a secondary dip at $z \approx 3.5$, where quasars have
similar colors to G/K stars in the $griz$ diagram.

The selection function for extended sources beyond $z = 2.2$ is very
low, as expected.  Radio source selection does not depend on color,
and thus the radio selection function shows no dependence on redshift
except at redshifts above 3 and $i > 19.1$, where the only
radio-detected objects are those selected by the $griz$ algorithm.

\begin{figure}
\epsscale{1.0}
\plotone{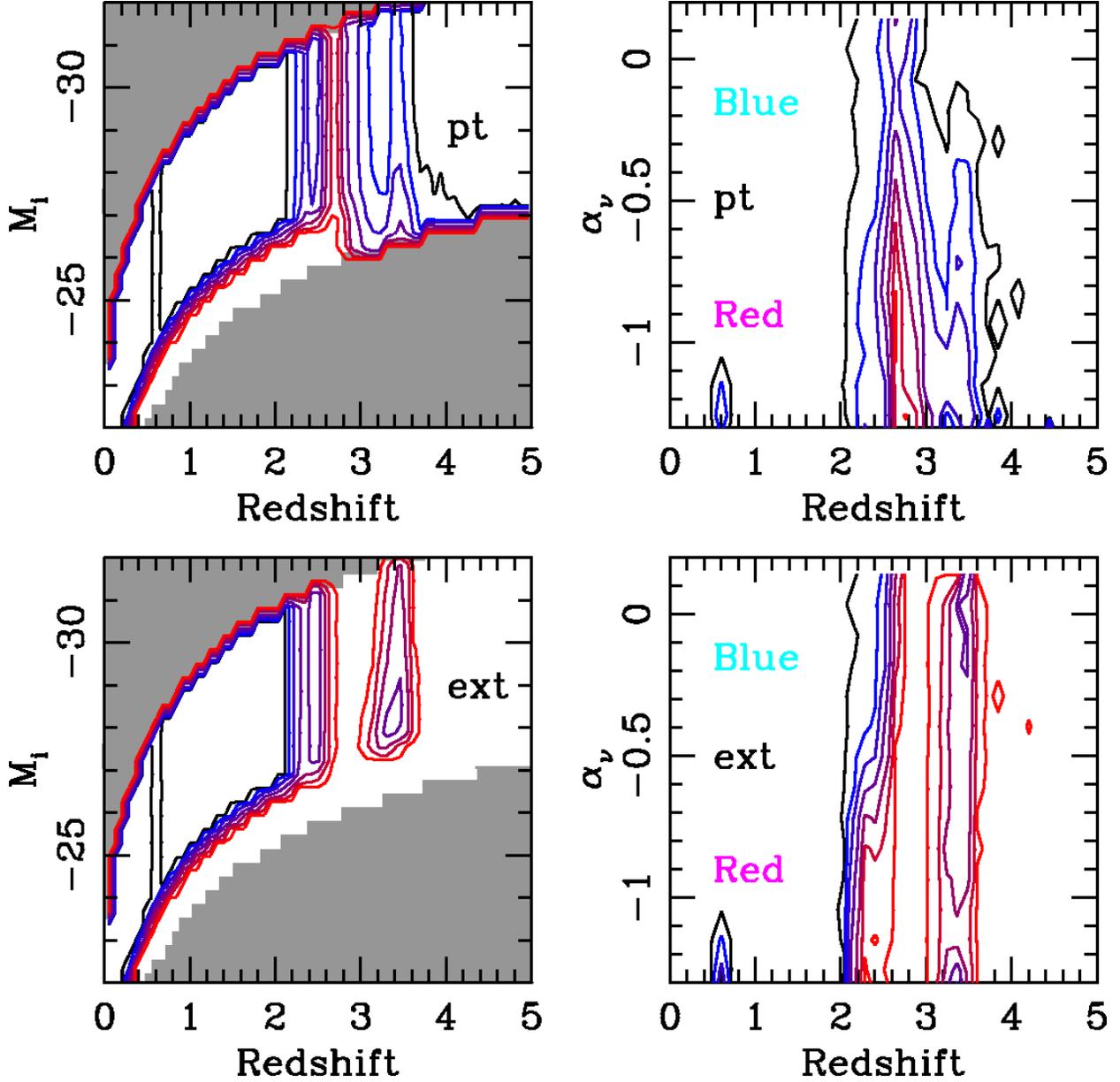}
\caption{Completeness as a function of redshift and $M_i$ ({\em left
panels}) and optical spectral index ($\alpha_{\nu}$; {\em right
panels}) for point (pt) and extended (ext) sources.  Contours are at
1, 10, 25, 50, 75, 90, and 99\% completeness.  The 99\% completeness
limit is given in black; the 1\% limit is red.  The left panels are
exactly the same as the left-hand panels in Figure~\ref{fig:fig6}
modulo a transformation of the axes.  Gray areas indicate regions of
parameter space not covered by the simulations.  We omit the radio
panel since it is featureless (aside from the flux limits).
\label{fig:fig7}}
\end{figure}

The left-hand panels of Figure~\ref{fig:fig7} shows the completeness
as a function of redshift and absolute $i$ magnitude.  These panels
are a simple transformation of variable from the left-hand panels in
the previous figure, but this presentation is helpful for comparison
with the luminosity function (\S~\ref{sec:lumfunct}).  Selection of
radio sources is independent of color, thus we exclude this panel in
Figure~\ref{fig:fig7}.  The right-hand panels of Figure~\ref{fig:fig7}
shows our completeness as a function of redshift and optical spectral
index ($\alpha_{\nu}$).  In the $2.2 < z < 3$ regime, intrinsically
bluer sources are much more likely to be selected than red sources,
and the redshift dependence of the selection function is seen to be
color dependent.

\begin{figure}
\epsscale{1.0}
\plotone{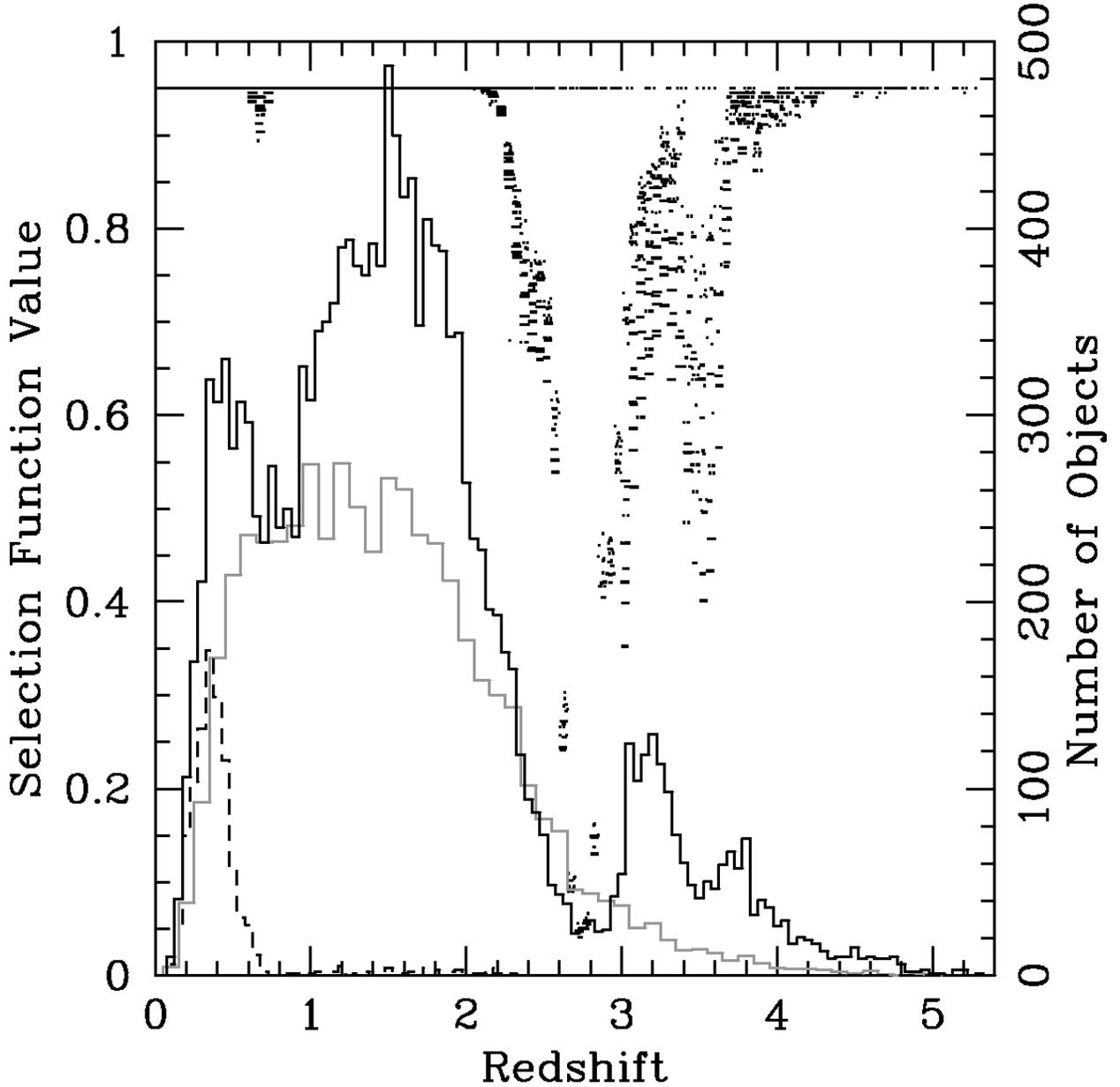}
\caption{Redshift distribution ({\em solid line}) of our main quasar
sample,
together with the correction for each quasar as derived from the
selection function for that quasar's redshift and magnitude
(Fig.~\ref{fig:fig6}).  The selection algorithm is quite complete for
$z<2.2$, but suffers from highly redshift-dependent incompleteness at
higher redshifts, reaching as low as $\sim$5\% at $z \approx 2.7$.
The line of $z\sim2.7$ quasars with selection function equal to 95\%
are radio-selected objects.  It is not strictly possible to correct
the redshift distribution of the full sample shown; however, the gray
curve shows the redshift distribution after applying an $i=19.1$
magnitude cut after correcting for emission line effects, applying the
selection function weights, and removing extended sources.  The
redshift distribution of extended sources (which may be
contaminated by host galaxy light) is shown by the dashed histogram.
\label{fig:fig8n}}
\end{figure}

Figure~\ref{fig:fig8n} overplots the redshift distribution of the
quasar sample and the value of the selection function for each quasar
(including radio sources).  The division between the low-redshift and
high-redshift branches of the target selection algorithm, with their
different magnitude limits, is at $z \sim 3$.  The fainter magnitude
limit for $z\ge3$ greatly exaggerates the discontinuity already
present because of the rapidly changing selection function at this
redshift.  The excess of quasars at $z\sim0.4$ is likely due to
uncorrected host galaxy (and emission line, see \S~\ref{sec:kcorrect})
flux which makes these quasars appear brighter than they should (and
thus are selected to a fainter intrinsic magnitude limit).

In the vicinity of $z \sim 2.7$, the selection function drops
precipitously, and is quite sensitive to such uncertain details of the
simulation as the mean and distribution function of the slope of the
UV continuum.  Indeed, Figure~\ref{fig:fig5} showed that the
simulations do not perfectly model the observed quasar color
distribution; these details mean that the large value of the
correction at this redshift is quite uncertain.  We therefore somewhat
arbitrarily place a lower limit of 0.333 on the selection function to
avoid over-correcting for incompleteness at these redshifts.  This
choice is based on our expectations regarding smoothness of the quasar
distribution with redshift in the absence of selection effects (e.g.,
see \S~\ref{sec:radiovscolor}).  Imposing this lower limit affects 285
quasars with $z \sim 2.7$, or 2\% of the full sample (22\% of quasars
with $2.2 \le z \le 3.0$).  The need for imposing this limit may be
indicative of quasars having somewhat bluer colors than the mean that
we assume for the simulations, bluer quasars being more likely to be
selected than redder quasars at $z\sim2.7$; see the upper right-hand
panel in Figure~\ref{fig:fig7}.

The changing magnitude limit near $z\sim3$ and the effects of emission
lines (see \S~\ref{sec:kcorrect}) make it impossible to simply correct
the raw redshift histogram in Figure~\ref{fig:fig8n}.  However, after
correcting for the selection effects discussed above, removing
extended sources, and limiting the sample to $i=19.1$ {\em after} applying
the emission line K-correction discussed in \S~\ref{sec:kcorrect}, we
find a rather smooth redshift distribution, which is shown by the gray
curve in Figure~\ref{fig:fig8n}.

\begin{figure}
\epsscale{1.0}
\plotone{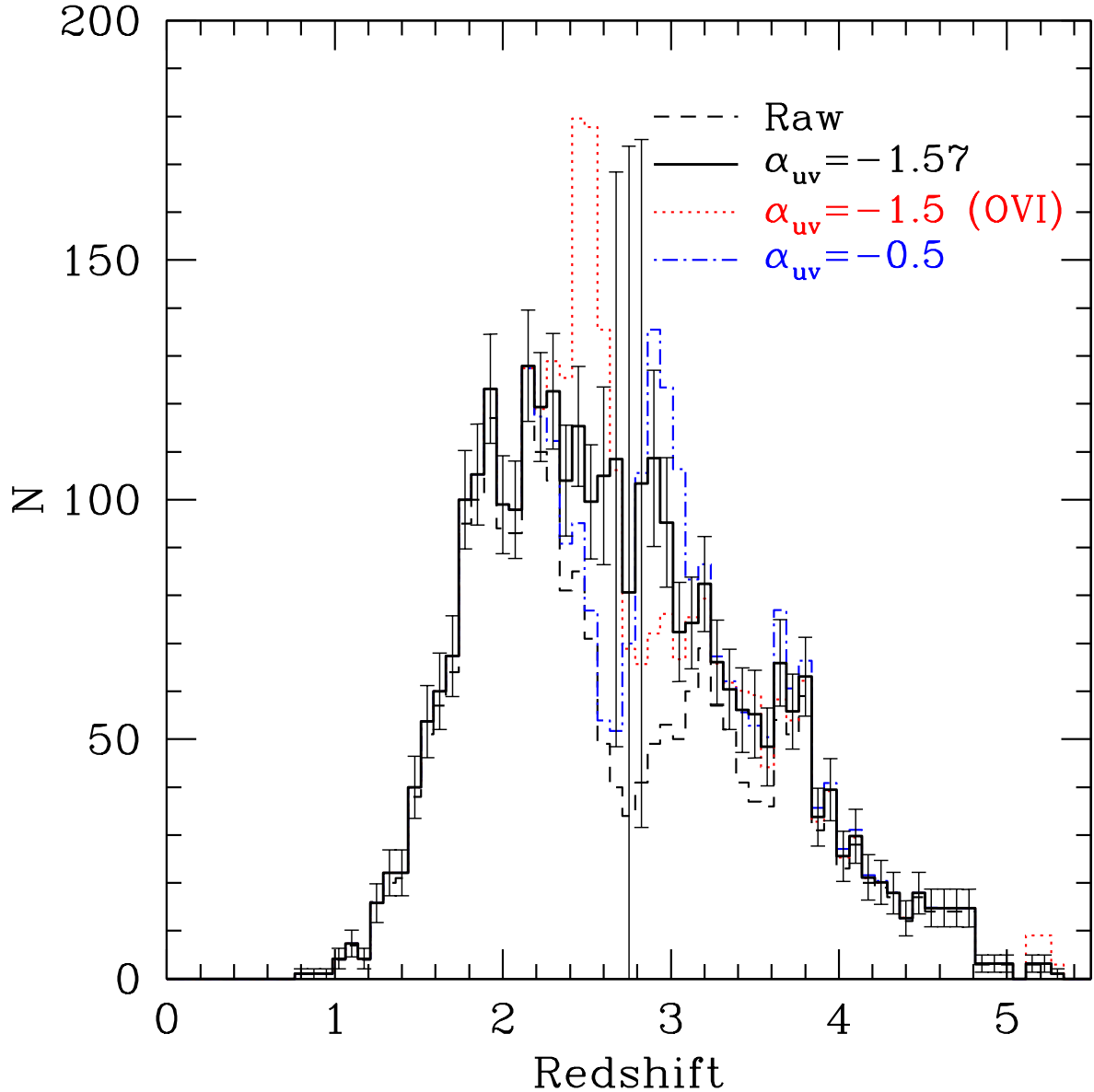}
\caption{Raw ({\em dashed line}) and corrected ({\em thick solid
line}) redshift histograms for $M_i<-27.6$.  This absolute magnitude
cut is the faintest for which the SDSS selection function is not
truncated by the apparent magnitude limits of the survey.  The dotted
red line and the dot-dash blue line show the corrected distribution
for two slightly different sets of simulated quasars (and the
resulting completeness corrections), see text for discussion.  The
corrected redshift distribution is much smoother than the observed
distribution, but is uncertain in the $z=2.2$--3 range.  The (Poisson)
error bars are conservative in the sense that they were determined
before imposing a floor of $0.333$ on the selection function.
\label{fig:fig9n}}
\end{figure}

More appropriate to the determination of the QLF is to examine the
redshift distribtution of an absolute magnitude portion of the sample.
We will see below that quasars with $M_i <-27.6$ (after applying a
non-standard $K$-correction, see \S~\ref{sec:kcorrect} below) fall
within our magnitude limits for $0.8 < z < 4.8$.  Thus
Figure~\ref{fig:fig9n} shows the raw ({\em dashed line}) and corrected
({\em thick solid line}) distribution of $M_i<-27.6$ quasars.  The
corrected histograms include the floor on the selection function of
0.333, which is applied throughout the rest of this work.  The error
bars in Figure~\ref{fig:fig9n} are derived from summing the squared
weights for each object ({\em without} the floor in the selection
function) in each bin; note the particularly large errors at $z
\approx 2.7$.

While the resulting redshift histogram is indeed quite smooth, we have
found that the results are quite sensitive to the details of the
simulations --- given the sharp gradients of the selection function
with redshift.  Small changes in the assumed UV spectral index
distribution in the simulated quasar spectra change the exact redshift
of the minimum of the selection function, producing dramatic changes
in the corrected redshift histogram.  Thus Figure~\ref{fig:fig9n} also
gives the corrected distribution for two other simulations to give the
reader an idea of the possible range of corrections.  The red dotted
line in Figure~\ref{fig:fig9n} shows a correction for a simulation
that also has a UV slope of $\sim-1.5$ (similar to \markcite{tzk+02}{Telfer} {et~al.} 2002),
but with the \ion{O}{6} equivalent width increased by 20\AA.  This
simulation produces a very strong, narrow peak near $z\sim2.5$ but
leaves a deficit near $z\sim2.8$.  Similarly, the blue dash-dot line
uses identical parameters as our principal simulation, but has a UV
spectral index of $-0.5$, following \markcite{skb+04}{Scott} {et~al.} (2004).  This simulation
produces a narrow peak near $z\sim3.0$, but leaves a hole at
$z\sim2.6$.  We will see in the next section that the set of
simulations that we have chosen (as illustrated by the solid black
line in Fig.~\ref{fig:fig9n}) is supported a posteriori by a
comparison between the radio and color selection.

\subsection{Radio Selection vs.\ Color Selection}
\label{sec:radiovscolor}

Quasars selected because of their radio properties provide an
independent probe of our quasar selection function.  These sources are
selected independent of their optical colors, and thus 100\% of such
objects (without fatal cosmetic defects) at, e.g., $z \approx 2.7$
with $i \le 19.1$ should be selected.  Thus the redshift dependence of
the ratio of color-selected to radio+color-selected quasars brighter
than $i=19.1$ can be used to check the selection function that we have
derived above.

\begin{figure}
\epsscale{1.0}
\plottwo{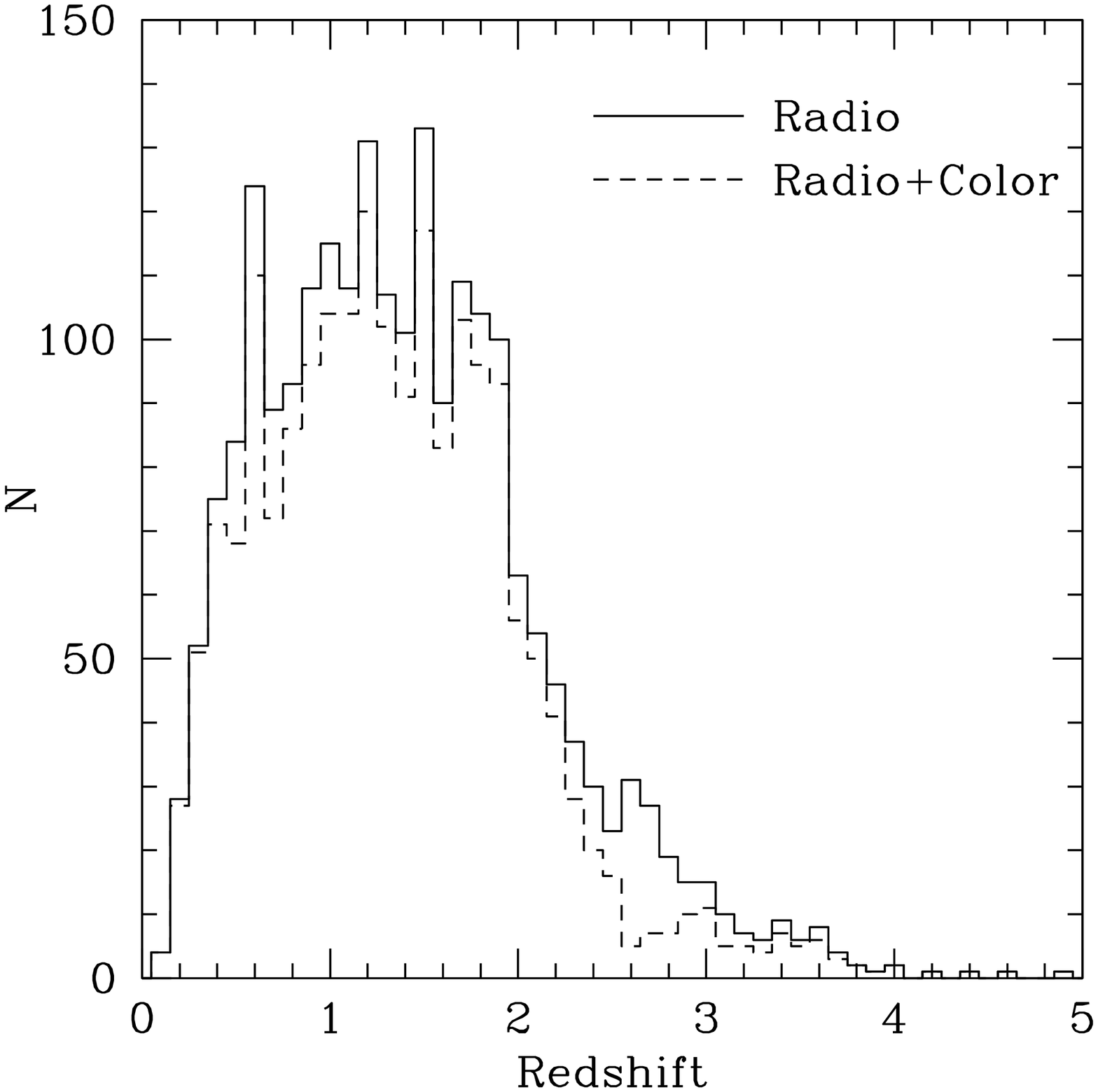}{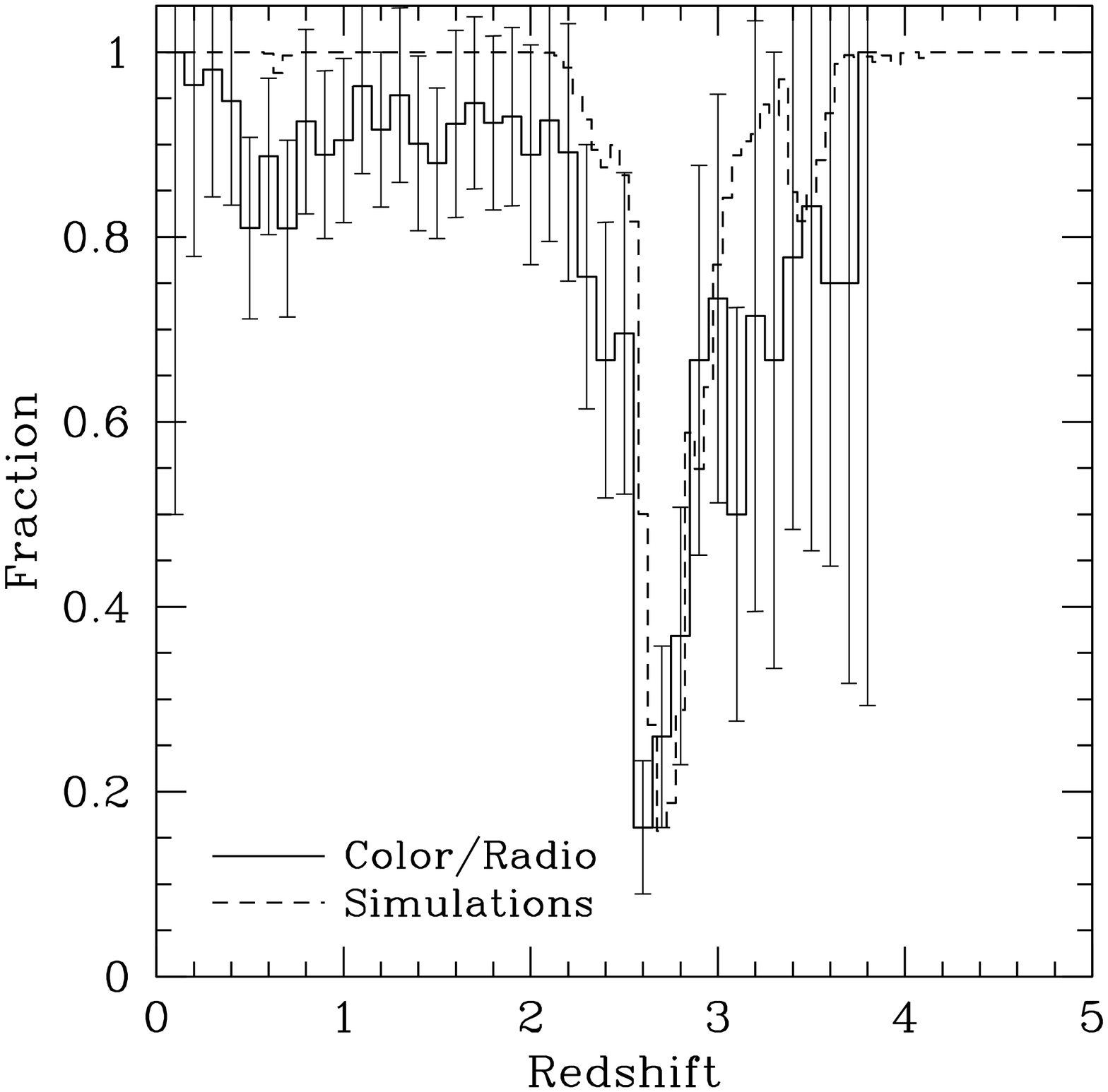}
\caption{{\em Left:} Redshift distribution of radio-selected ({\em
solid line}) and radio+color-selected ({\em dashed line}) unresolved
DR3 quasars.  Note the deficit of color-selected objects at $z\sim2.7$
where SDSS color selection is difficult.  {\em Right:} Color-selection
completeness as a function of redshift determined from the ratio of
color- to radio-selected $i \le 19.1$ quasars ({\em solid}) and
determined from the simulations ({\em dashed}).  The depth and
position of the dip at $z = 2.7$ are in reasonable agreement between
the two determinations.  The errorbars are Poisson.  The fractions
derived from the simulations have {\em not} been corrected for the 5\%
cosmetic defect incompleteness.
\label{fig:fig10}}
\end{figure}

The left-hand panel of Figure~\ref{fig:fig10} compares the redshift
distribution of radio-selected quasars to those that were {\em both}
radio and color selected (using the full DR3Q sample). The radio
selection is much smoother, and does not show deficits at $z = 2.7$
and $z = 3.5$.  The right-hand panel compares the ratio of these two
curves to the results of the simulations shown in
Figure~\ref{fig:fig6}; the agreement is good, especially in the
redshift of the minimum of the selection function, giving us
confidence that we have modeled the selection function reasonably.
The alternative selection functions have redshift minima which are
offset from that of the observed radio-selected ratio.

In principle, we could use the radio-selected quasars to determine the
selection function and drop the simulation-based selection function we
derived in \S~\ref{sec:completeness}.  However, we do not do so for
several reasons:
\begin{itemize} 
\item The radio sample is relatively small (2174 quasars), and in
particular, there are not enough radio-selected quasars at $z\gtrsim4$
to accurately determine our completeness at high redshift, or to use
smaller redshift bins than are shown in Figure~\ref{fig:fig10},
or to explore the selection function simultaneously in redshift and
magnitude.
\item The SDSS quasar selection algorithm only targets radio sources
(explicitly) as quasar candidates to $i=19.1$, whereas the color
selection goes to $i=20.2$ for $z>3$.
\item Several authors have suggested that radio-detected quasars are
systematically redder than are radio-quiet quasars --- even after
accounting for the ability of radio selection to uncover dust-reddened
quasars \markcite{imk+02}(e.g., {Ivezi{\' c}} {et~al.} 2002).
\item Finally, the radio and optical redshift distributions may be
intrinsically different.
\end{itemize}
Hence, we use the radio selection only to check our simulation-based
selection function determination a posteriori; we have found that the
selection function is reasonable.

\section{Number Counts}
\label{sec:number_counts}

\begin{figure}
\epsscale{1.0}
\plotone{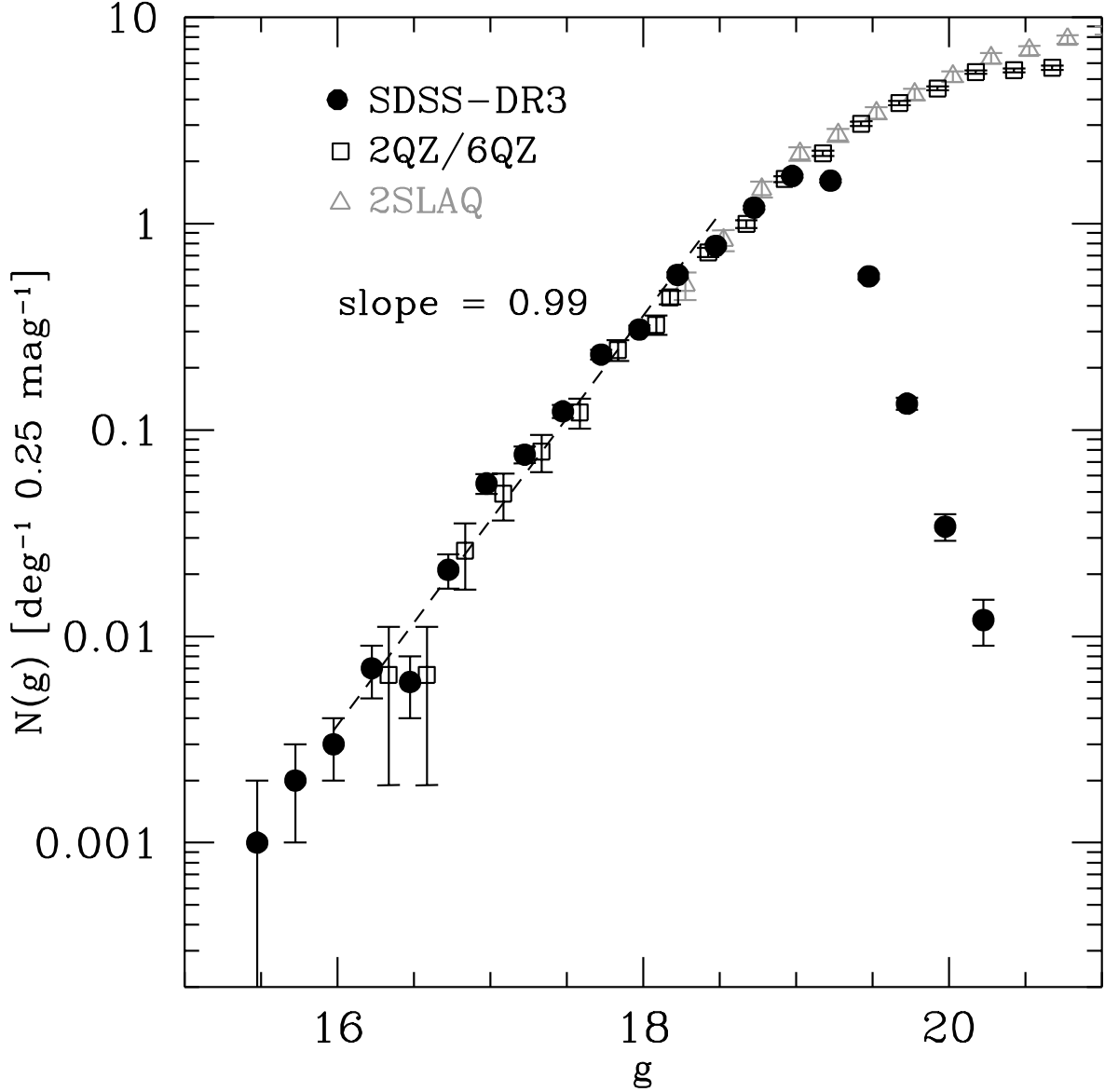}
\caption{Differential $g$-band number counts for quasars matching the
selection criteria of 2QZ ($0.4<z<2.1$; $M_g<-22.5$ with a
$K$-correction using a fixed $\alpha_\nu=-0.5$).  2QZ/6QZ data are
given by open squares (assuming $b_J\approx g$); 2SLAQ data are given
by open gray triangles.  The fall-off at faint magnitudes in the
SDSS-DR3 sample is due to the $i$-band limiting magnitude of the
survey.  Also shown is a power-law fit to the bright end of the
SDSS-DR3 sample; it has a slope of $0.99\pm0.12$.
\label{fig:fig11n}} 
\end{figure}

\begin{figure}
\epsscale{1.0}
\plotone{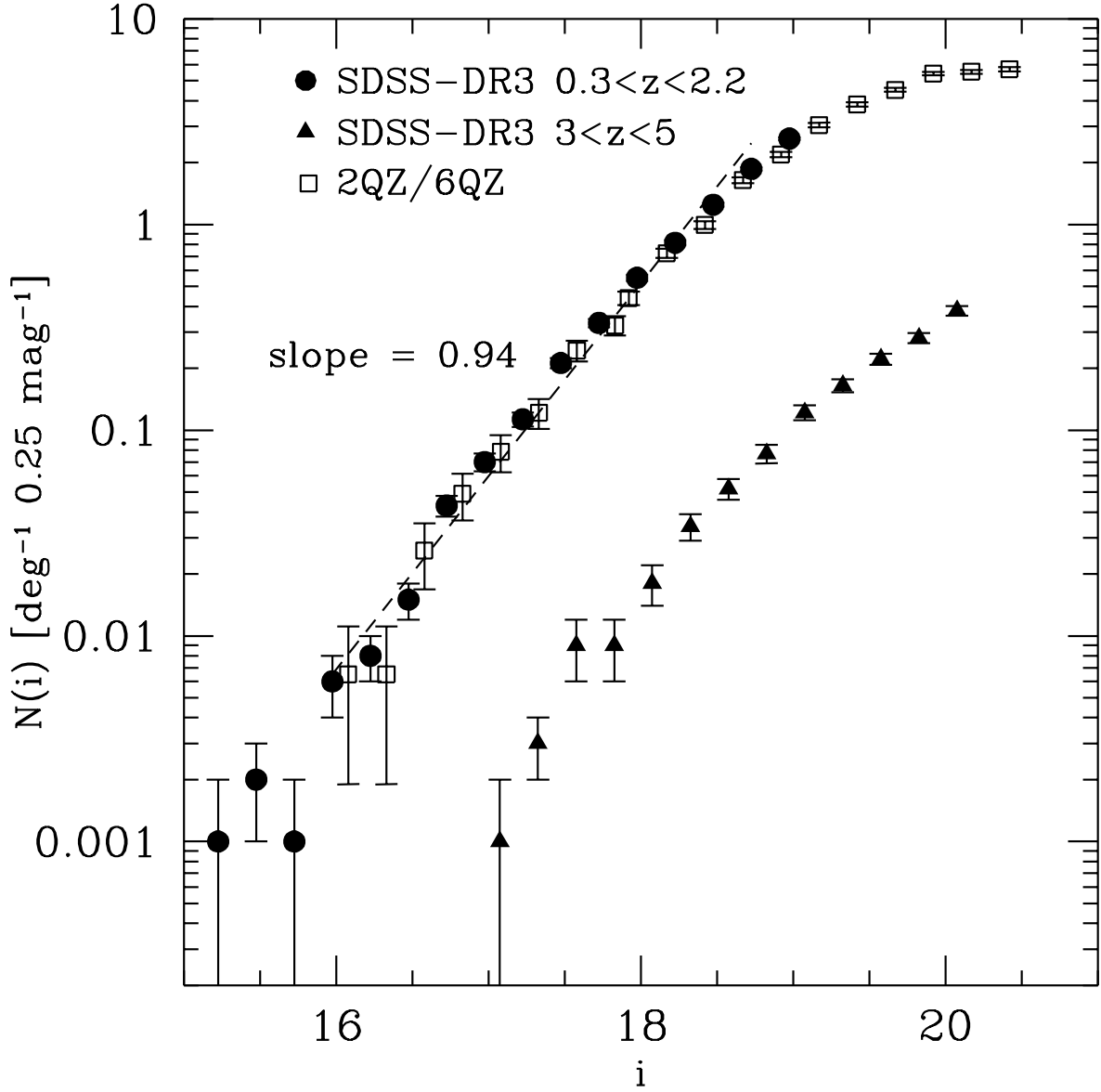}
\caption{Differential $i$-band number counts with $0.3<z<2.2$ ({\em
filled circles}) and $3<z<5$ ({\em filled triangles}) [both have
$M_i<-22.5$ $K$-corrected to $z = 0$ with a fixed $\alpha_\nu=-0.5$].
2QZ/6QZ data are given by open squares and have been converted to $i$
according to $i=g-0.255$.  A power-law fit over the range shown by the
dashed line has a slope of $0.94\pm0.09$.
\label{fig:fig12n}} 
\end{figure}

\begin{figure}
\epsscale{1.0}
\plotone{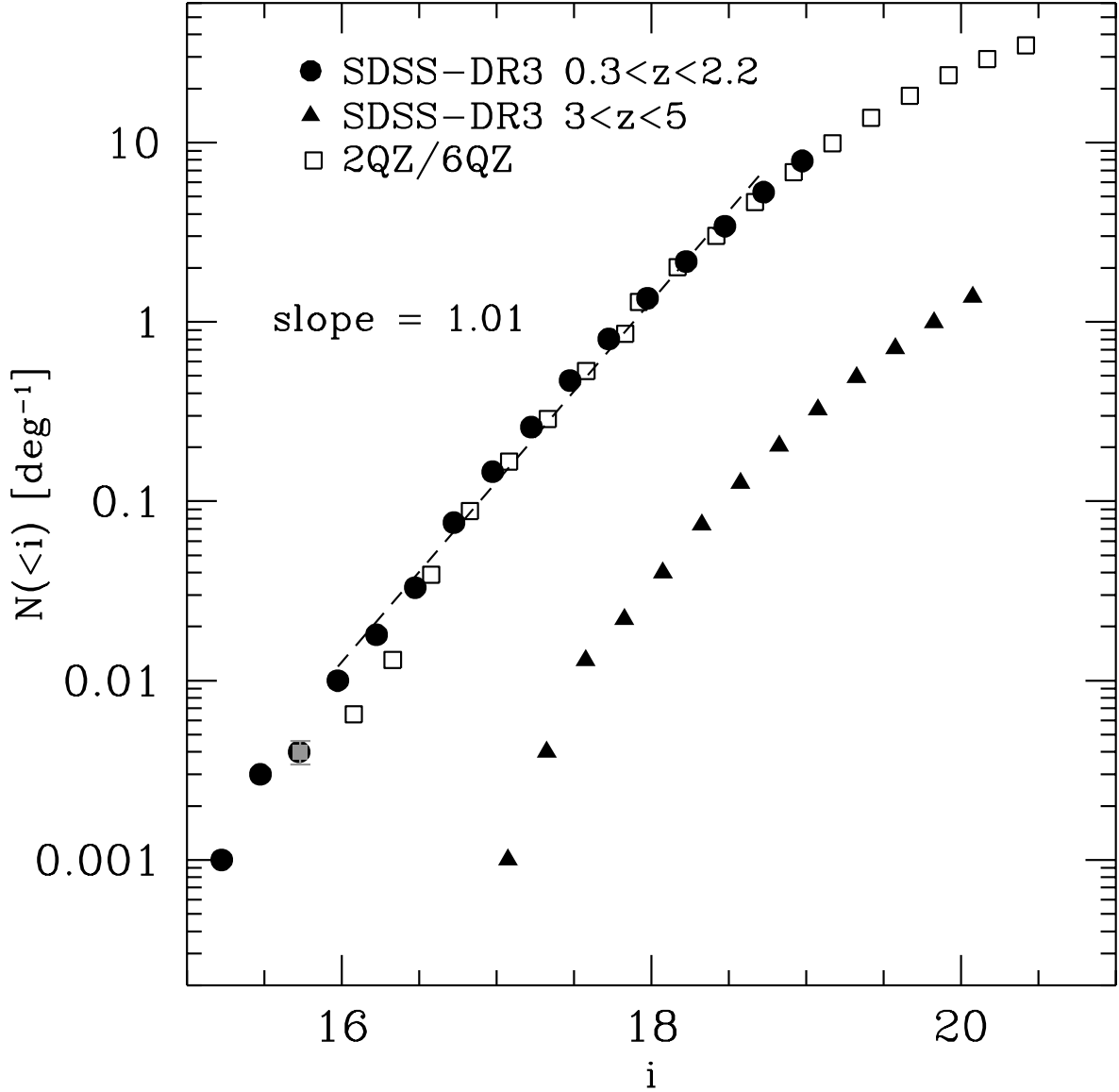}
\caption{Cumulative $i$-band number counts with $0.3<z<2.2$ ({\em
filled circles}) and $3<z<5$ ({\em filled triangles}) [both have
$M_i<-22.5$ and fixed $\alpha_\nu=-0.5$].  2QZ/6QZ data are given by
open squares and have been converted to $i$ according to $i=g-0.255$.
The gray point is the cumulative density of PG quasars \markcite{sg83}({Schmidt} \& {Green} 1983)
with $0.3<z<2.2$ and $M_i<-22.5$ assuming $i=B-0.14-0.287$.  A
power-law fit over the range shown by the dashed line has a slope of
$1.01\pm0.07$.  Since the errors are correlated, error bars are not
shown, but their approximate size can be determined from
Figure~\ref{fig:fig12n}.
\label{fig:fig13n}} 
\end{figure}

The differential number counts distribution for our statistical sample
of quasars is shown in Figure~\ref{fig:fig11n}.  We use the $g$-band,
limited to $0.4<z<2.1$ and $M_g<-22.5$ ($\alpha_{\nu}=-0.5$) in order
to mimic the final 2QZ/6QZ sample (open squares; \markcite{csb+04}{Croom} {et~al.} 2004) and
the 2SLAQ sample (open triangles; \markcite{rca+05}{Richards} {et~al.} 2005).  At the bright
end ($16<g<18.5$) we find a slope of $0.99\pm0.12$.
Figure~\ref{fig:fig12n} presents a similar analysis in the $i$-band
for $0.3<z<2.2$ and $M_i<-22.5$ (using $M_i[z=0]$ --- see next section
--- and $\alpha_{\nu}=-0.5$) and also for $3<z<5$.  In the range
$16<i<19$, the $0.3<z<2.2$ slope is $0.94\pm0.09$.  The cumulative
$i$-band number counts are shown in Figure~\ref{fig:fig13n}.  A
least-squares fit between $i=16$ and $i=19$ yields a slope of
$1.01\pm0.07$.\footnote{Note the discrepancy with the cumulative
number counts analysis by \markcite{vsr+05}{Vanden Berk} {et~al.} (2005), where the counts of bright
quasars appear to be overestimated.}  The counts have been corrected
for cosmetic defect incompleteness and for the redshift- and
magnitude-dependent color-selection incompleteness as discussed above.
The number counts are also given in tabular form for $0.3<z<2.2$ in
Table~\ref{tab:tab02} and $3<z<5$ in Table~\ref{tab:tab03}.  $N(i)$ is
the number of quasars per 0.25 mag per square degree.  $N(<i)$ is the
number of quasars per square degree brighter than magnitude $i$.
$N_Q$ is the number of observed quasars in each 0.25 mag bin.
$N_{Q\; \rm cor}$ is the number of quasars after correcting for selection
effects.  For $z<2.2$ those corrections are negligible.

The cumulative number counts shown in Figure~\ref{fig:fig13n} include
a data point derived from the 114 quasars found by the Palomar-Green
(PG) Bright Quasar Survey \markcite{sg83}(BQS; {Schmidt} \& {Green} 1983) over an area of
10,714~deg$^2$.  We convert the BQS $B$ Vega magnitudes to $i$-band AB
magnitudes as $i = B - 0.14 -0.287$, where the first correction term
shifts from the Vega to the AB system, and the second corrects for the
different effective wavelengths of the two filters (7470\,\AA\ and
4400\,\AA), assuming an $\alpha_{\nu}=-0.5$ power law.  The PG data
point agrees well with our SDSS number counts relationship, seemingly
in contrast with previous claims of incompleteness in the PG survey
\markcite{wp85,gmf+92,wcb+00}({Wampler} \& {Ponz} 1985; {Goldschmidt} {et~al.} 1992; {Wisotzki} {et~al.} 2000).  However, both the SDSS and PG points at
this magnitude fall below an extrapolation from fainter data points.
A combination of incompleteness due to the large photometric errors in
the BQS photometry \markcite{jsr+05}({Jester} {et~al.} 2005) and complicated Eddington bias
corrections due to the steep local slope of the number counts (as
compared to the global average slope) may reconcile the reported
incompleteness of the PG sample with the agreement with our number
counts \markcite{jsr+05}({Jester} {et~al.} 2005).

\section{$K$-Corrections and the Calculation of Luminosities}
\label{sec:kcorrect}

In order to compare luminosity functions at different redshifts, we
must correct our photometry for the effects of redshift on the portion
of the spectrum sampled by a given filter.  We will use {\em
continuum} luminosities throughout as a measure of the energy output
of the central engine, subtracting the contribution of emission lines
to the observed flux.  This section describes the determination of the
$K$-correction, which brings the observed magnitudes to a
common effective rest-frame bandpass.

The sign convention of the $K$-correction, $K(z)$, is defined by
\markcite{os68}{Oke} \& {Sandage} (1968) as $m_{\rm intrinsic} = m_{\rm observed} - K(z)$.  
The $K$-correction itself depends on the object's SED and is given by
Equation~4 in \markcite{os68}{Oke} \& {Sandage} (1968) or equivalently by Equation~8 in
\markcite{hbb+02}{Hogg} {et~al.} (2002).

While the $K$-correction depends on the overall quasar SED, we will
find it useful to consider the component due to the continuum ($K_{\rm
cont}$) and emission lines ($K_{\rm em}$) separately.  This
distinction will allow us to quantify the sensitivity of the
$K$-correction on the continuum slope, $\alpha_\nu$, and to correct
luminosities for the contribution of emission lines.  In particular,
the $K$-correction to $z=0$ for a power-law continuum is given by
$K_{\rm cont} = -2.5(1+\alpha_\nu)\log_{10}(1+z)$, where the first
term corrects for the effective narrowing of the filter width with
redshift.  As emission lines make quasars appear brighter relative to
the continuum, $K_{\rm em}$ will be used to {\em subtract} the
emission-line contribution from the observed fluxes.

\begin{figure}
\epsscale{1.0}
\plotone{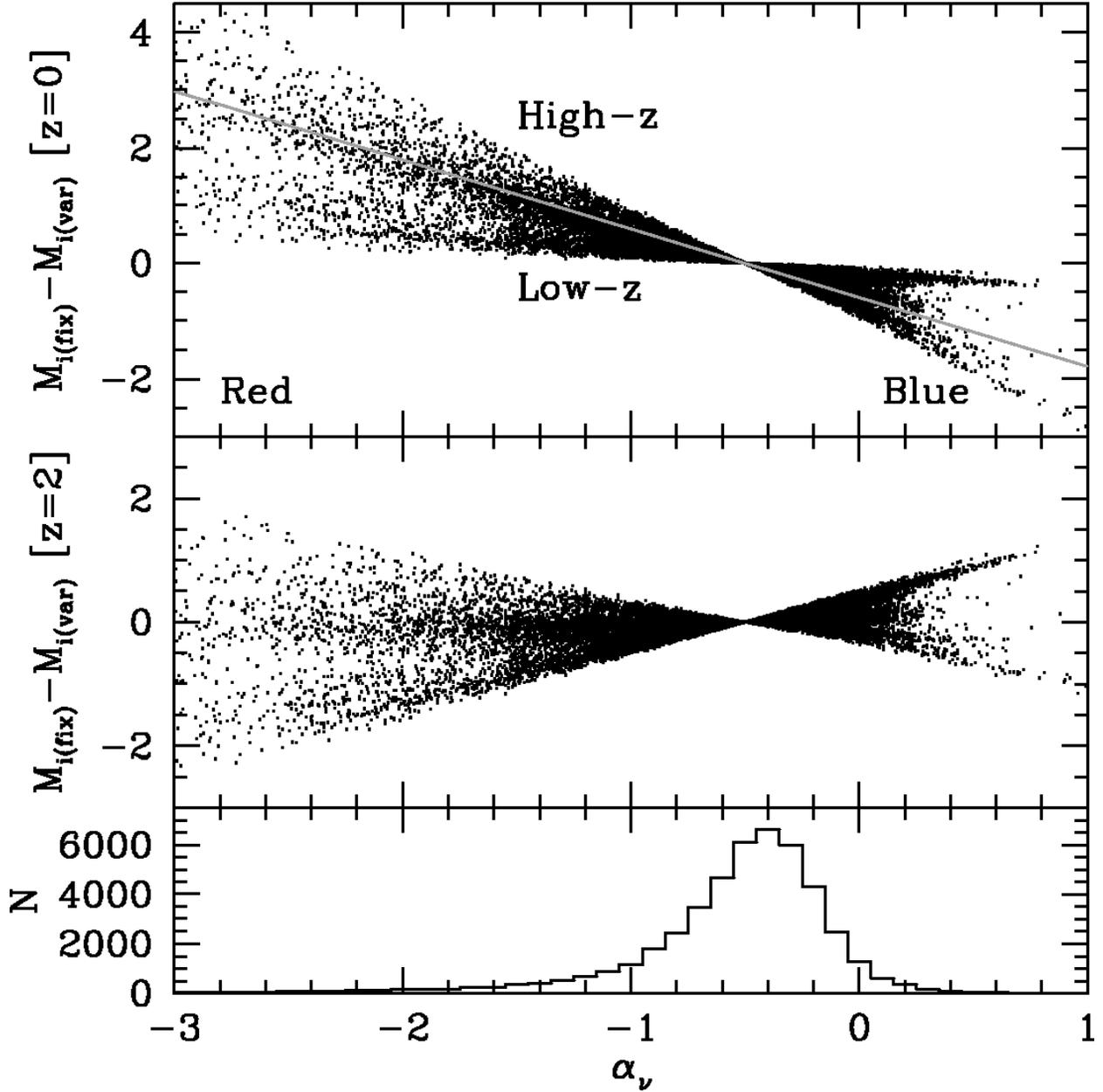}
\caption{Comparison of DR3Q $K$-corrected absolute $i$ magnitudes
computed using both a fixed and a photometrically-derived spectral
index.  The top panel gives the difference for a $K$-correction
normalized to $z=0$.  Note that the bluest and reddest objects at high
redshift incur significant errors when using a fixed spectral index
for all objects.  Moving the zero-point of the $K$-correction to $z=2$
rotates these points about the $z=2$ line (shown in gray in the top
panel) as can be seen in the middle panel, significantly reducing the
systematic error incurred by extrapolating the wrong spectral index to
high redshift.  Note that these corrections are large only for objects
whose spectral indices deviate significantly from the assumed spectral
index of $\alpha_{\nu}=-0.5$; the distribution is shown in the bottom
panel.
\label{fig:fig14n}}
\end{figure}

$K$-corrections are traditionally defined relative to redshift zero;
that is, the photometry of all quasars are referenced to a bandpass
measuring rest-frame optical light.  However, low-redshift quasars are
very rare and for the vast majority of the quasars in our sample the
$K$-correction has to extrapolate the observed SED far into the
observed infrared.  This is problematic, especially because there is a
wide range of continuum slopes of quasars.  This is illustrated in the
top panel of Figure~\ref{fig:fig14n}, which shows the difference
between the $K_{\rm cont}$ for quasars in our sample, using the
canonical $\alpha_\nu = -0.5$, and that determined from the observed
relative $g-i$ color (see the discussion by \markcite{rhv+03}{Richards} {et~al.} 2003):
$\alpha_\nu = -0.5 - \Delta(g-i)/0.508$.\footnote{Quasars with
power-law continua of $\alpha_\nu = 0$ and $\alpha_\nu = -1$ have
$\Delta(g-i)$ that differ by 0.508.}  At redshift 5, the difference
in luminosity between the bluest and reddest objects is more than a
factor of 100!

Therefore, in this paper, we follow \markcite{wis00}{Wisotzki} (2000) and \markcite{bhb+03}{Blanton} {et~al.} (2003a)
and $K$-correct to a redshift closer to the median redshift of our
sample.  In most of what follows we will use a continuum
$K$-correction calculated for an $\alpha_\nu = -0.5$ power law.  Our
determination of absolute magnitude will be pegged to that of the $i$
band for a quasar at $z = 2$, i.e., with an effective rest wavelength
of $\sim2500$\AA\ (but we will need to account for the changing size of
the bandpass at $z=2$ in order to determine the $2500$\AA\ luminosity,
see Eq.~4).  This is a redshift close to that of the peak of luminous
quasar activity.  2500\AA\ is also the canonical wavelength used to
determine the spectral index between optical and X-rays in quasars
\markcite{tab+79,sbs+05}({Tananbaum} {et~al.} 1979; {Strateva} {et~al.} 2005).  $K$-correcting to closer to the median
redshift of the sample significantly reduces the systematic error
incurred by assuming a constant $\alpha_\nu = -0.5$, as illustrated in
the middle panel of Figure~\ref{fig:fig14n}.

\begin{figure}
\epsscale{1.0} 
\plotone{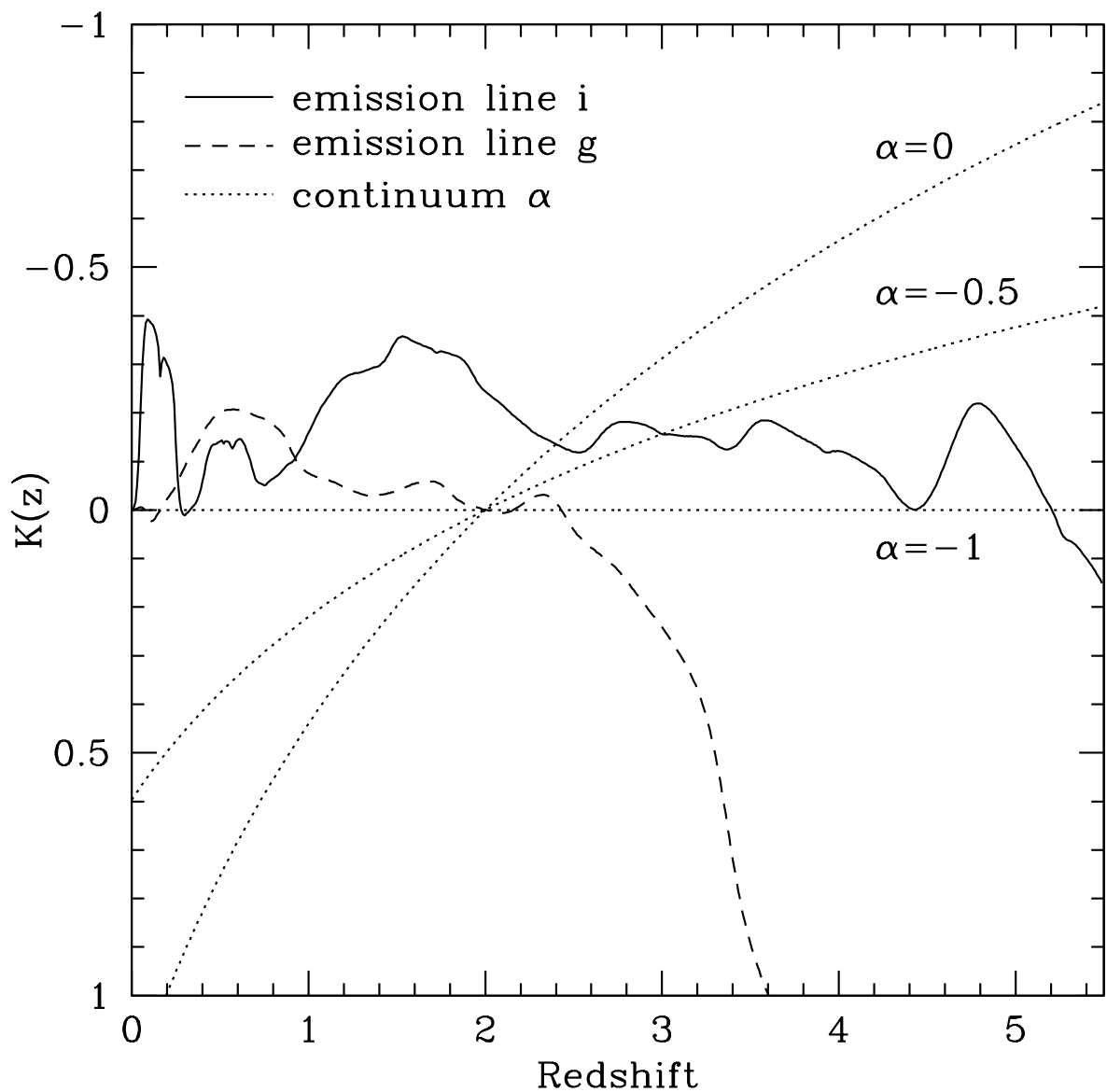}
\caption{Continuum ({\em dotted}) and emission line ($i$: {\em solid};
$g$: {\em dashed}) $K$-corrections.  The continuum $K$-corrections are
zero at $z=2$ by definition.  For comparison, the $i$-band
$K$-correction to $z=0$ for $\alpha_{\nu}=-0.5$ would be more negative
by 0.596 mag.
\label{fig:fig15n}}
\end{figure}

For the emission-line $K$-correction we proceed as follows.  We first
construct a variance weighted mean quasar spectrum from the 16,713 DR1
quasars from \markcite{sfh+03}{Schneider} {et~al.} (2003) using the algorithm of \markcite{vrb+01}{Vanden Berk} {et~al.} (2001).
This spectrum is well-fit from Ly$\alpha$ to H$\beta$ by a power law
with $\alpha_\nu=-0.436$, in good agreement with the \markcite{vrb+01}{Vanden Berk} {et~al.} (2001)
composite spectrum.  This spectrum shows an appreciably flatter slope
longwards of H$\beta$ due to contamination from stellar light from
quasar hosts at low redshift; this is subtracted off (essentially
extrapolating the $\alpha_\nu$ power law to longer wavelengths).
Next, the continuum is subtracted from the spectrum and the resulting
emission line spectrum is convolved with the SDSS filter curves to
create an emission-line-only $K$-correction, $K_{\rm em}$, for each of
the filters.  Figure~\ref{fig:fig15n} shows $K_{\rm em}$ for both the
$g$ and $i$ passbands along with $K_{\rm cont}$ for three choices of
spectral index.  For $z\gtrsim2.5$ $K_{\rm em}$ for the $g$-band is
essentially meaningless as a result of contamination of the continuum
by Ly$\alpha$ forest absorption and our uncertainty of the intrinsic
continuum shape at wavelengths shorter than Ly$\alpha$ emission (see
the discussion in \S~\ref{sec:sim}).  The Ly$\alpha$ forest does not
enter the $i$ band until much higher redshift, which is part of the
motivation for setting the SDSS quasar selection flux limit in the $i$
band.  Table~\ref{tab:tab04} gives our $z=2$ normalized $i$-band
$K$-correction vector as a function of redshift for a power-law
continuum with $\alpha_{\nu}=-0.5$ and the emission line
$K$-correction as discussed above.  Comparison of the gray and black
curves in Figure~\ref{fig:fig8n} demonstrates that the emission line
component of the K-corrections has a significant impact upon the true
magnitude limit of the sample.

Note that we have {\em not} corrected the photometry for the presence
of host galaxies.  At most redshifts, we restrict ourselves to
absolute magnitudes that are several magnitudes brighter than $L^*$
for galaxies, thus the contribution of host galaxies is likely to be
small.  We plan to explore this issue in more detail in the future
using information on the stellar component from the spectra, when we
examine the continuity between quasars and Seyferts at low redshifts.

Our usage of a $z=2$ K-corrected $i$-band magnitude is non-standard
and it is useful to have a conversion between this and more commonly
used magnitudes.  First, the conversion between $M_i(z=0)$ and
$M_i(z=2)$ is given by
\begin{equation}
M_i(z=0) = M_i(z=2) + 2.5(1+\alpha_{\nu})\log(1+2) = M_i(z=2) + 0.596.
\end{equation}
For reference, $M_i(z=2)=-27.2$ for 3C~273.  Then to convert from $i$
to $g$ we must account for both the slope of the spectrum between $i$
and $g$ and the average emission line flux in the $g$-band as follows:
\begin{equation}
M_g(z=0) = M_i(z=0) + 2.5\alpha_{\nu}\log\left(\frac{4670\,\rm
  \AA}{7471\,\rm \AA}\right) - 0.187 = M_i(z=0) + 0.255 - 0.187,
\end{equation}
where $\alpha_{\nu}=-0.5$ makes quasars 0.255 mag fainter in g than i
and emission lines contribute an average of 0.187 mag over $0.3<z<2.2$
(including 0.114 mag of flux from emission lines at $z=0$).
Vega-based photometric systems require an additional correction term:
$B-g$ is $\sim0.14$ \markcite{fig+96}({Fukugita} {et~al.} 1996).  \markcite{rca+05}{Richards} {et~al.} (2005) empirically find
that $g-b_J \sim -0.045$, thus theoretically we expect
$M_{b_J}-M_i(z=2)$ to be $\sim0.71$.  Empirically, we determine
$M_{b_J}-M_i(z=2)=0.66\pm0.31$ from a sample of 1046 quasars detected
by both SDSS and 2QZ with $0.3<z<2.2$ and $i<19.1$.  Finally, assuming
a power-law spectral index, the conversion between a monochromatic
1450\AA\ absolute magnitude and $M_i(z=2)$ is
\begin{equation}
M_{1450} = M_i(z=2) + 0.596 + 2.5\alpha_{\nu}\log\left(\frac{1450\,\rm
  \AA}{7471\,\rm \AA}\right) = M_i(z=2) + 1.486.
\end{equation}

\begin{figure}
\epsscale{1.0} 
\plotone{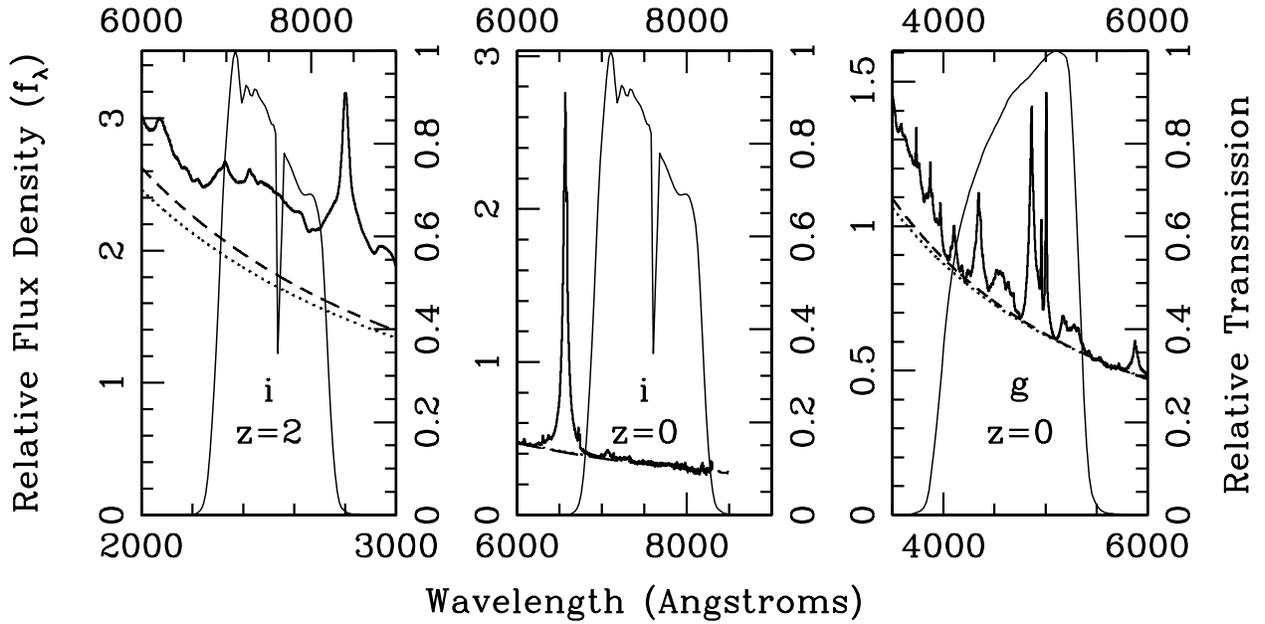}
\caption{Composite quasar spectrum for three different bandpasses: $i$
at $z=2$, $i$ at $z=0$, and $g$ at $z=0$.  The composite spectrum is
shown as the thick solid line (relative flux given on the left axes),
the filter curves are the thin solid lines (relative transmission
given on the right axes).  The bottom axes are rest wavelength, the
top axes are observed wavelength.  Dashed and dotted lines are for
$\alpha_{\nu}=-0.436$ and $\alpha_{\nu}=-0.5$, respectively.  Our
$K$-corrected absolute magnitudes are defined using the $z=2$
$i$-bandpass shown in the left-hand panel --- after excluding the
emission line component above the $\alpha_{\nu}=-0.436$ continuum.
\label{fig:fig16n}}
\end{figure}

Figure~\ref{fig:fig16n} illustrates the difference between our
$K$-corrections and a more standard $K$-correction in terms of the
quasar SED.  The left-hand panel shows the composite quasar spectrum
discussed above at $z=2$ with the $i$-band filter curve overplotted.
Our $M_i(z=2)$ absolute magnitude is defined using this bandpass,
excluding the mean emission line component (above the continuum level
shown by the dashed line).  The middle panel shows the composite
spectrum at $z=0$ relative to the $i$ bandpass; there is essentially
no emission line flux, thus the emission line component of the
$K$-correction is naturally zero.  The right-hand panel shows a more
traditional bandpass for absolute magnitudes, the $g$ bandpass at
$z=0$.  Traditional systems include the emission line component in the
absolute magnitude definition by {\em defining} the $z=0$
$K$-correction to be zero.

Finally, to assist in converting between magnitudes and luminosity, we
give the conversion from $M_i(z=2)$ to 2500\,\AA\ luminosity density
in cgs units ($\rm erg\,s^{-1}\,Hz^{-1}$) following \markcite{og83}{Oke} \& {Gunn} (1983):
\begin{equation}
\log\left({L_{2500\,{\rm \AA}} \over 4\,\pi d^2}\right) =
-0.4[M_i(z=2) + 48.60 + 2.5\log(1+2)]
\end{equation}
where $d = 10\,\rm pc = 3.08 \times 10^{19}\,cm$ and the last term on
the right hand side corrects for the size of the $z=2$ bandpass
relative to a $z=0$ bandpass (and is needed to convert $M_i(z=2)$ to
physical units in the rest frame).  Our correction for $K_{\rm em}$
means that, on average, this luminosity measures the continuum only
and is roughly a nuclear luminosity.  In particular, it excludes the
{\em average} contribution from the Balmer continuum and \ion{Fe}{2}
complexes to the 2500\,\AA\ bandpass.

\section{Luminosity Function}
\label{sec:lumfunct}

We compute the quasar luminosity function by two methods: by binning
the quasars in redshift and luminosity and by using a
maximum-likelihood (ML) fit to a parameterized form.  The input for
both of these QLFs is the homogeneous statistical sample of $15,343$
quasars drawn from 1622 deg$^2$.  These objects are given in
Table~\ref{tab:tab5}, which includes the object name, redshift,
$i$-band magnitude (dereddened), $M_i(z=2)$, the relative color
$\Delta(g-i)$, and the value of the selection function (Cor).

\subsection{The Binned QLF}

\begin{figure}
\epsscale{1.0}
\plotone{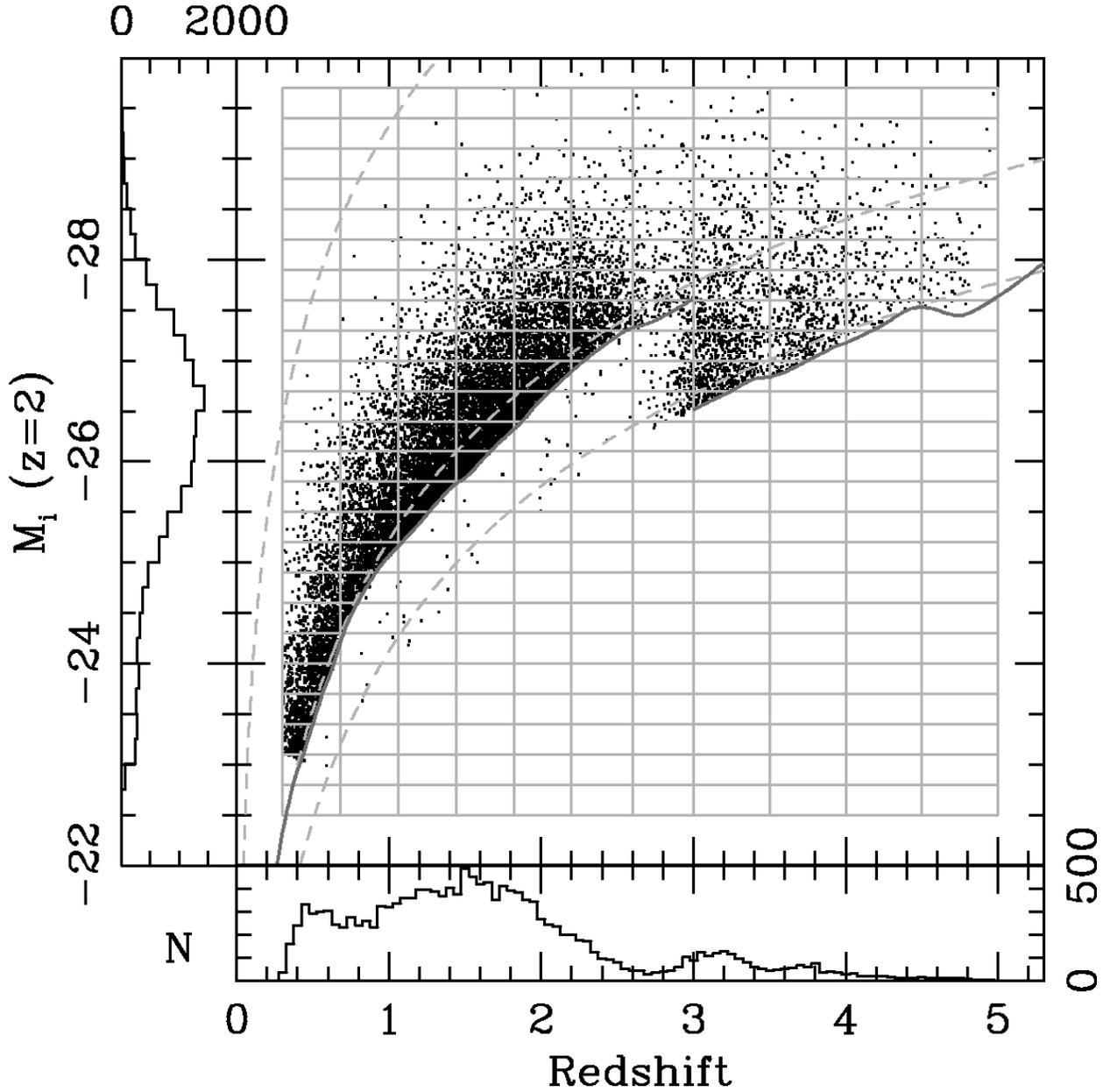}
\caption{Absolute magnitude, $M_i(z=2)$, of the complete sample versus
redshift.  The solid light gray lines show the bins that are used in
computing the (binned) luminosity function.  Dashed light gray curves
show the $i=15.0$, $i=19.1$, and $i=20.2$ magnitude limits of the
survey (without emission line $K$-corrections).  The difference
between the dashed light gray lines and the solid dark gray lines
shows the effect of the emission line $K$-correction.  $z\lesssim3$
quasars with $i>19.1$ were selected by the high-$z$ ($griz$) branch of
the algorithm and clearly do not represent a complete sample; they are
not used in the determination of the QLF.  The bottom and side panels
show the marginal distributions in redshift and absolute magnitude,
respectively.
\label{fig:fig17n}} 
\end{figure}

Figure~\ref{fig:fig17n} plots absolute magnitude as a function of
redshift for our sample, and also shows the bins in which the
luminosity function will be calculated. The edges of the redshift bins
are 0.30, 0.68, 1.06, 1.44, 1.82, 2.20, 2.6, 3.0, 3.5, 4.0, 4.5, and
5.0.  The $M_i$ bins start at $-22.5$ and are in increments of 0.3
mag.  The dashed light gray curves are the limiting apparent
magnitudes of the survey expressed as a luminosity as a function of
redshift, while the solid dark gray curves show the effect of
subtracting the emission-line component, $K_{\rm em}(z)$.

\begin{figure}
\epsscale{1.0}
\plotone{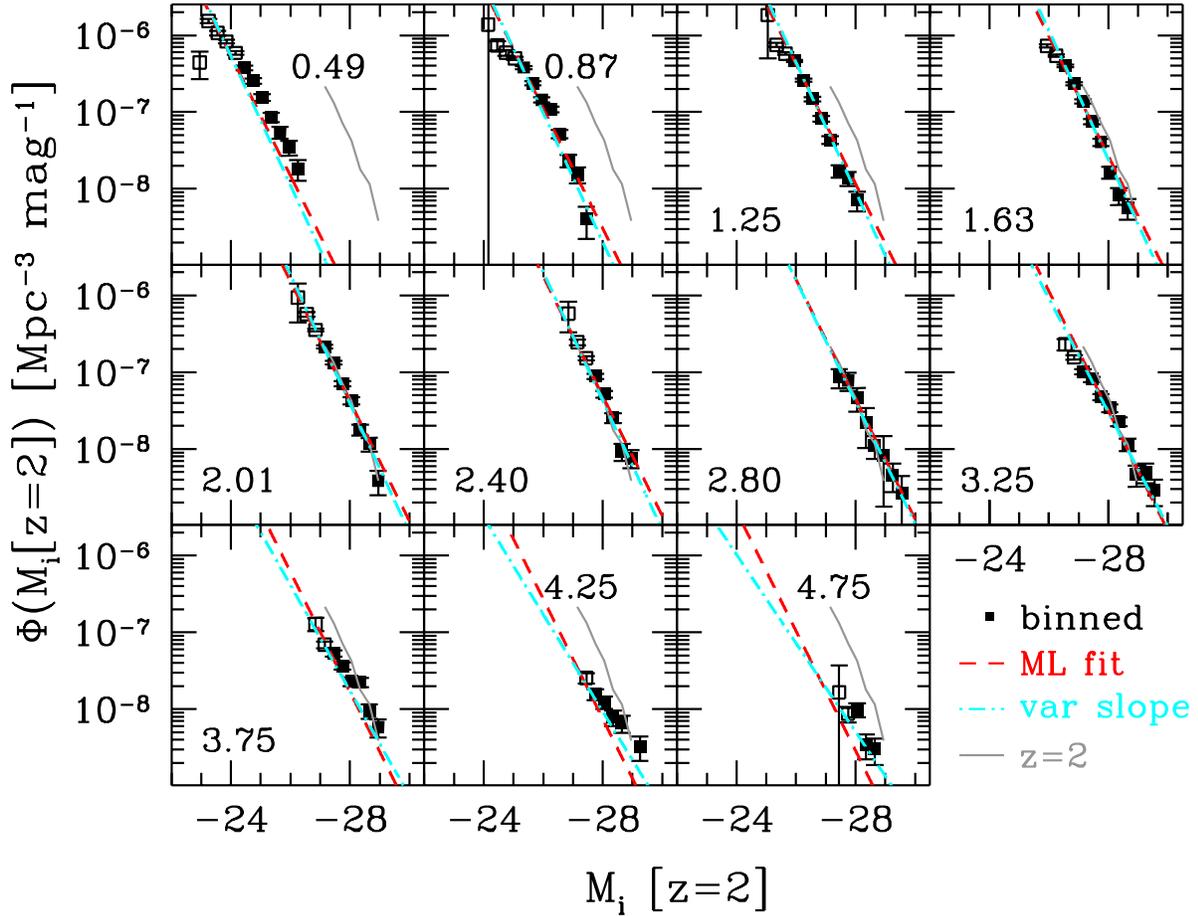}
\caption{$M_i(z=2)$ luminosity function (roughly a 2500\AA\
luminosity, see Eq.~4).  The redshift of each slice is given in the
upper right hand or lower left hand corner of each panel.  The points
show the binned luminosity function using a $1/V_{max}$ method; open
points are incomplete bins.  The $z=2.01$ curve ({\em gray}) is
reproduced in each panel for the sake of comparison.  The red dashed
line is our best fit maximum likelihood parameterization assuming a
constant slope with redshift, while the cyan dot-dashed line allows
for a slope change at high redshift.  Corrections for cosmetic defects
and color selection as a function of redshift and magnitude have all
been applied.  Note that there is almost no overlap in absolute
magnitude between the highest- and lowest-redshift bins.
\label{fig:fig18n}} 
\end{figure}

To compute the binned QLF, we use the \markcite{pc00}{Page} \& {Carrera} (2000) implementation of
the $1/V_{\rm a}$ method \markcite{sch68,ab80}({Schmidt} 1968; {Avni} \& {Bahcall} 1980) to correct for bins which
intersect the apparent magnitude limits (i.e., incomplete bins).  The
resulting $i$-band QLF is shown by the black points in
Figure~\ref{fig:fig18n}; the error bars are given by Poisson
statistics.  The binned QLF is also given in Table~\ref{tab:tab6},
which lists the redshift and absolute magnitude, the log of the space
density ($\Phi$) in Mpc$^{-3}$ mag$^{-1}$, the error in $\Phi$, an
indicator of whether or not the bin is filled, the mean redshift of
the quasars in the bin, the number of quasars in the bin and the
weighted number of quasars in the bin.  Filled points in
Figure~\ref{fig:fig18n} represent complete bins (i.e., those that lie
completely above the completeness limit shown in the
Fig.~\ref{fig:fig17n}), whereas open points are those bins for which we
have applied a correction for incomplete coverage of the bin.
Corrections for cosmetic defects and color selection as a function of
redshift and magnitude have been applied.

\subsection{Choosing a Maximum Likelihood Form}

We have also determined the luminosity function as derived from a
maximum likelihood analysis which requires no binning.  The likelihood
function is calculated using 
Equation 22 of \markcite{fss+01}{Fan} {et~al.} (2001) (see also \markcite{mar85}{Marshall} 1985), and is 
maximized using Powell's method
\markcite{ptv+92}({Press} {et~al.} 1992).  The maximum likelihood solution is shown by the red
dashed line in Figure~\ref{fig:fig18n}; our choice of parameterization
is discussed below.

The quasar luminosity function is often parameterized by a standard
double power-law form \markcite{pei95,pet97,bsc+00,csb+04}(e.g., {Pei} 1995; {Peterson} 1997; {Boyle} {et~al.} 2000; {Croom} {et~al.} 2004):
\begin{equation}
\Phi(M,z) = \frac{\Phi(M^*)}{10^{0.4(\alpha+1)(M-M^*)} + 10^{0.4(\beta+1)(M-M^*)}}.
\end{equation}
where $\Phi(M,z)dM$ is the number of quasars per unit comoving volume
at redshift, $z$, with absolute magnitudes between $M-dM/2$ and
$M+dM/2$.  This is the standard form for deep quasar surveys, which
generally find a flatter slope fainter than some characteristic
luminosity.  However, the SDSS quasar survey, while covering a very
large area of sky, is actually quite shallow.  The limiting magnitude
is such that, at most redshifts, the SDSS does not observe objects
fainter than the ``break'' characteristic luminosity and a double
power-law form is not justified.  While there is some curvature in the
shape of the QLF and our low redshift data would be better fit by a
double power law form, overall the luminosity coverage is not broad
enough to justify the added parameter.

\markcite{bsc+00}{Boyle} {et~al.} (2000) and \markcite{csb+04}{Croom} {et~al.} (2004) use the 2QZ data to show that the
redshift evolution of luminous $z\lesssim2.2$ quasars can be
parameterized by pure luminosity evolution (PLE), whereby $M^*$ is
either a quadratic in redshift or an exponential function of look-back
time.  However, neither of these forms is appropriate for a QLF that
extends to higher redshifts.  The space density is seen to fall for
$z>2.5$ \markcite{osm82,ssg95,kdd+95,fss+01}({Osmer} 1982; {Schmidt} {et~al.} 1995; {Kennefick} {et~al.} 1995; {Fan} {et~al.} 2001) while the exponential form
rises with $z$.  The quadratic form properly falls with $z$, but it
assumes a fall that is symmetric with the rise from $z=0$ to $z\sim2$,
while the data exhibit a less steep decline in redshift (the decline
is {\em much} steeper for $z>2$ when considering look-back time).  As
such we cannot use the traditional PLE parameterizations.

Thus a hybrid form is required.  Indeed, X-ray surveys have recently
begun to use a luminosity dependent density evolution (LDDE)
parameterization to describe the X-ray QLF \markcite{sg83,Ueda03}({Schmidt} \& {Green} 1983; {Ueda} {et~al.} 2003).  Such
a parameterization allows the redshift of the peak quasar density to
change as a function of luminosity (as seems to be required by the
X-ray data).

\begin{figure}
\epsscale{1.0}
\plotone{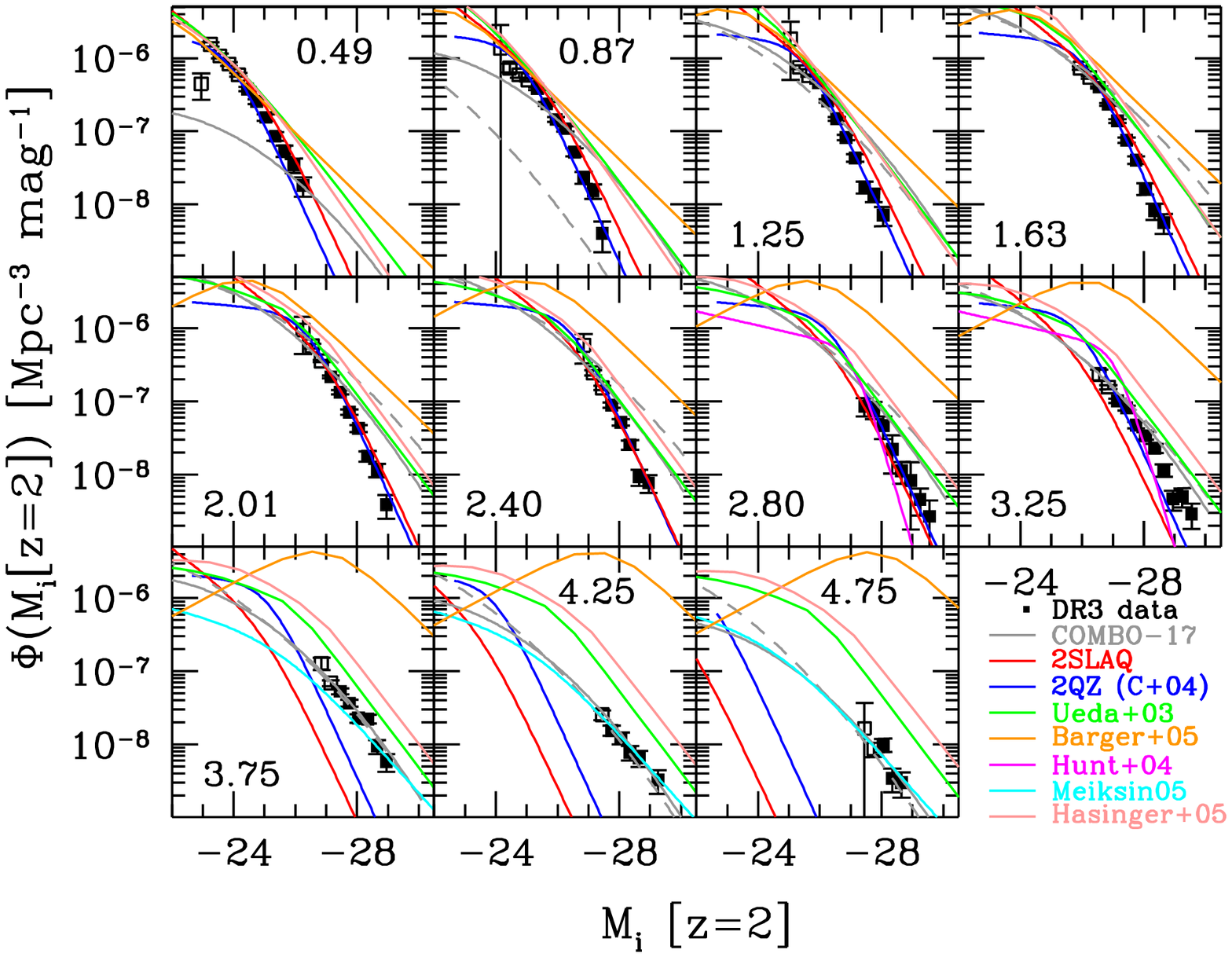}
\caption{Comparison with other QLFs used/derived by COMBO-17
(\markcite{wwb+03}{Wolf} {et~al.} 2003; PLE in solid gray, PDE in dashed gray), 2SLAQ
(\markcite{rca+05}{Richards} {et~al.} 2005; red, PLE), 2QZ (\markcite{csb+04}{Croom} {et~al.} 2004; blue, PLE),
\markcite{Ueda03}{Ueda} {et~al.} (2003) in green (X-ray, LDDE), \markcite{bcm+05}{Barger} {et~al.} (2005) in orange
(X-ray, PLE), \markcite{hsa+04}{Hunt} {et~al.} (2004) in magenta, \markcite{mei05}{Meiksin} (2005) in cyan, and
\markcite{hms05}{Hasinger} {et~al.} (2005, LDDE) in pink.  See \markcite{rca+05}{Richards} {et~al.} (2005) for the conversion between
X-ray and optical luminosity functions.  All of the parameterizations
are extended considerably beyond the data that were used to construct
them; this presentation merely emphasizes the difficulty of
parameterizing such a large range in luminosity and redshift.  The
\markcite{Ueda03}{Ueda} {et~al.} (2003) parameterization appears to do the best over the full
redshift range, but does not follow the observed slope change with
redshift.
\label{fig:fig19n}}
\end{figure}

Figure~\ref{fig:fig19n} illustrates the complexity of choosing a
functional form for the QLF for a luminous quasar sample spanning
$0<z<5$ by showing what happens when one naively extrapolates the
forms used or derived by \markcite{wwb+03}{Wolf} {et~al.} (2003), \markcite{rca+05}{Richards} {et~al.} (2005),
\markcite{Ueda03}{Ueda} {et~al.} (2003), \markcite{csb+04}{Croom} {et~al.} (2004), \markcite{bcm+05}{Barger} {et~al.} (2005), \markcite{hsa+04}{Hunt} {et~al.} (2004),
\markcite{mei05}{Meiksin} (2005), and \markcite{hms05}{Hasinger} {et~al.} (2005).  We have followed \markcite{rca+05}{Richards} {et~al.} (2005) to
convert between X-ray and optical luminosity functions.  All of the
parameterizations are extended well beyond the data that were used to
construct them and there is a considerable degree of ambiguity in the
transformations, thus this comparison is intended to be illustrative
only.  This presentation is not meant to belittle the accuracy of
these models, but rather to show that existing parameterizations do
not generally provide an accurate description of the redshift
evolution of luminous quasars simultaneously at low {\em and} high
redshift.  In particular, surveys with relatively small areas are
expected to predict QLF slopes that are too flat when extrapolated to
higher luminosities at low redshift.  This flattening occurs as a
result of the intrinsic flattening of the QLF at luminosities where
the majority of the sources are found (deep, pencil-beam surveys
having fainter mean luminosities than shallow, wide-area surveys).

\subsection{The Maximum Likelihood QLF}

In this work we have chosen to determine the maximum likelihood
solution with respect to a PLE form similar to that of \markcite{wwb+03}{Wolf} {et~al.} (2003),
specifically
\begin{equation}
\Phi = \Phi^* 10^{A_1\mu}
\label{eq:Phiform}
\end{equation}
where
\begin{equation}
\mu = M - (M^* + B_1\xi + B_2\xi^2 + B_3\xi^3)
\end{equation}
and
\begin{equation}
\xi = \log\left(\frac{1+z}{1+z_{\rm ref}}\right).
\end{equation}
$\Phi^*$, $A_1$, $B_1$, $B_2$, and $B_3$ are free parameters.  $z_{\rm
ref}$ has been set to 2.45 and $M^*$ has been set to $-26$.  This form
differs from \markcite{wwb+03}{Wolf} {et~al.} (2003) only in that it lacks a second order $A$
term, which is not justified by the dynamic range of our data.  At any
given redshift, this luminosity function is a single power law:
$\Phi(L)\propto L^{\beta}$, where $\beta = -(2.5A_1+1)$.  For single
power-law LF, there is no difference between PLE and PDE as there is
no characteristic scale in luminosity.  The best fit values and their
uncertainties from our maximum likelihood analysis are given in the
first row of Table~\ref{tab:tab7}.  Our value of $A_1$ corresponds to
a bright end slope in the \markcite{csb+04}{Croom} {et~al.} (2004) parameterization of
$\beta=-2.95\;(A_1=0.78)$.  This form does reasonably well at
describing the overall redshift and luminosity evolution of the
quasars in our sample; see the dashed red line in
Figure~\ref{fig:fig18n}.  However, the $\chi^2$ of this ML fit (as
compared to the binned QLF) is 394 for 69 degrees of freedom ---
suggesting that a more accurate parameterization is still needed.

\begin{figure}
\epsscale{1.0} 
\plotone{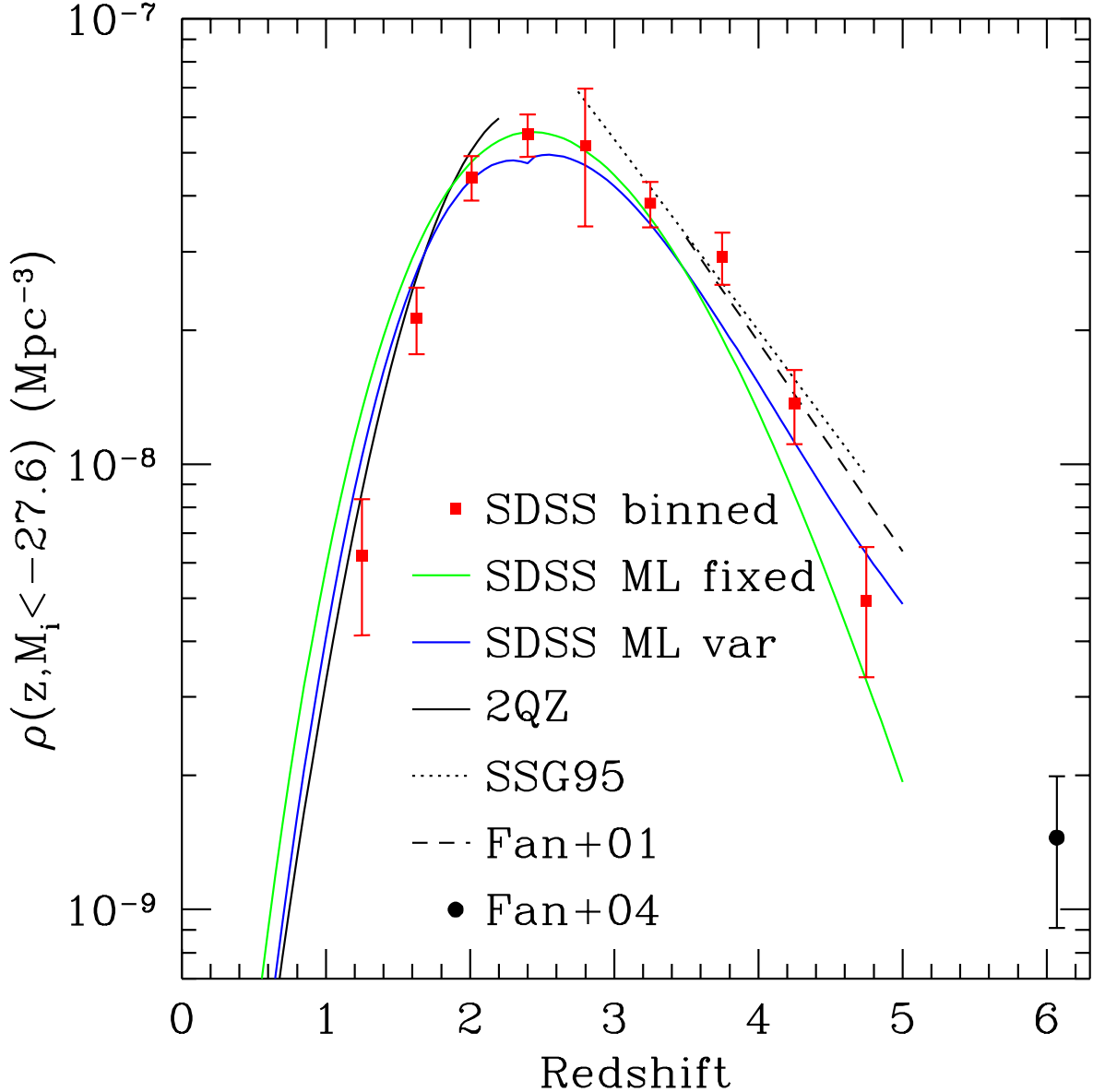}
\caption{Integrated $i$-band luminosity function to $M_i(z=2)=-27.6$.
The solid black line is from 2QZ \markcite{bsc+00}({Boyle} {et~al.} 2000).  The red points are
from the binned SDSS-DR3 QLF.  The green and blue lines are from the
fixed slope and variable high-redshift slope maximum likelihood (ML)
parameterizations of the SDSS-DR3 QLF, respectively.  The dashed and
dotted lines are from \markcite{fss+01}{Fan} {et~al.} (2001, Fan+01) and
\markcite{ssg95}{Schmidt} {et~al.} (1995, SSG95).  The $z\sim6$ point from
\markcite{fhr+04}{Fan} {et~al.} (2004, Fan+04), converted to our units and cosmology, is
shown by the solid black circle.  We caution that our ML fits should
not be used beyond $z=5$ as they are cubic fits and quickly diverge
beyond the limits of our data.
\label{fig:fig20n}} 
\end{figure}

Figure~\ref{fig:fig20n} shows the space density of luminous quasars
(i.e., the integral of the QLF).  This shows the familiar peak at $z
\approx 2.5$; at much lower and higher redshifts, luminous quasars are
very rare indeed.  The exact redshift of the peak is uncertain, a
situation exacerbated by the large and uncertain incompleteness in our
sample at $z \approx 2.7$.  Note the good agreement between our space
density evolution and that of previous papers.  In particular,
extrapolating the $z=3$--5 trend reveals good agreement with the
$z=6$ point from \markcite{fhr+04}{Fan} {et~al.} (2004), but we caution that our
functional forms should not be used beyond the $z\sim5$ limits of our
data as they are cubic fits that diverge quickly.

\subsection{Redshift Evolution of the Slope}
\label{sec:zevolution}

\begin{figure}
\epsscale{1.0}
\plotone{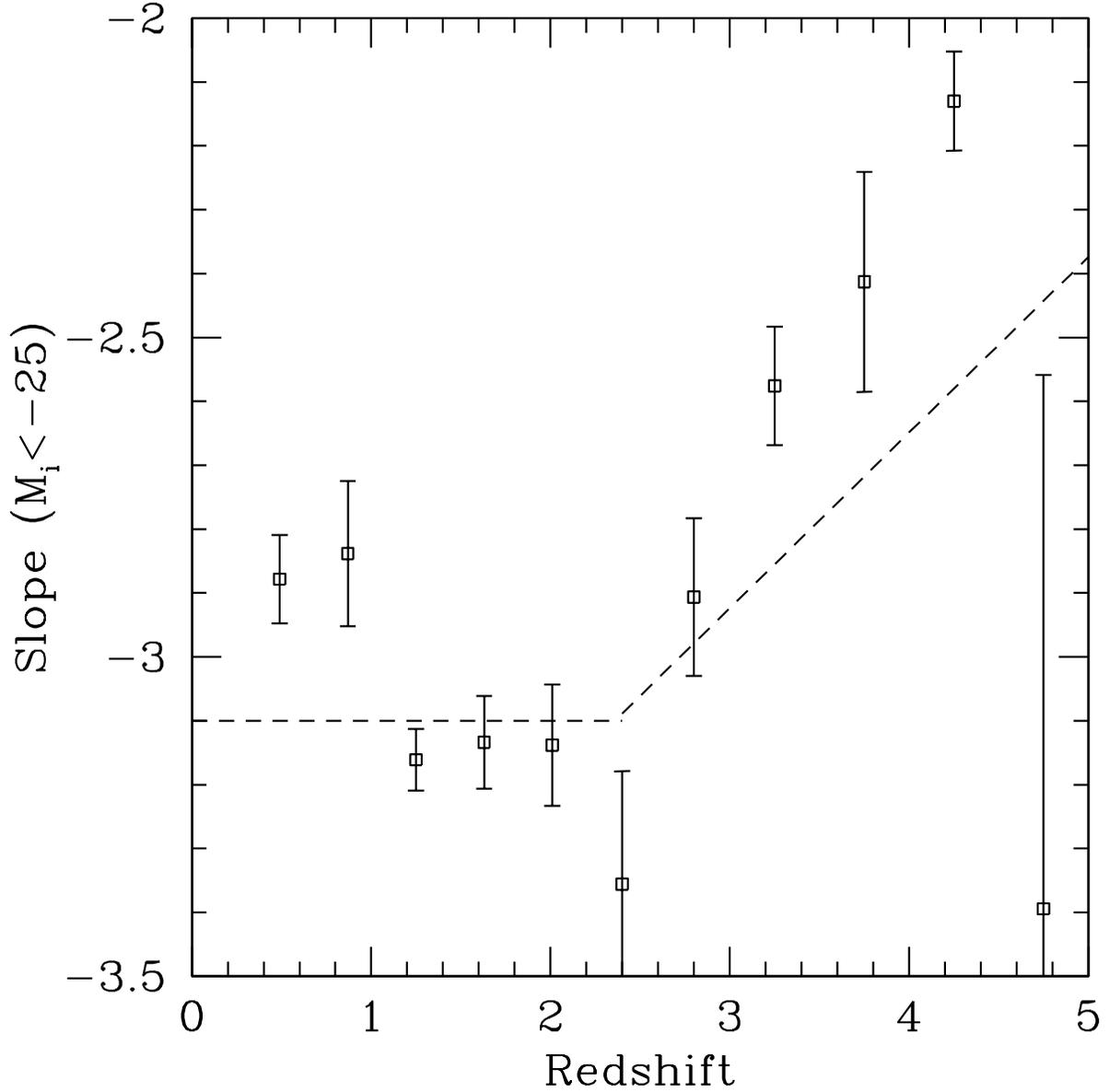}
\caption{Slope of the binned QLF as a function of redshift determined
from a linear least squares fit to the (complete) $M_i(z=2)<-25$
points.  The slope of the luminosity function significantly flattens
with redshift at $z > 3$ (the seemingly discrepant point at $z = 4.75$
was determined from only three luminosity bins, and has a large
uncertainty).  The dashed line shows the best fit constant slope for
$z\le2.4$ and the best fit redshift-dependent slope for $z>2.4$.
\label{fig:fig21n}}
\end{figure}

\markcite{ssg95}{Schmidt} {et~al.} (1995) and \markcite{fss+01}{Fan} {et~al.} (2001) showed that the slope of the $z>4$
QLF has a value of $\beta=-2.5$, much shallower than seen for $z<2.2$
quasars, which typically exhibit a slope of $\beta\sim-3.3$
\markcite{csb+04}({Croom} {et~al.} 2004).  Indeed this flattening is also apparent in our data.
If we fit a line to the $M_i<-25$ binned LF data (to avoid the
curvature at the faint end at low redshift) as a function of redshift,
we find the slopes given in Figure~\ref{fig:fig21n}.  For $z\le2.4$, the
slopes are roughly constant to within the errors; a ML fit yields
$\beta=-3.1\;(A_1=0.84)$.  However, at higher redshifts, if we ignore the poorly
constrained slope at $z=4.75$, there appears to be a flattening with
redshift.

Therefore, we have also attempted to allow for variation in slope in
our functional form.  To accomplish this we add an $A_2(z-2.45)$ term
to the exponent in Equation~\ref{eq:Phiform} above, such that 
\begin{equation}
\Phi = \Phi^* 10^{\mu[A_1 + A_2(z-2.45)]}.
\end{equation}
Since the change in slope at high redshift does not appear to extend
to lower redshifts, we allow the slope to vary linearly only for
$z>2.4$, fixing the slope to $\beta=-3.1$ for lower redshifts.  The
second and third rows of Table~\ref{tab:tab7} show the resulting best
fit values of the free parameters, which show that $\beta$ flattens to
$\gtrsim-2.37$ by $z=5$.  The result of this parameterization is shown
by the dot-dashed cyan line in Figure~\ref{fig:fig18n} and by the
solid blue line in Figure~\ref{fig:fig20n}.

Adding these two parameters ($A_2$, and the explicit decision of the
redshift at which to break the functional form) reduces the $\chi^2$
by 123, a highly significant change.  However, the $\chi^2$ per degree
of freedom is still 4; this functional form does not fit the data
perfectly. Figure~\ref{fig:fig18n} reveals that much of the excess
$\chi^2$ comes from the poor fit at $z<1$ (where we are probing faint
enough to see unmodeled curvature in the QLF and possibly host galaxy
contamination).
A more appropriate measure of the improvement of the fit is the amount
by which the quantity that is being minimized changes.  A $1\sigma$
change in a single variable will change the maximum likelihood
parameter by unity, whereas our change of parameterization reduces the
value by 102, thus the added complexity in the parameterization is
justified.

Finally, we reiterate the point made by \markcite{wis00}{Wisotzki} (2000) that the
measured slope is sensitive to the extrapolation of the
$K$-correction.  $K$-corrections normalized to $z=0$ and using a fixed
spectral index will cause the slope of the high-redshift QLF to appear
steeper than it should since the presumed absolute magnitude
distribution is narrower than the true distribution.  Our use of a
$z=2$ normalized $K$-correction helps to alleviate this problem and
highlights the slope change at high redshift.  Gravitational lensing
can also change the observed slope of the high-redshift QLF
\markcite{sef92}({Schneider}, {Ehlers}, \& {Falco} 1992); however, \markcite{rsp+04}{Richards} {et~al.} (2004b) and \markcite{rhp+06}{Richards} {et~al.} (2006) have used
{\em Hubble Space Telescope} imaging of $z>4$ SDSS quasars to put
limits on this effect.

\section{Discussion and Conclusions}
\label{sec:conclusions}

One of the most interesting results to come out of recent AGN surveys
is the evidence in favor of ``cosmic downsizing,'' wherein the peak of
AGN activity occurs at higher redshifts for more luminous objects than
less luminous objects \markcite{cbb+03,Ueda03,mer04,bcm+05}({Cowie} {et~al.} 2003; {Ueda} {et~al.} 2003; {Merloni} 2004; {Barger} {et~al.} 2005).  Comparison
of X-ray, infrared, and optical surveys requires careful consideration
of the fact that many groups find that the ratio of obscured (type 2)
to unobscured (type 1) AGN is inversely correlated with AGN luminosity
(e.g., \markcite{law91,Ueda03,hao05}{Lawrence} 1991; {Ueda} {et~al.} 2003; {Hao} {et~al.} 2005a; but see \markcite{tu05}{Treister} \& {Urry} 2005).
Ignoring this effect and examining the most uniform luminous sample
that we can form over the largest redshift range ($M_i<-27.6$),
Figure~\ref{fig:fig20n} shows that the peak in type 1 quasar activity
occurs between $z=2.2$ and $z=2.8$.  Unfortunately, this redshift
range is the least sensitive in the SDSS and subject to large error,
see Figure~\ref{fig:fig9n}.  A substantial observing campaign for
$z\sim2.5$ quasars that are buried in the stellar locus (i.e., a
sample with close to unity selection function in this redshift region)
is needed to resolve this issue.  To this end \markcite{Chiu04}{Chiu} (2004) and
\markcite{jiang05}{Jiang et al.} (2006) describe complete (i.e., not sparsely sampled) surveys
of quasars in the mid-$z$ range to address this problem.  In addition,
near-IR selected samples such as can be obtained from {\em Spitzer
Space Telescope} photometry should be able to better isolate the peak
redshift of luminous type 1 quasars \markcite{Brown05}({Brown} {et~al.} 2006).

Our most interesting result is the flattening of the slope of the QLF
with increasing redshift.  This flattening has been demonstrated
before using small samples of high-$z$ quasars \markcite{ssg95,fss+01}({Schmidt} {et~al.} 1995; {Fan} {et~al.} 2001),
but never so robustly and over such a large redshift range as with
these data.  While there is little overlap in luminosity between the
lowest and highest redshift data (deeper surveys at high redshift are
clearly needed), previous constraints on the QLF and the presumption
that the QLF will be well-behaved outside of the regions explored
(e.g., that the slope does not get {\em steeper} for faint
high-redshift quasars), suggests that the slope change is due to
redshift and not luminosity.  Small area samples such as the most
sensitive hard X-ray surveys \markcite{Ueda03,bcm+05}({Ueda} {et~al.} 2003; {Barger} {et~al.} 2005) and the COMBO-17
survey \markcite{wwb+03}({Wolf} {et~al.} 2003) primarily probe the low-luminosity end of the QLF,
where the slope is flatter, thus it is not surprising that they
systematically find flatter slopes (see Fig.~\ref{fig:fig19n}).  Our
confirmation of the flattening of the high-redshift slope has
significant consequences in terms of our understanding of the
formation and evolution of active galaxies, particularly in light of
the popularity of recent models invoking kinetic and radiative AGN
feedback in the evolution of galaxies
\markcite{sr98,fab99,wl03,hhc+05}(e.g., {Silk} \& {Rees} 1998; {Fabian} 1999; {Wyithe} \& {Loeb} 2003; {Hopkins} {et~al.} 2005a).

A particularly interesting explanation for the steepness of the
low-redshift QLF comes from the model of \markcite{wl03}{Wyithe} \& {Loeb} (2003).  In their
scenario, the predicted slope at low redshift is much flatter than the
observed slope (see Fig.~1 of \markcite{wl03}{Wyithe} \& {Loeb} 2003) for the most luminous
quasars.  They argue that the so-called ``break'' between less and
more luminous quasars for $z\lesssim2$ is the result of ``the
inability of gas to cool inside massive dark matter halos'', thus
preventing the formation of $v_c\gtrsim500\,{\rm km\,s^{-1}}$ galaxies
and their resulting luminous quasars in the most massive dark matter
halos.  Such an idea may be in conflict with the work of
\markcite{hhc+05z}{Hopkins et al.} (2005b) who find that the break in the QLF occurs naturally in
their models.  In their case the break occurs at the maximum of a
(roughly) log normal peak luminosity distribution with more luminous
quasars accreting near Eddington and less luminous quasars perhaps
accreting at lower rates.  However, in both the \markcite{wl03}{Wyithe} \& {Loeb} (2003) and
\markcite{hhc+05z}{Hopkins et al.} (2005b) PLE models the predicted slopes for luminous quasars
are actually {\em steeper} at high redshift than at low redshift (see
Fig.~1 in \markcite{wl03}{Wyithe} \& {Loeb} 2003 and Fig.~11 in \markcite{hhc+05z}{Hopkins et al.} 2005b) --- {\em
opposite of that which we observe}.  In the case of \markcite{hhc+05z}{Hopkins et al.} (2005b),
the evolution to high redshift is determined simply by adjusting the
break luminosity of their $0<z<3$ model to fit the existing high
redshift data.  Thus, to match our observed flattening at high
redshift, \markcite{hhc+05z}{Hopkins et al.} (2005b) will either need to change their model or
(at the very least) the details of its extrapolation to higher
redshift.  To explain a flatter QLF slope at high redshift in their
model, one would need to invoke a broader distribution of quasar peak
luminosities at high redshifts than low, with relatively more
low-luminosity objects at low redshift than high.  Thus our
observations and future observations of even fainter type 1 quasars at
$z>3$ provide an important litmus tests for models of galaxy
evolution.

While our analysis was based on over 15,000 quasars, to form this
uniform sample we were forced to drop over half of the objects in the
DR3 Quasar Catalog because of inhomogeneity in the selection
algorithms.  Data Release 5 of the SDSS, which is planned to occur in
mid-2006, will contain more than twice as many ``new" quasars as are
in our DR3 uniform sample.  We can define appreciably larger samples
yet, using the photometric selection and photometric redshift
techniques of \markcite{rng+04}{Richards} {et~al.} (2004a) and \markcite{wrs+04}{Weinstein} {et~al.} (2004); such methods will
result in a sample approaching a million quasars probing appreciably
further down the luminosity function, albeit at the price of less
certain redshifts.  We will also connect the low-luminosity end of the
quasar luminosity function to that of Seyfert galaxies measured from
the SDSS galaxy sample \markcite{hst+05}({Hao} {et~al.} 2005b) and explore the continuity of
the AGN population at low redshifts and luminosities.  We will explore
the dependence of the luminosity function on color, to determine, for
example, whether the luminosity function of intrinsically red quasars
\markcite{rhv+03}({Richards} {et~al.} 2003) has the same slope and varies with redshift the same
way that blue quasars do; differences could indicate correlations of
color with a physical parameter such as accretion rate, mass, or
orientation, or possible redshift- or luminosity-dependent dust
obscuration.  Finally, a number of groups are carrying out deeper
spectroscopic quasar surveys based on deep SDSS photometry (see
\markcite{rca+05,jiang05}{Richards} {et~al.} 2005; {Jiang et al.} 2006), and we can look forward to a yet more
comprehensive view of the quasar luminosity function in a few years'
time.

\acknowledgements

Funding for the creation and distribution of the SDSS Archive has been
provided by the Alfred P. Sloan Foundation, the Participating
Institutions, the National Aeronautics and Space Administration, the
National Science Foundation, the U.S. Department of Energy, the
Japanese Monbukagakusho, and the Max Planck Society. The SDSS Web site
is http://www.sdss.org/.  The SDSS is managed by the Astrophysical
Research Consortium (ARC) for the Participating Institutions. The
Participating Institutions are The University of Chicago, Fermilab,
the Institute for Advanced Study, the Japan Participation Group, The
Johns Hopkins University, the Korean Scientist Group, Los Alamos
National Laboratory, the Max-Planck-Institute for Astronomy (MPIA),
the Max-Planck-Institute for Astrophysics (MPA), New Mexico State
University, University of Pittsburgh, University of Portsmouth,
Princeton University, the United States Naval Observatory, and the
University of Washington.  We thank the referee for suggestions on
shortening the paper and comparison with radio work.  DPS and DVB
acknowledge the support of NSF grant AST-0307582.  XF acknowledges
supports from NSF grant AST-0307384, a Sloan Research Fellowship and a
Packard Fellowship for Science and Engineering.  GTR and MAS
acknowledge the support of NSF grant AST-0307409.  GTR acknowledges
support from a Gordon and Betty Moore Fellowship in data intensive
sciences.  GTR thanks Michael Weinstein and Michael Brown for
assistance with code development, Takamitsu Miyaji for helping with
optical to X-ray QLF comparisons, and Scott Croom for providing
$M_{b_J}$ values for 2QZ quasars.

\appendix
\section*{Appendix}
\label{sec:SQL}

The following query, in SQL (structured query language), was used to
construct the sample of DR3 quasar candidates selected from the TARGET
photometry from the SDSS CAS.  We note that it is {\em not} possible
to do this query using the public DR3 CAS as the ``region''
information that is needed for simultaneously determining the target
selection version and area thereof is lacking in that database.
However, this information will be available for SDSS Data Release 5.

\smallskip

\noindent{-- {\it query the table that includes all targeted objects}\\
SELECT * FROM Target as t\\
-- {\it match to the tables with geometry and version information}\\
inner join Region as r on t.regionid $=$ r.regionid\\
inner join TargetInfo as ti on t.targetid $=$ ti.targetid\\
-- {\it extract the TARGET photometry}\\
inner join TARGDR3..photoTag as p on ti.targetobjid $=$ p.objid\\
-- {\it match to objects with spectroscopy}\\
left outer join specObj as s on s.targetid $=$ t.targetid\\
WHERE (\\
-- {\it restrict sample to target selection version v3\_1\_0}\\
$\phm{xx}$ r.regionid in (\\
$\phm{xxxx}$	select b.boxid\\
$\phm{xxxx}$	from region2box b, tilinggeometry g\\
$\phm{xxxx}$	where b.boxtype $=$ 'SECTOR'\\
$\phm{xxxxxx}$          and b.regiontype $=$ 'TIPRIMARY'\\
$\phm{xxxxxx}$          and b.id $=$ g.tilinggeometryid\\
$\phm{xxxx}$	group by b.boxid\\
$\phm{xxxx}$	having min(g.targetversion) $\ge$ 'v3\_1\_0'\\
$\phm{xx}$ ) AND\\
$\phm{xx}$ -- {\it include only ``primary'' objects}\\
$\phm{xx}$  ( (p.mode $=$ 1) AND ((p.status \& 0x10) $>$ 0) AND ((p.status \& 0x2000) $>$ 0) )\\
$\phm{xx}$ -- {\it include only explicit quasar targets}\\
$\phm{xx}$  AND ((p.primTarget \& 0x0000001f) $>$ 0)\\
)}
\medskip 

\noindent The restriction on {\tt region} (defined by the intersection
of various survey geometrical constraints, such as the intersection of
an imaging scan and a spectroscopic plate) limits the sample to
targets selected with version v3\_1\_0 and later of the targeting
algorithm.  The restrictions on {\tt mode} and {\tt status} restrict
the sample to the ``primary'' observations of each object (ignoring
any repeat ``secondary'' observations that exist) that are within the
nominal DR3 footprint (some areas with DR3 imaging formally belong to
later data releases due to geometrical definitions).  The bitwise AND
restriction on the {\tt primTarget} values returns only quasar
candidates selected by the main quasar selection algorithm
\markcite{rfn+02}({Richards} {et~al.} 2002).

The joins\footnote{The intersection of two database tables.  An inner
join returns only objects that exist in both tables.  A left outer
join returns output for each object in the ``left'' table, regardless
of whether it has a match in the ``right'' table.} with {\tt
TargetInfo} and {\tt TARGDR3..photoTag} are necessary to extract the
TARGET (as opposed to BEST) photometry that will be used in our
analysis.  The join on {\tt specObj} allows the extraction of
spectroscopic parameters from the database.

\clearpage

 

\clearpage

\clearpage

\clearpage


\begin{deluxetable}{ccccc}
\tabletypesize{\small}
\tablewidth{0pt}
\tablecaption{Quasar Selection Function \label{tab:tab01}}
\tablehead{
\colhead{$i$ mag} &
\colhead{$z_{\rm em}$} &
\colhead{point} &
\colhead{radio} &
\colhead{extended}
}
\startdata
15.0 & 0.00 & 1.000 & 1.000 & 1.000 \\
15.0 & 0.05 & 1.000 & 1.000 & 1.000 \\
15.0 & 0.10 & 1.000 & 1.000 & 1.000 \\
15.0 & 0.15 & 1.000 & 1.000 & 1.000 \\
15.0 & 0.20 & 1.000 & 1.000 & 1.000 \\
15.0 & 0.25 & 1.000 & 1.000 & 1.000 \\
15.0 & 0.30 & 1.000 & 1.000 & 1.000 \\
15.0 & 0.35 & 1.000 & 1.000 & 1.000 \\
15.0 & 0.40 & 1.000 & 1.000 & 1.000 \\
15.0 & 0.45 & 1.000 & 1.000 & 1.000 \\
15.0 & 0.50 & 1.000 & 1.000 & 1.000 \\
15.0 & 0.55 & 1.000 & 1.000 & 1.000 \\
15.0 & 0.60 & 1.000 & 1.000 & 1.000 \\
15.0 & 0.65 & 0.980 & 1.000 & 0.975 \\
15.0 & 0.70 & 0.995 & 1.000 & 0.990 \\
15.0 & 0.75 & 1.000 & 1.000 & 1.000 \\
15.0 & 0.80 & 1.000 & 1.000 & 1.000 \\
15.0 & 0.85 & 1.000 & 1.000 & 1.000 \\
15.0 & 0.90 & 1.000 & 1.000 & 1.000 \\
15.0 & 0.95 & 1.000 & 1.000 & 1.000 \\
\enddata
\tablecomments{[Full table to appear in the on-line edition.]}
\end{deluxetable}


\begin{deluxetable}{lccrccr}
\tabletypesize{\footnotesize}
\tablewidth{0pt}
\tablecaption{Quasar Number Counts ($0.3<z<2.2$) \label{tab:tab02}}
\tablehead{
\colhead{mag} &
\colhead{N($g$)} &
\colhead{N($<g$)} &
\colhead{${\rm N_Q}$} &
\colhead{N($i$)} &
\colhead{N($<i$)} &
\colhead{${\rm N_Q}$}
}
\startdata
15.475 & $0.00\pm0.00$  & $0.00\pm0.00$ &    2 & $0.00\pm0.00$  & $0.00\pm0.00$ &    1 \\
15.725 & $0.00\pm0.00$  & $0.00\pm0.00$ &    3 & $0.00\pm0.00$  & $0.00\pm0.00$ &    3 \\
15.975 & $0.00\pm0.00$  & $0.01\pm0.00$ &    4 & $0.00\pm0.00$  & $0.00\pm0.00$ &    2 \\
16.225 & $0.01\pm0.00$  & $0.01\pm0.00$ &   11 & $0.01\pm0.00$  & $0.01\pm0.00$ &   10 \\
16.475 & $0.01\pm0.00$  & $0.02\pm0.00$ &   10 & $0.01\pm0.00$  & $0.02\pm0.00$ &   12 \\
16.725 & $0.02\pm0.00$  & $0.04\pm0.00$ &   32 & $0.01\pm0.00$  & $0.03\pm0.00$ &   23 \\
16.975 & $0.06\pm0.01$  & $0.10\pm0.01$ &   85 & $0.04\pm0.01$  & $0.08\pm0.01$ &   67 \\
17.225 & $0.08\pm0.01$  & $0.17\pm0.01$ &  117 & $0.07\pm0.01$  & $0.15\pm0.01$ &  107 \\
17.475 & $0.12\pm0.01$  & $0.29\pm0.01$ &  190 & $0.11\pm0.01$  & $0.26\pm0.01$ &  174 \\
17.725 & $0.23\pm0.01$  & $0.53\pm0.01$ &  357 & $0.21\pm0.01$  & $0.47\pm0.01$ &  327 \\
17.975 & $0.31\pm0.01$  & $0.83\pm0.01$ &  472 & $0.33\pm0.01$  & $0.80\pm0.01$ &  511 \\
18.225 & $0.56\pm0.02$  & $1.40\pm0.02$ &  868 & $0.55\pm0.02$  & $1.35\pm0.02$ &  849 \\
18.475 & $0.78\pm0.02$  & $2.18\pm0.02$ & 1202 & $0.82\pm0.02$  & $2.17\pm0.02$ & 1257 \\
18.725 & $1.19\pm0.03$  & $3.37\pm0.03$ & 1839 & $1.25\pm0.03$  & $3.42\pm0.03$ & 1923 \\
18.975 & $1.70\pm0.03$  & $5.07\pm0.03$ & 2620 & $1.86\pm0.04$  & $5.28\pm0.03$ & 2870 \\
19.225 & $1.61\pm0.03$  & $6.68\pm0.03$ & 2484 & $2.62\pm0.04$  & $7.90\pm0.04$ & 4028 \\
19.475 & $0.56\pm0.02$  & $7.24\pm0.02$ &  855 & \ldots  & \ldots &  900 \\
19.725 & $0.13\pm0.01$  & $7.37\pm0.01$ &  206 & \ldots  & \ldots &  216 \\
19.975 & $0.03\pm0.01$  & $7.41\pm0.00$ &   53 & \ldots  & \ldots &   55 \\
20.225 & $0.01\pm0.00$  & $7.42\pm0.00$ &   19 & \ldots  & \ldots &   20 \\
\enddata
\end{deluxetable}

\begin{deluxetable}{lccrr}
\tabletypesize{\footnotesize}
\tablewidth{0pt}
\tablecaption{Quasar Number Counts ($3<z<5$) \label{tab:tab03}}
\tablehead{
\colhead{$i$} &
\colhead{N($i$)} &
\colhead{N($<i$)} &
\colhead{${\rm N_Q}$} &
\colhead{${\rm N_Q}$ cor}
}
\startdata
17.075 & $0.00\pm0.00$  & $0.00\pm0.00$ &    1 &  1.1 \\
17.325 & $0.00\pm0.00$  & $0.00\pm0.00$ &    4 &  4.5 \\
17.575 & $0.01\pm0.00$  & $0.01\pm0.00$ &   12 & 14.2 \\
17.825 & $0.01\pm0.00$  & $0.02\pm0.00$ &   12 & 14.2 \\
18.075 & $0.02\pm0.00$  & $0.04\pm0.00$ &   26 & 29.7 \\
18.325 & $0.03\pm0.01$  & $0.07\pm0.00$ &   47 & 55.8 \\
18.575 & $0.05\pm0.01$  & $0.13\pm0.01$ &   72 & 83.8 \\
18.825 & $0.08\pm0.01$  & $0.20\pm0.01$ &  106 & 124.7 \\
19.075 & $0.12\pm0.01$  & $0.33\pm0.01$ &  162 & 197.5 \\
19.325 & $0.17\pm0.01$  & $0.49\pm0.01$ &  206 & 268.4 \\
19.575 & $0.22\pm0.01$  & $0.71\pm0.01$ &  275 & 360.2 \\
19.825 & $0.28\pm0.02$  & $0.99\pm0.01$ &  322 & 455.4 \\
20.075 & $0.38\pm0.02$  & $1.37\pm0.02$ &  410 & 617.5 \\
\enddata
\end{deluxetable}

\begin{deluxetable}{cc}
\tabletypesize{\small}
\tablewidth{0pt}
\tablecaption{$K$-Corrections \label{tab:tab04}}
\tablehead{
\colhead{$z_{\rm em}$} &
\colhead{$K$-correction}
}
\startdata
0.00 & $0.596$ \\
0.01 & $0.587$ \\
0.02 & $0.569$ \\
0.03 & $0.531$ \\
0.04 & $0.462$ \\
0.05 & $0.372$ \\
0.06 & $0.268$ \\
0.07 & $0.203$ \\
0.08 & $0.170$ \\
0.09 & $0.157$ \\
\enddata
\tablecomments{$i$-band $K$-corrections, including both the emission
line and continuum ($\alpha_{\nu}=-0.5$, normalized at $z=2$)
components.  [Full table to appear in the on-line edition.]}
\end{deluxetable}

\begin{deluxetable}{lrrrrr}
\tabletypesize{\small}
\tablewidth{0pt}
\tablecaption{Statistical Quasar Sample \label{tab:tab5}}
\tablehead{
\colhead{Name} &
\colhead{$z_{\rm em}$} &
\colhead{$i$} &
\colhead{$M_i(z=2)$} &
\colhead{$\Delta(g-i)$} &
\colhead{Cor.} \\
\colhead{(SDSS J)} &
\colhead{} &
\colhead{} &
\colhead{} &
\colhead{} &
\colhead{} 
}
\startdata
$000009.26+151754.5$ & 1.199 & 19.08 & $-25.40$ & $0.25$ & 0.95 \\
$000009.38+135618.4$ & 2.240 & 18.18 & $-27.86$ & $0.32$ & 0.93 \\
$000011.41+145545.6$ & 0.460 & 19.09 & $-23.21$ & $0.01$ & 0.95 \\
$000013.14+141034.6$ & 0.949 & 19.05 & $-25.02$ & $-0.08$ & 0.95 \\
$000024.02+152005.4$ & 0.989 & 18.99 & $-25.15$ & $-0.11$ & 0.95 \\
\enddata
\tablecomments{[Full table to appear in the on-line edition.]}
\end{deluxetable}

\begin{deluxetable}{lrrrcrrr}
\tabletypesize{\footnotesize}
\tablewidth{0pt}
\tablecaption{Binned Quasar Luminosity Function\label{tab:tab6}}
\tablehead{
\colhead{$z$} & 
\colhead{$M_i(z=2)$} & 
\colhead{$\log\Phi$} & 
\colhead{$\sigma_{\Phi}$} & 
\colhead{Fill} &
\colhead{$\overline{z}$} &
\colhead{$N_{\rm Q}$} &
\colhead{$N_{\rm Q}$ cor} \\
\colhead{} & 
\colhead{} & 
\colhead{} & 
\colhead{($\times10^{-9}$)} & 
\colhead{} & 
\colhead{} & 
\colhead{} & 
\colhead{}
}
\startdata
0.49 & $-26.25$ & $-7.74$ & 5.48 & 1 & 0.59 & 11 & 11.6 \\
0.49 & $-25.95$ & $-7.46$ & 7.55 & 1 & 0.58 & 21 & 22.2 \\
0.49 & $-25.65$ & $-7.26$ & 9.47 & 1 & 0.58 & 33 & 34.8 \\
0.49 & $-25.35$ & $-7.07$ & 11.90 & 1 & 0.55 & 52 & 54.9 \\
0.49 & $-25.05$ & $-6.81$ & 16.01 & 1 & 0.56 & 94 & 99.3 \\
0.49 & $-24.75$ & $-6.59$ & 20.52 & 1 & 0.58 & 154 & 163.0 \\
0.49 & $-24.45$ & $-6.42$ & 24.94 & 1 & 0.57 & 228 & 241.0 \\
0.49 & $-24.15$ & $-6.23$ & 35.98 & 0 & 0.54 & 337 & 358.2 \\
0.49 & $-23.85$ & $-6.08$ & 43.96 & 0 & 0.51 & 358 & 377.2 \\
0.49 & $-23.55$ & $-5.97$ & 75.48 & 0 & 0.45 & 311 & 331.3 \\
0.49 & $-23.25$ & $-5.81$ & 102.81 & 0 & 0.41 & 290 & 307.2 \\
0.49 & $-22.95$ & $-6.35$ & 174.24 & 0 & 0.41 & 39 & 46.9 \\
0.87 & $-27.45$ & $-8.39$ & 1.81 & 1 & 0.95 & 5 & 5.3 \\
0.87 & $-27.15$ & $-7.81$ & 3.53 & 1 & 0.95 & 19 & 20.0 \\
0.87 & $-26.85$ & $-7.63$ & 4.36 & 1 & 0.95 & 29 & 30.5 \\
0.87 & $-26.55$ & $-7.29$ & 6.48 & 1 & 0.94 & 64 & 67.4 \\
0.87 & $-26.25$ & $-6.96$ & 9.38 & 1 & 0.91 & 134 & 141.2 \\
0.87 & $-25.95$ & $-6.84$ & 10.84 & 1 & 0.92 & 179 & 188.5 \\
0.87 & $-25.65$ & $-6.62$ & 13.87 & 1 & 0.92 & 293 & 308.6 \\
0.87 & $-25.35$ & $-6.42$ & 17.57 & 1 & 0.93 & 470 & 495.1 \\
0.87 & $-25.05$ & $-6.30$ & 24.49 & 0 & 0.89 & 492 & 520.0 \\
0.87 & $-24.75$ & $-6.23$ & 51.36 & 0 & 0.78 & 284 & 307.6 \\
0.87 & $-24.45$ & $-6.13$ & 100.79 & 0 & 0.73 & 122 & 133.7 \\
0.87 & $-24.15$ & $-5.86$ & 1446.59 & 0 & 0.74 & 9 & 13.8 \\
1.25 & $-28.05$ & $-8.15$ & 2.06 & 1 & 1.31 & 12 & 12.6 \\
1.25 & $-27.75$ & $-7.86$ & 2.85 & 1 & 1.32 & 23 & 24.2 \\
1.25 & $-27.45$ & $-7.76$ & 3.20 & 1 & 1.27 & 29 & 30.5 \\
1.25 & $-27.15$ & $-7.36$ & 5.07 & 1 & 1.30 & 73 & 76.8 \\
1.25 & $-26.85$ & $-7.08$ & 7.00 & 1 & 1.30 & 139 & 146.3 \\
1.25 & $-26.55$ & $-6.83$ & 9.37 & 1 & 1.29 & 249 & 262.1 \\
1.25 & $-26.25$ & $-6.58$ & 12.44 & 1 & 1.29 & 439 & 462.1 \\
1.25 & $-25.95$ & $-6.33$ & 16.62 & 1 & 1.28 & 783 & 824.2 \\
1.25 & $-25.65$ & $-6.24$ & 22.56 & 0 & 1.23 & 733 & 773.5 \\
1.25 & $-25.35$ & $-6.12$ & 48.01 & 0 & 1.13 & 360 & 382.8 \\
1.25 & $-25.05$ & $-5.74$ & 1332.39 & 0 & 1.09 & 12 & 14.6 \\
1.63 & $-28.65$ & $-8.25$ & 1.70 & 1 & 1.66 & 11 & 11.6 \\
1.63 & $-28.35$ & $-8.08$ & 2.06 & 1 & 1.67 & 16 & 16.8 \\
1.63 & $-28.05$ & $-7.78$ & 2.91 & 1 & 1.68 & 32 & 33.7 \\
1.63 & $-27.75$ & $-7.39$ & 4.57 & 1 & 1.66 & 79 & 83.2 \\
1.63 & $-27.45$ & $-7.12$ & 6.23 & 1 & 1.67 & 147 & 154.7 \\
1.63 & $-27.15$ & $-6.86$ & 8.43 & 1 & 1.66 & 269 & 283.2 \\
1.63 & $-26.85$ & $-6.63$ & 10.94 & 1 & 1.65 & 453 & 476.8 \\
1.63 & $-26.55$ & $-6.40$ & 14.35 & 1 & 1.66 & 779 & 820.0 \\
1.63 & $-26.25$ & $-6.26$ & 19.09 & 0 & 1.62 & 892 & 940.9 \\
1.63 & $-25.95$ & $-6.14$ & 38.86 & 0 & 1.52 & 477 & 506.0 \\
2.01 & $-28.95$ & $-8.41$ & 1.37 & 1 & 1.99 & 8 & 8.4 \\
2.01 & $-28.65$ & $-7.94$ & 2.37 & 1 & 2.02 & 24 & 25.3 \\
2.01 & $-28.35$ & $-7.75$ & 2.94 & 1 & 2.02 & 37 & 39.0 \\
2.01 & $-28.05$ & $-7.37$ & 4.53 & 1 & 2.01 & 88 & 92.7 \\
2.01 & $-27.75$ & $-7.15$ & 5.86 & 1 & 2.03 & 147 & 155.0 \\
2.01 & $-27.45$ & $-6.87$ & 8.03 & 1 & 2.02 & 276 & 291.0 \\
2.01 & $-27.15$ & $-6.67$ & 10.22 & 1 & 2.01 & 447 & 471.1 \\
2.01 & $-26.85$ & $-6.45$ & 13.97 & 0 & 1.99 & 648 & 682.7 \\
2.01 & $-26.55$ & $-6.24$ & 40.12 & 0 & 1.90 & 430 & 464.3 \\
2.01 & $-26.25$ & $-6.03$ & 495.10 & 0 & 1.87 & 30 & 39.4 \\
2.40 & $-28.95$ & $-8.12$ & 2.07 & 1 & 2.44 & 14 & 17.9 \\
2.40 & $-28.65$ & $-8.03$ & 2.30 & 1 & 2.41 & 17 & 21.8 \\
2.40 & $-28.35$ & $-7.59$ & 3.71 & 1 & 2.37 & 48 & 59.7 \\
2.40 & $-28.05$ & $-7.29$ & 5.28 & 1 & 2.37 & 98 & 121.2 \\
2.40 & $-27.75$ & $-7.05$ & 7.01 & 1 & 2.37 & 164 & 208.1 \\
2.40 & $-27.45$ & $-6.82$ & 9.50 & 0 & 2.38 & 265 & 356.0 \\
2.40 & $-27.15$ & $-6.61$ & 22.11 & 0 & 2.30 & 246 & 309.0 \\
2.40 & $-26.85$ & $-6.24$ & 249.51 & 0 & 2.25 & 37 & 48.8 \\
2.80 & $-29.55$ & $-8.57$ & 1.76 & 1 & 2.70 & 3 & 6.2 \\
2.80 & $-29.25$ & $-8.34$ & 1.84 & 1 & 2.82 & 7 & 10.6 \\
2.80 & $-28.95$ & $-8.08$ & 6.51 & 1 & 2.82 & 10 & 19.2 \\
2.80 & $-28.65$ & $-7.95$ & 3.68 & 1 & 2.80 & 12 & 25.9 \\
2.80 & $-28.35$ & $-7.65$ & 11.94 & 1 & 2.83 & 24 & 51.5 \\
2.80 & $-28.05$ & $-7.33$ & 15.75 & 1 & 2.81 & 50 & 107.9 \\
2.80 & $-27.75$ & $-7.11$ & 20.63 & 1 & 2.82 & 75 & 180.9 \\
2.80 & $-27.45$ & $-7.07$ & 24.17 & 1 & 2.79 & 73 & 199.8 \\
3.25 & $-29.55$ & $-8.54$ & 1.10 & 1 & 3.35 & 7 & 8.2 \\
3.25 & $-29.25$ & $-8.30$ & 1.45 & 1 & 3.25 & 12 & 14.2 \\
3.25 & $-28.95$ & $-8.33$ & 1.40 & 1 & 3.20 & 11 & 13.1 \\
3.25 & $-28.65$ & $-7.94$ & 2.18 & 1 & 3.25 & 28 & 32.5 \\
3.25 & $-28.35$ & $-7.64$ & 3.12 & 1 & 3.23 & 54 & 64.6 \\
3.25 & $-28.05$ & $-7.47$ & 3.89 & 1 & 3.25 & 77 & 95.9 \\
3.25 & $-27.75$ & $-7.32$ & 4.72 & 1 & 3.24 & 103 & 134.3 \\
3.25 & $-27.45$ & $-7.09$ & 6.47 & 1 & 3.21 & 161 & 228.8 \\
3.25 & $-27.15$ & $-6.99$ & 7.48 & 1 & 3.22 & 191 & 287.9 \\
3.25 & $-26.85$ & $-6.80$ & 10.40 & 0 & 3.20 & 234 & 377.5 \\
3.25 & $-26.55$ & $-6.63$ & 39.58 & 0 & 3.11 & 52 & 103.7 \\
3.75 & $-28.95$ & $-8.24$ & 1.56 & 1 & 3.69 & 14 & 15.7 \\
3.75 & $-28.65$ & $-8.02$ & 2.03 & 1 & 3.67 & 22 & 25.7 \\
3.75 & $-28.35$ & $-7.65$ & 3.06 & 1 & 3.77 & 54 & 60.5 \\
3.75 & $-28.05$ & $-7.64$ & 3.11 & 1 & 3.76 & 55 & 62.1 \\
3.75 & $-27.75$ & $-7.44$ & 4.07 & 1 & 3.74 & 83 & 99.3 \\
3.75 & $-27.45$ & $-7.27$ & 4.96 & 1 & 3.76 & 119 & 144.2 \\
3.75 & $-27.15$ & $-7.16$ & 6.59 & 0 & 3.72 & 117 & 158.4 \\
3.75 & $-26.85$ & $-6.89$ & 25.48 & 0 & 3.59 & 27 & 54.6 \\
4.25 & $-29.25$ & $-8.48$ & 1.16 & 1 & 4.25 & 8 & 8.4 \\
4.25 & $-28.65$ & $-8.18$ & 1.64 & 1 & 4.22 & 16 & 16.9 \\
4.25 & $-28.35$ & $-8.11$ & 1.79 & 1 & 4.19 & 19 & 20.0 \\
4.25 & $-28.05$ & $-7.91$ & 2.26 & 1 & 4.21 & 30 & 31.8 \\
4.25 & $-27.75$ & $-7.81$ & 2.53 & 1 & 4.24 & 38 & 40.2 \\
4.25 & $-27.45$ & $-7.59$ & 3.84 & 0 & 4.16 & 44 & 47.1 \\
4.75 & $-28.65$ & $-8.52$ & 1.14 & 1 & 4.66 & 7 & 7.4 \\
4.75 & $-28.35$ & $-8.46$ & 1.22 & 1 & 4.71 & 8 & 8.4 \\
4.75 & $-28.05$ & $-8.00$ & 2.07 & 1 & 4.66 & 23 & 24.2 \\
4.75 & $-27.75$ & $-8.06$ & 1.94 & 0 & 4.66 & 20 & 21.1 \\
4.75 & $-27.45$ & $-7.78$ & 20.57 & 0 & 4.62 & 6 & 8.3 \\
\enddata
\tablecomments{Columns are 1) redshift, 2) $M_i(z=2)$, 3) $\Phi$ [Mpc$^{-3}$ mag$^{-1}$], 4) $\sigma_{\Phi}$, 5) an indicator of whether the bin is completely covered by data (1 if yes, 0 if no), 6) the mean redshift of quasars in the bin, 7) the number of quasars in the bin, and 8) the corrected number of quasars in that bin after applying the selection function.}
\end{deluxetable}

\begin{deluxetable}{lcccccccccc}
\tabletypesize{\scriptsize}
\tablewidth{0pt}
\tablecaption{Summary of maximum likelihood fits.\label{tab:tab7}}
\tablehead{
\colhead{Form} & 
\colhead{$A_1$} & 
\colhead{$A_2$} & 
\colhead{$B_1$} & 
\colhead{$B_2$} & 
\colhead{$B_3$} & 
\colhead{$M^*$} & 
\colhead{$z_{\rm ref}$} & 
\colhead{$\log\Phi^*$} & 
\colhead{$\chi^2$} & 
\colhead{$\nu$}
}
\startdata
Fixed Power Law & 0.78$\pm$0.01 & \ldots & 0.10$\pm$0.04 & 27.35$\pm$0.10 & 19.27$\pm$0.25 & $-$26 & 2.45 & $-$5.75 & 394 & 69 \\
Variable Power Law ($z>2.4$) & 0.83$\pm$0.01 & $-$0.11$\pm$0.01 & 1.43$\pm$0.04 & 36.63$\pm$0.10 & 34.39$\pm$0.26 & $-$26 & 2.45 & $-$5.70 & 271 & 67 \\
Variable Power Law ($z\le2.4$) & 0.84 & 0.00 & 1.43$\pm$0.04 & 36.63$\pm$0.10 & 34.39$\pm$0.26 & $-$26 & 2.45 & $-$5.70 & 271 & 67 \\
\enddata
\tablecomments{The fixed power-law model is given by Eqs.~6--8. See Eq.~9 and \S~\ref{sec:zevolution} for the variable power-law model.  $\Phi$ is in units of Mpc$^{-3}$ mag$^{-1}$. $M^*$ and $z_{\rm ref}$ are not free parameters, rather they are defined to have the values indicated.}
\end{deluxetable}


\begin{thebibliography}{}

\bibitem[{Abazajian}, {Adelman-McCarthy}, {Ag{\"  u}eros}, {Allam}, {Anderson}, {Anderson}, {Annis}, {Bahcall}, {Baldry},  {Bastian}, {Berlind}, {Bernardi}, {Blanton}, {Bochanski}, {Boroski},  {Brewington}, {Briggs}, {Brinkmann}, {Brunner}, {Budav{\' a}ri}, {Carey},  {Castander}, {Connolly}, {Covey}, {Csabai}, {Dalcanton}, {Doi}, {Dong},  {Eisenstein}, {Evans}, {Fan}, {Finkbeiner}, {Friedman}, {Frieman},  {Fukugita}, {Gillespie}, {Glazebrook}, {Gray}, {Grebel}, {Gunn}, {Gurbani},  {Hall}, {Hamabe}, {Harbeck}, {Harris}, {Harris}, {Harvanek}, {Hawley},  {Hayes}, {Heckman}, {Hendry}, {Hennessy}, {Hindsley}, {Hogan}, {Hogg},  {Holmgren}, {Holtzman}, {Ichikawa}, {Ichikawa}, {Ivezi{\' c}}, {Jester},  {Johnston}, {Jorgensen}, {Juri{\' c}}, {Kent}, {Kleinman}, {Knapp},  {Kniazev}, {Kron}, {Krzesinski}, {Lamb}, {Lampeitl}, {Lee}, {Lin}, {Long},  {Loveday}, {Lupton}, {Mannery}, {Margon}, {Mart{\'{\i}}nez-Delgado},  {Matsubara}, {McGehee}, {McKay}, {Meiksin}, {M{\' e}nard}, {Munn}, {Nash},  {Neilsen}, {Newberg}, {Newman}, {Nichol}, {Nicinski}, {Nieto-Santisteban},  {Nitta}, {Okamura}, {O'Mullane}, {Owen}, {Padmanabhan}, {Pauls}, {Peoples},  {Pier}, {Pope}, {Pourbaix}, {Quinn}, {Raddick}, {Richards}, {Richmond},  {Rix}, {Rockosi}, {Schlegel}, {Schneider}, {Schroeder}, {Scranton},  {Sekiguchi}, {Sheldon}, {Shimasaku}, {Silvestri}, {Smith}, {Smol{\v c}i{\'  c}}, {Snedden}, {Stebbins}, {Stoughton}, {Strauss}, {SubbaRao}, {Szalay},  {Szapudi}, {Szkody}, {Szokoly}, {Tegmark}, {Teodoro}, {Thakar}, {Tremonti},  {Tucker}, {Uomoto}, {Vanden Berk}, {Vandenberg}, {Vogeley}, {Voges}, {Vogt},  {Walkowicz}, {Wang}, {Weinberg}, {West}, {White}, {Wilhite}, {Xu}, {Yanny},  {Yasuda}, {Yip}, {Yocum}, {York}, {Zehavi}, {Zibetti}, \& {Zucker} 2005]{aaa+05}
{Abazajian}, K., {Adelman-McCarthy}, J.~K., {Ag{\" u}eros}, M.~A., {Allam},  S.~S., {Anderson}, K.~S.~J., {Anderson}, S.~F., {Annis}, J., {Bahcall},  N.~A., {et al.} 2005, \aj, 129, 1755

\bibitem[{Abazajian}, {Adelman-McCarthy}, {Ag{\"  u}eros}, {Allam}, {Anderson}, {Anderson}, {Annis}, {Bahcall}, {Baldry},  {Bastian}, {Berlind}, {Bernardi}, {Blanton}, {Bochanski}, {Boroski},  {Briggs}, {Brinkmann}, {Brunner}, {Budav{\' a}ri}, {Carey}, {Carliles},  {Castander}, {Connolly}, {Csabai}, {Doi}, {Dong}, {Eisenstein}, {Evans},  {Fan}, {Finkbeiner}, {Friedman}, {Frieman}, {Fukugita}, {Gal}, {Gillespie},  {Glazebrook}, {Gray}, {Grebel}, {Gunn}, {Gurbani}, {Hall}, {Hamabe},  {Harris}, {Harris}, {Harvanek}, {Heckman}, {Hendry}, {Hennessy}, {Hindsley},  {Hogan}, {Hogg}, {Holmgren}, {Ichikawa}, {Ichikawa}, {Ivezi{\' c}}, {Jester},  {Johnston}, {Jorgensen}, {Kent}, {Kleinman}, {Knapp}, {Kniazev}, {Kron},  {Krzesinski}, {Kunszt}, {Kuropatkin}, {Lamb}, {Lampeitl}, {Lee}, {Leger},  {Li}, {Lin}, {Loh}, {Long}, {Loveday}, {Lupton}, {Malik}, {Margon},  {Matsubara}, {McGehee}, {McKay}, {Meiksin}, {Munn}, {Nakajima}, {Nash},  {Neilsen}, {Newberg}, {Newman}, {Nichol}, {Nicinski}, {Nieto-Santisteban},  {Nitta}, {Okamura}, {O'Mullane}, {Ostriker}, {Owen}, {Padmanabhan},  {Peoples}, {Pier}, {Pope}, {Quinn}, {Richards}, {Richmond}, {Rix}, {Rockosi},  {Schlegel}, {Schneider}, {Scranton}, {Sekiguchi}, {Seljak}, {Sergey},  {Sesar}, {Sheldon}, {Shimasaku}, {Siegmund}, {Silvestri}, {Smith}, {Smol{\v  c}i{\' c}}, {Snedden}, {Stebbins}, {Stoughton}, {Strauss}, {SubbaRao},  {Szalay}, {Szapudi}, {Szkody}, {Szokoly}, {Tegmark}, {Teodoro}, {Thakar},  {Tremonti}, {Tucker}, {Uomoto}, {Vanden Berk}, {Vandenberg}, {Vogeley},  {Voges}, {Vogt}, {Walkowicz}, {Wang}, {Weinberg}, {West}, {White}, {Wilhite},  {Xu}, {Yanny}, {Yasuda}, {Yip}, {Yocum}, {York}, {Zehavi}, {Zibetti}, \&  {Zucker} 2004]{aaa+04}
{Abazajian}, K., {Adelman-McCarthy}, J.~K., {Ag{\" u}eros}, M.~A., {Allam},  S.~S., {Anderson}, K.~S.~J., {Anderson}, S.~F., {Annis}, J., {Bahcall},  N.~A., {et al.} 2004, \aj, 128, 502

\bibitem[{Abazajian}, {Adelman-McCarthy}, {Ag{\"  u}eros}, {Allam}, {Anderson}, {Annis}, {Bahcall}, {Baldry}, {Bastian},  {Berlind}, {Bernardi}, {Blanton}, {Blythe}, {Bochanski}, {Boroski},  {Brewington}, {Briggs}, {Brinkmann}, {Brunner}, {Budav{\' a}ri}, {Carey},  {Carr}, {Castander}, {Chiu}, {Collinge}, {Connolly}, {Covey}, {Csabai},  {Dalcanton}, {Dodelson}, {Doi}, {Dong}, {Eisenstein}, {Evans}, {Fan},  {Feldman}, {Finkbeiner}, {Friedman}, {Frieman}, {Fukugita}, {Gal},  {Gillespie}, {Glazebrook}, {Gonzalez}, {Gray}, {Grebel}, {Grodnicki}, {Gunn},  {Gurbani}, {Hall}, {Hao}, {Harbeck}, {Harris}, {Harris}, {Harvanek},  {Hawley}, {Heckman}, {Helmboldt}, {Hendry}, {Hennessy}, {Hindsley}, {Hogg},  {Holmgren}, {Holtzman}, {Homer}, {Hui}, {Ichikawa}, {Ichikawa}, {Inkmann},  {Ivezi{\' c}}, {Jester}, {Johnston}, {Jordan}, {Jordan}, {Jorgensen},  {Juri{\' c}}, {Kauffmann}, {Kent}, {Kleinman}, {Knapp}, {Kniazev}, {Kron},  {Krzesi{\' n}ski}, {Kunszt}, {Kuropatkin}, {Lamb}, {Lampeitl}, {Laubscher},  {Lee}, {Leger}, {Li}, {Lidz}, {Lin}, {Loh}, {Long}, {Loveday}, {Lupton},  {Malik}, {Margon}, {McGehee}, {McKay}, {Meiksin}, {Miknaitis}, {Moorthy},  {Munn}, {Murphy}, {Nakajima}, {Narayanan}, {Nash}, {Neilsen}, {Newberg},  {Newman}, {Nichol}, {Nicinski}, {Nieto-Santisteban}, {Nitta}, {Odenkirchen},  {Okamura}, {Ostriker}, {Owen}, {Padmanabhan}, {Peoples}, {Pier}, {Pindor},  {Pope}, {Quinn}, {Rafikov}, {Raymond}, {Richards}, {Richmond}, {Rix},  {Rockosi}, {Schaye}, {Schlegel}, {Schneider}, {Schroeder}, {Scranton},  {Sekiguchi}, {Seljak}, {Sergey}, {Sesar}, {Sheldon}, {Shimasaku}, {Siegmund},  {Silvestri}, {Sinisgalli}, {Sirko}, {Smith}, {Smol{\v c}i{\' c}}, {Snedden},  {Stebbins}, {Steinhardt}, {Stinson}, {Stoughton}, {Strateva}, {Strauss},  {SubbaRao}, {Szalay}, {Szapudi}, {Szkody}, {Tasca}, {Tegmark}, {Thakar},  {Tremonti}, {Tucker}, {Uomoto}, {Vanden Berk}, {Vandenberg}, {Vogeley},  {Voges}, {Vogt}, {Walkowicz}, {Weinberg}, {West}, {White}, {Wilhite},  {Willman}, {Xu}, {Yanny}, {Yarger}, {Yasuda}, {Yip}, {Yocum}, {York},  {Zakamska}, {Zehavi}, {Zheng}, {Zibetti}, \& {Zucker} 2003]{aaa+03}
{Abazajian}, K., {Adelman-McCarthy}, J.~K., {Ag{\" u}eros}, M.~A., {Allam},  S.~S., {Anderson}, S.~F., {Annis}, J., {Bahcall}, N.~A., {Baldry}, I.~K., {et al.} 2003, \aj, 126, 2081

\bibitem[{Adelman-McCarthy}, {Ag{\"u}eros},  {Allam}, {Anderson}, {Anderson}, {Annis}, {Bahcall}, {Baldry}, {Barentine},  {Berlind}, {Bernardi}, {Blanton}, {Boroski}, {Brewington}, {Brinchmann},  {Brinkmann}, {Brunner}, {Budav{\'a}ri}, {Carey}, {Carr}, {Castander},  {Connolly}, {Csabai}, {Czarapata}, {Dalcanton}, {Doi}, {Dong}, {Eisenstein},  {Evans}, {Fan}, {Finkbeiner}, {Friedman}, {Frieman}, {Fukugita}, {Gillespie},  {Glazebrook}, {Gray}, {Grebel}, {Gunn}, {Gurbani}, {de Haas}, {Hall},  {Harris}, {Harvanek}, {Hawley}, {Hayes}, {Hendry}, {Hennessy}, {Hindsley},  {Hirata}, {Hogan}, {Hogg}, {Holmgren}, {Holtzman}, {Ichikawa}, {Ivezi{\'c}},  {Jester}, {Johnston}, {Jorgensen}, {Juri{\'c}}, {Kent}, {Kleinman}, {Knapp},  {Kniazev}, {Kron}, {Krzesinski}, {Kuropatkin}, {Lamb}, {Lampeitl}, {Lee},  {Leger}, {Lin}, {Long}, {Loveday}, {Lupton}, {Margon},  {Mart{\'{\i}}nez-Delgado}, {Mandelbaum}, {Matsubara}, {McGehee}, {McKay},  {Meiksin}, {Munn}, {Nakajima}, {Nash}, {Neilsen}, {Newberg}, {Newman},  {Nichol}, {Nicinski}, {Nieto-Santisteban}, {Nitta}, {O'Mullane}, {Okamura},  {Owen}, {Padmanabhan}, {Pauls}, {Peoples}, {Pier}, {Pope}, {Pourbaix},  {Quinn}, {Richards}, {Richmond}, {Rockosi}, {Schlegel}, {Schneider},  {Schroeder}, {Scranton}, {Seljak}, {Sheldon}, {Shimasaku}, {Smith}, {Smol{\v  c}i{\'c}}, {Snedden}, {Stoughton}, {Strauss}, {SubbaRao}, {Szalay},  {Szapudi}, {Szkody}, {Tegmark}, {Thakar}, {Tucker}, {Uomoto}, {Vanden Berk},  {Vandenberg}, {Vogeley}, {Voges}, {Vogt}, {Walkowicz}, {Weinberg}, {West},  {White}, {Xu}, {Yanny}, {Yocum}, {York}, {Zehavi}, {Zibetti}, \&  {Zucker} 2006]{adel+05}
{Adelman-McCarthy}, J.~K., {Ag{\"u}eros}, M.~A., {Allam}, S.~S., {Anderson},  K.~S.~J., {Anderson}, S.~F., {Annis}, J., {Bahcall}, N.~A., {Baldry}, I.~K., {et al.} 2006, \apjs, 162, 38

\bibitem[{Anderson}, {Voges}, {Margon}, {Tr{\"  u}mper}, {Ag{\" u}eros}, {Boller}, {Collinge}, {Homer}, {Stinson}, {Strauss},  {Annis}, {G{\' o}mez}, {Hall}, {Nichol}, {Richards}, {Schneider}, {Vanden  Berk}, {Fan}, {Ivezi{\' c}}, {Munn}, {Newberg}, {Richmond}, {Weinberg},  {Yanny}, {Bahcall}, {Brinkmann}, {Fukugita}, \& {York} 2003]{avm+03}
{Anderson}, S.~F., {Voges}, W., {Margon}, B., {Tr{\" u}mper}, J., {Ag{\"  u}eros}, M.~A., {Boller}, T., {Collinge}, M.~J., {Homer}, L., {et al.} 2003, \aj, 126, 2209

\bibitem[{Avni} \& {Bahcall} 1980]{ab80}
{Avni}, Y. \& {Bahcall}, J.~N. 1980, \apj, 235, 694

\bibitem[{Barger}, {Cowie}, {Mushotzky}, {Yang},  {Wang}, {Steffen}, \& {Capak} 2005]{bcm+05}
{Barger}, A.~J., {Cowie}, L.~L., {Mushotzky}, R.~F., {Yang}, Y., {Wang}, W.-H.,  {Steffen}, A.~T., \& {Capak}, P. 2005, \aj, 129, 578

\bibitem[{Becker}, {White}, \& {Helfand} 1995]{bwh95}
{Becker}, R.~H., {White}, R.~L., \& {Helfand}, D.~J. 1995, \apj, 450, 559

\bibitem[{Begelman} 2004]{beg04}
{Begelman}, M.~C. 2004, in Coevolution of Black Holes and Galaxies, ed. L.~Ho  (Cambridge University Press), 375

\bibitem[{Blanton}, {Hogg}, {Bahcall},  {Brinkmann}, {Britton}, {Connolly}, {Csabai}, {Fukugita}, {Loveday},  {Meiksin}, {Munn}, {Nichol}, {Okamura}, {Quinn}, {Schneider}, {Shimasaku},  {Strauss}, {Tegmark}, {Vogeley}, \& {Weinberg} 2003a]{bhb+03}
{Blanton}, M.~R., {Hogg}, D.~W., {Bahcall}, N.~A., {Brinkmann}, J., {Britton},  M., {Connolly}, A.~J., {Csabai}, I., {Fukugita}, M., {et al.} 2003a, \apj, 592, 819

\bibitem[{Blanton}, {Lin}, {Lupton},  {Maley}, {Young}, {Zehavi}, \& {Loveday} 2003b]{blm+03}
{Blanton}, M.~R., {Lin}, H., {Lupton}, R.~H., {Maley}, F.~M., {Young}, N.,  {Zehavi}, I., \& {Loveday}, J. 2003b, \aj, 125, 2276

\bibitem[{Boyle}, {Shanks}, {Croom}, {Smith}, {Miller},  {Loaring}, \& {Heymans} 2000]{bsc+00}
{Boyle}, B.~J., {Shanks}, T., {Croom}, S.~M., {Smith}, R.~J., {Miller}, L.,  {Loaring}, N., \& {Heymans}, C. 2000, \mnras, 317, 1014

\bibitem[{Boyle}, {Shanks}, \& {Peterson} 1988]{bsp88}
{Boyle}, B.~J., {Shanks}, T., \& {Peterson}, B.~A. 1988, \mnras, 235, 935

\bibitem[{Brown}, {Brand}, {Dey}, {Jannuzi}, {Cool}, {Le  Floc'h}, {Kochanek}, {Armus}, {Bian}, {Higdon}, {Higdon}, {Papovich},  {Rieke}, {Rieke}, {Smith}, {Soifer}, \& {Weedman} 2006]{Brown05}
{Brown}, M.~J.~I., {Brand}, K., {Dey}, A., {Jannuzi}, B.~T., {Cool}, R., {Le  Floc'h}, E., {Kochanek}, C.~S., {Armus}, L., {et al.} 2006, \apj, 638, 88

\bibitem[{Chiu} 2004]{Chiu04}
{Chiu}, K. 2004, \baas, 205, 167.04

\bibitem[{Collinge}, {Strauss}, {Hall}, {Ivezi{\'  c}}, {Munn}, {Schlegel}, {Zakamska}, {Anderson}, {Harris}, {Richards},  {Schneider}, {Voges}, {York}, {Margon}, \& {Brinkmann} 2005]{csh+05}
{Collinge}, M.~J., {Strauss}, M.~A., {Hall}, P.~B., {Ivezi{\' c}}, {\v Z}.,  {Munn}, J.~A., {Schlegel}, D.~J., {Zakamska}, N.~L., {Anderson}, S.~F., {et al.} 2005, \aj, 129, 2542

\bibitem[{Cowie}, {Barger}, {Bautz}, {Brandt}, \&  {Garmire} 2003]{cbb+03}
{Cowie}, L.~L., {Barger}, A.~J., {Bautz}, M.~W., {Brandt}, W.~N., \& {Garmire},  G.~P. 2003, \apjl, 584, L57

\bibitem[{Croom}, {Smith}, {Boyle}, {Shanks}, {Miller},  {Outram}, \& {Loaring} 2004]{csb+04}
{Croom}, S.~M., {Smith}, R.~J., {Boyle}, B.~J., {Shanks}, T., {Miller}, L.,  {Outram}, P.~J., \& {Loaring}, N.~S. 2004, \mnras, 349, 1397

\bibitem[{Di Matteo}, {Croft}, {Springel}, \&  {Hernquist} 2003]{dcs+03}
{Di Matteo}, T., {Croft}, R.~A.~C., {Springel}, V., \& {Hernquist}, L. 2003,  \apj, 593, 56

\bibitem[{Di Matteo}, {Springel}, \&  {Hernquist} 2005]{dsh05}
{Di Matteo}, T., {Springel}, V., \& {Hernquist}, L. 2005, \nat, 433, 604

\bibitem[{Elvis}, {Wilkes}, {McDowell}, {Green},  {Bechtold}, {Willner}, {Oey}, {Polomski}, \& {Cutri} 1994]{ewm+94}
{Elvis}, M., {Wilkes}, B.~J., {McDowell}, J.~C., {Green}, R.~F., {Bechtold},  J., {Willner}, S.~P., {Oey}, M.~S., {Polomski}, E., {et al.} 1994,  \apjs, 95, 1

\bibitem[{Fabian} 1999]{fab99}
{Fabian}, A.~C. 1999, \mnras, 308, L39

\bibitem[{Fan} 1999]{fan99}
{Fan}, X. 1999, \aj, 117, 2528

\bibitem[{Fan}, {Hennawi}, {Richards}, {Strauss},  {Schneider}, {Donley}, {Young}, {Annis}, {Lin}, {Lampeitl}, {Lupton}, {Gunn},  {Knapp}, {Brandt}, {Anderson}, {Bahcall}, {Brinkmann}, {Brunner}, {Fukugita},  {Szalay}, {Szokoly}, \& {York} 2004]{fhr+04}
{Fan}, X., {Hennawi}, J.~F., {Richards}, G.~T., {Strauss}, M.~A., {Schneider},  D.~P., {Donley}, J.~L., {Young}, J.~E., {Annis}, J., {et al.} 2004,  \aj, 128, 515

\bibitem[{Fan}, {Strauss}, {Schneider}, {Gunn}, {Lupton},  {Becker}, {Davis}, {Newman}, {Richards}, {White}, {Anderson}, {Annis},  {Bahcall}, {Brunner}, {Csabai}, {Hennessy}, {Hindsley}, {Fukugita}, {Kunszt},  {Ivezi{\' c}}, {Knapp}, {McKay}, {Munn}, {Pier}, {Szalay}, \&  {York} 2001]{fss+01}
{Fan}, X., {Strauss}, M.~A., {Schneider}, D.~P., {Gunn}, J.~E., {Lupton},  R.~H., {Becker}, R.~H., {Davis}, M., {Newman}, J.~A., {et al.} 2001, \aj,  121, 54

\bibitem[{Ferrarese} \& {Merritt} 2000]{fm00}
{Ferrarese}, L. \& {Merritt}, D. 2000, \apjl, 539, L9

\bibitem[{Finlator}, {Ivezi{\' c}}, {Fan}, {Strauss},  {Knapp}, {Lupton}, {Gunn}, {Rockosi}, {Anderson}, {Csabai}, {Hennessy},  {Hindsley}, {McKay}, {Nichol}, {Schneider}, {Smith}, {York}, \& {the SDSS  Collaboration} 2000]{fif+00}
{Finlator}, K., {Ivezi{\' c}}, {\v Z}., {Fan}, X., {Strauss}, M.~A., {Knapp},  G.~R., {Lupton}, R.~H., {Gunn}, J.~E., {Rockosi}, C.~M., {et al.} 2000, \aj, 120, 2615

\bibitem[{Fukugita}, {Ichikawa}, {Gunn}, {Doi},  {Shimasaku}, \& {Schneider} 1996]{fig+96}
{Fukugita}, M., {Ichikawa}, T., {Gunn}, J.~E., {Doi}, M., {Shimasaku}, K., \&  {Schneider}, D.~P. 1996, \aj, 111, 1748

\bibitem[{Gebhardt}, {Bender}, {Bower}, {Dressler},  {Faber}, {Filippenko}, {Green}, {Grillmair}, {Ho}, {Kormendy}, {Lauer},  {Magorrian}, {Pinkney}, {Richstone}, \& {Tremaine} 2000]{gbb+00}
{Gebhardt}, K., {Bender}, R., {Bower}, G., {Dressler}, A., {Faber}, S.~M.,  {Filippenko}, A.~V., {Green}, R., {Grillmair}, C., {et al.} 2000, \apjl, 539, L13

\bibitem[{Goldschmidt}, {Miller}, {La Franca}, \&  {Cristiani} 1992]{gmf+92}
{Goldschmidt}, P., {Miller}, L., {La Franca}, F., \& {Cristiani}, S. 1992,  \mnras, 256, 65P

\bibitem[{Granato}, {De Zotti}, {Silva}, {Bressan}, \&  {Danese} 2004]{gds+04}
{Granato}, G.~L., {De Zotti}, G., {Silva}, L., {Bressan}, A., \& {Danese}, L.  2004, \apj, 600, 580

\bibitem[{Gunn}, {Carr}, {Rockosi}, {Sekiguchi}, {Berry},  {Elms}, {de Haas}, {Ivezi{\' c} }, {Knapp}, {Lupton}, {Pauls}, {Simcoe},  {Hirsch}, {Sanford}, {Wang}, {York}, {Harris}, {Annis}, {Bartozek},  {Boroski}, {Bakken}, {Haldeman}, {Kent}, {Holm}, {Holmgren}, {Petravick},  {Prosapio}, {Rechenmacher}, {Doi}, {Fukugita}, {Shimasaku}, {Okada}, {Hull},  {Siegmund}, {Mannery}, {Blouke}, {Heidtman}, {Schneider}, {Lucinio}, \&  {Brinkman} 1998]{gcr+98}
{Gunn}, J.~E., {Carr}, M., {Rockosi}, C., {Sekiguchi}, M., {Berry}, K., {Elms},  B., {de Haas}, E., {Ivezi{\' c} }, {\v Z}., {et al.} 1998, \aj,  116, 3040

\bibitem[{Gunn et al.} 2005]{gunn+05}
{Gunn et al.} 2005, \aj, submitted

\bibitem[{Haas}, {Siebenmorgen}, {Leipski}, {Ott},  {Cunow}, {Meusinger}, {M{\" u}ller}, {Chini}, \& {Schartel} 2004]{hsl+04}
{Haas}, M., {Siebenmorgen}, R., {Leipski}, C., {Ott}, S., {Cunow}, B.,  {Meusinger}, H., {M{\" u}ller}, S.~A.~H., {Chini}, R., {et al.} 2004, \aap, 419, L49

\bibitem[{Hao}, {Strauss}, {Fan},  {Tremonti}, {Schlegel}, {Heckman}, {Kauffmann}, {Blanton}, {Gunn}, {Hall},  {Ivezi{\' c}}, {Knapp}, {Krolik}, {Lupton}, {Richards}, {Schneider},  {Strateva}, {Zakamska}, {Brinkmann}, \& {Szokoly} 2005a]{hao05}
{Hao}, L., {Strauss}, M.~A., {Fan}, X., {Tremonti}, C.~A., {Schlegel}, D.~J.,  {Heckman}, T.~M., {Kauffmann}, G., {Blanton}, M.~R., {et al.} 2005a, \aj, 129,  1795

\bibitem[{Hao}, {Strauss}, {Tremonti},  {Schlegel}, {Heckman}, {Kauffmann}, {Blanton}, {Fan}, {Gunn}, {Hall},  {Ivezi{\' c}}, {Knapp}, {Krolik}, {Lupton}, {Richards}, {Schneider},  {Strateva}, {Zakamska}, {Brinkmann}, {Brunner}, \& {Szokoly} 2005b]{hst+05}
{Hao}, L., {Strauss}, M.~A., {Tremonti}, C.~A., {Schlegel}, D.~J., {Heckman},  T.~M., {Kauffmann}, G., {Blanton}, M.~R., {Fan}, X., {et al.} 2005b, \aj, 129, 1783

\bibitem[{Hasinger}, {Miyaji}, \& {Schmidt} 2005]{hms05}
{Hasinger}, G., {Miyaji}, T., \& {Schmidt}, M. 2005, \aap, 441, 417

\bibitem[{Heckman} 1980]{hec80}
{Heckman}, T.~M. 1980, \aap, 87, 152

\bibitem[{Heckman}, {Kauffmann}, {Brinchmann},  {Charlot}, {Tremonti}, \& {White} 2004]{hkb+04}
{Heckman}, T.~M., {Kauffmann}, G., {Brinchmann}, J., {Charlot}, S., {Tremonti},  C., \& {White}, S.~D.~M. 2004, \apj, 613, 109

\bibitem[{Hewett}, {Foltz}, \& {Chaffee} 1993]{hfc93}
{Hewett}, P.~C., {Foltz}, C.~B., \& {Chaffee}, F.~H. 1993, \apjl, 406, L43

\bibitem[{Hogg} 1999]{hog99}
{Hogg}, D.~W. 1999, astro-ph/9905116

\bibitem[{Hogg}, {Baldry}, {Blanton}, \&  {Eisenstein} 2002]{hbb+02}
{Hogg}, D.~W., {Baldry}, I.~K., {Blanton}, M.~R., \& {Eisenstein}, D.~J. 2002,  astro-ph/0210394

\bibitem[{Hogg}, {Finkbeiner}, {Schlegel}, \&  {Gunn} 2001]{hfs+01}
{Hogg}, D.~W., {Finkbeiner}, D.~P., {Schlegel}, D.~J., \& {Gunn}, J.~E. 2001,  \aj, 122, 2129

\bibitem[{Hopkins}, {Hernquist}, {Cox}, {Di Matteo},  {Martini}, {Robertson}, \& {Springel} 2005a]{hhc+05}
{Hopkins}, P.~F., {Hernquist}, L., {Cox}, T.~J., {Di Matteo}, T., {Martini},  P., {Robertson}, B., \& {Springel}, V. 2005a, \apj, 630, 705

\bibitem[{Hopkins}, {Strauss}, {Hall}, {Richards},  {Cooper}, {Schneider}, {Vanden Berk}, {Jester}, {Brinkmann}, \&  {Szokoly} 2004]{hsh+04}
{Hopkins}, P.~F., {Strauss}, M.~A., {Hall}, P.~B., {Richards}, G.~T., {Cooper},  A.~S., {Schneider}, D.~P., {Vanden Berk}, D.~E., {Jester}, S., {et al.} 2004, \aj, 128, 1112

\bibitem[{Hopkins et al.} 2005b]{hhc+05z}
{Hopkins et al.} 2005b, astro-ph/0506398

\bibitem[{Hubeny}, {Agol}, {Blaes}, \&  {Krolik} 2000]{hab+00}
{Hubeny}, I., {Agol}, E., {Blaes}, O., \& {Krolik}, J.~H. 2000, \apj, 533, 710

\bibitem[{Hunt}, {Steidel}, {Adelberger}, \&  {Shapley} 2004]{hsa+04}
{Hunt}, M.~P., {Steidel}, C.~C., {Adelberger}, K.~L., \& {Shapley}, A.~E. 2004,  \apj, 605, 625

\bibitem[{Ivezi{\' c}}, {Lupton}, {Schlegel},  {Boroski}, {Adelman-McCarthy}, {Yanny}, {Kent}, {Stoughton}, {Finkbeiner},  {Padmanabhan}, {Rockosi}, {Gunn}, {Knapp}, {Strauss}, {Richards},  {Eisenstein}, {Nicinski}, {Kleinman}, {Krzesinski}, {Newman}, {Snedden},  {Thakar}, {Szalay}, {Munn}, {Smith}, {Tucker}, \& {Lee} 2004]{ils+04}
{Ivezi{\' c}}, {\v Z}., {Lupton}, R.~H., {Schlegel}, D., {Boroski}, B.,  {Adelman-McCarthy}, J., {Yanny}, B., {Kent}, S., {Stoughton}, C., {et al.} 2004, Astronomische Nachrichten, 325, 583

\bibitem[{Ivezi{\' c}}, {Menou}, {Knapp},  {Strauss}, {Lupton}, {Vanden Berk}, {Richards}, {Tremonti}, {Weinstein},  {Anderson}, {Bahcall}, {Becker}, {Bernardi}, {Blanton}, {Eisenstein}, {Fan},  {Finkbeiner}, {Finlator}, {Frieman}, {Gunn}, {Hall}, {Kim}, {Kinkhabwala},  {Narayanan}, {Rockosi}, {Schlegel}, {Schneider}, {Strateva}, {SubbaRao},  {Thakar}, {Voges}, {White}, {Yanny}, {Brinkmann}, {Doi}, {Fukugita},  {Hennessy}, {Munn}, {Nichol}, \& {York} 2002]{imk+02}
{Ivezi{\' c}}, {\v Z}., {Menou}, K., {Knapp}, G.~R., {Strauss}, M.~A.,  {Lupton}, R.~H., {Vanden Berk}, D.~E., {Richards}, G.~T., {Tremonti}, C., {et al.} 2002,  \aj, 124, 2364

\bibitem[{Jester}, {Schneider}, {Richards}, {Green},  {Schmidt}, {Hall}, {Strauss}, {Vanden Berk}, {Stoughton}, {Gunn},  {Brinkmann}, {Kent}, {Smith}, {Tucker}, \& {Yanny} 2005]{jsr+05}
{Jester}, S., {Schneider}, D.~P., {Richards}, G.~T., {Green}, R.~F., {Schmidt},  M., {Hall}, P.~B., {Strauss}, M.~A., {Vanden Berk}, D.~E., {et al.} 2005, \aj, 130, 873

\bibitem[{Jiang et al.} 2006]{jiang05}
{Jiang et al.} 2006, submitted

\bibitem[{Kennefick}, {Djorgovski}, \& {de  Carvalho} 1995]{kdd+95}
{Kennefick}, J.~D., {Djorgovski}, S.~G., \& {de Carvalho}, R.~R. 1995, \aj,  110, 2553

\bibitem[{Koo} \& {Kron} 1988]{kk88}
{Koo}, D.~C. \& {Kron}, R.~G. 1988, \apj, 325, 92

\bibitem[{Lawrence} 1991]{law91}
{Lawrence}, A. 1991, \mnras, 252, 586

\bibitem[{Lupton}, {Gunn}, {Ivezi{\' c}}, {Knapp},  {Kent}, \& {Yasuda} 2001]{lgi+01}
{Lupton}, R.~H., {Gunn}, J.~E., {Ivezi{\' c}}, Z., {Knapp}, G.~R., {Kent}, S.,  \& {Yasuda}, N. 2001, in ASP Conf. Ser. 238: Astronomical Data Analysis  Software and Systems X, Vol.~10, 269

\bibitem[{Lupton}, {Gunn}, \& {Szalay} 1999]{lgs+99}
{Lupton}, R.~H., {Gunn}, J.~E., \& {Szalay}, A.~S. 1999, \aj, 118, 1406

\bibitem[{Lupton}, {Ivezi{\' c}}, {Gunn}, {Knapp},  {Strauss}, \& {Yasuda} 2002]{lig+02}
{Lupton}, R.~H., {Ivezi{\' c}}, {\v Z}., {Gunn}, J.~E., {Knapp}, G., {Strauss},  M.~A., \& {Yasuda}, N. 2002, in Survey and Other Telescope Technologies and  Discoveries. Edited by Tyson, J. Anthony; Wolff, Sidney. Proceedings of the  SPIE, Volume 4836, 350

\bibitem[{Marshall} 1985]{mar85}
{Marshall}, H.~L. 1985, \apj, 299, 109

\bibitem[{Meiksin} 2005]{mei05}
{Meiksin}, A. 2005, \mnras, 356, 596

\bibitem[{Merloni} 2004]{mer04}
{Merloni}, A. 2004, \mnras, 353, 1035

\bibitem[{Oke} \& {Gunn} 1983]{og83}
{Oke}, J.~B. \& {Gunn}, J.~E. 1983, \apj, 266, 713

\bibitem[{Oke} \& {Sandage} 1968]{os68}
{Oke}, J.~B. \& {Sandage}, A. 1968, \apj, 154, 21

\bibitem[{Osmer} 1982]{osm82}
{Osmer}, P.~S. 1982, \apj, 253, 28

\bibitem[{Page} \& {Carrera} 2000]{pc00}
{Page}, M.~J. \& {Carrera}, F.~J. 2000, \mnras, 311, 433

\bibitem[{Pei} 1995]{pei95}
{Pei}, Y.~C. 1995, \apj, 438, 623

\bibitem[{Peterson} 1997]{pet97}
{Peterson}, B.~M. 1997, {An Introduction to Active Galactic Nuclei} (Cambridge  University Press)

\bibitem[{Pier}, {Munn}, {Hindsley}, {Hennessy}, {Kent},  {Lupton}, \& {Ivezi{\' c}} 2003]{pmh+03}
{Pier}, J.~R., {Munn}, J.~A., {Hindsley}, R.~B., {Hennessy}, G.~S., {Kent},  S.~M., {Lupton}, R.~H., \& {Ivezi{\' c}}, {\v Z}. 2003, \aj, 125, 1559

\bibitem[{Press}, {Teukolsky}, {Vetterling}, \&  {Flannery} 1992]{ptv+92}
{Press}, W.~H., {Teukolsky}, S.~A., {Vetterling}, W.~T., \& {Flannery}, B.~P.  1992, {Numerical recipes in C. The art of scientific computing} (Cambridge:  University Press)

\bibitem[{Richards}, {Croom}, {Anderson},  {Bland-Hawthorn}, {Boyle}, {De Propris}, {Drinkwater}, {Fan}, {Gunn},  {Ivezi{\' c}}, {Jester}, {Loveday}, {Meiksin}, {Miller}, {Myers}, {Nichol},  {Outram}, {Pimbblet}, {Roseboom}, {Ross}, {Schneider}, {Shanks}, {Sharp},  {Stoughton}, {Strauss}, {Szalay}, {Vanden Berk}, \& {York} 2005]{rca+05}
{Richards}, G.~T., {Croom}, S.~M., {Anderson}, S.~F., {Bland-Hawthorn}, J.,  {Boyle}, B.~J., {De Propris}, R., {Drinkwater}, M.~J., {Fan}, X., {et al.} 2005, \mnras, 360, 839

\bibitem[{Richards}, {Fan}, {Newberg}, {Strauss},  {Vanden Berk}, {Schneider}, {Yanny}, {Boucher}, {Burles}, {Frieman}, {Gunn},  {Hall}, {Ivezi{\' c}}, {Kent}, {Loveday}, {Lupton}, {Rockosi}, {Schlegel},  {Stoughton}, {SubbaRao}, \& {York} 2002]{rfn+02}
{Richards}, G.~T., {Fan}, X., {Newberg}, H.~J., {Strauss}, M.~A., {Vanden  Berk}, D.~E., {Schneider}, D.~P., {Yanny}, B., {Boucher}, A., {et al.} 2002, \aj, 123, 2945

\bibitem[{Richards}, {Fan}, {Schneider}, {Vanden  Berk}, {Strauss}, {York}, {Anderson}, {Anderson}, {Annis}, {Bahcall},  {Bernardi}, {Briggs}, {Brinkmann}, {Brunner}, {Burles}, {Carey}, {Castander},  {Connolly}, {Crocker}, {Csabai}, {Doi}, {Finkbeiner}, {Friedman}, {Frieman},  {Fukugita}, {Gunn}, {Hindsley}, {Ivezi{\' c}}, {Kent}, {Knapp}, {Lamb},  {Leger}, {Long}, {Loveday}, {Lupton}, {McKay}, {Meiksin}, {Merrelli}, {Munn},  {Newberg}, {Newcomb}, {Nichol}, {Owen}, {Pier}, {Pope}, {Richmond},  {Rockosi}, {Schlegel}, {Siegmund}, {Smee}, {Snir}, {Stoughton}, {Stubbs},  {SubbaRao}, {Szalay}, {Szokoly}, {Tremonti}, {Uomoto}, {Waddell}, {Yanny}, \&  {Zheng} 2001]{rfs+01}
{Richards}, G.~T., {Fan}, X., {Schneider}, D.~P., {Vanden Berk}, D.~E.,  {Strauss}, M.~A., {York}, D.~G., {Anderson}, J.~E., {Anderson}, S.~F., {et al.} 2001, \aj, 121, 2308

\bibitem[{Richards}, {Haiman}, {Pindor}, {Strauss},  {Fan}, {Eisenstein}, {Schneider}, {Bahcall}, {Brinkmann}, \&  {Fukugita} 2006]{rhp+06}
{Richards}, G.~T., {Haiman}, Z., {Pindor}, B., {Strauss}, M.~A., {Fan}, X.,  {Eisenstein}, D., {Schneider}, D.~P., {Bahcall}, N.~A., {et al.} 2006, \aj, 131, 49

\bibitem[{Richards}, {Hall}, {Vanden Berk},  {Strauss}, {Schneider}, {Weinstein}, {Reichard}, {York}, {Knapp}, {Fan},  {Ivezi{\' c}}, {Brinkmann}, {Budav{\' a}ri}, {Csabai}, \& {Nichol} 2003]{rhv+03}
{Richards}, G.~T., {Hall}, P.~B., {Vanden Berk}, D.~E., {Strauss}, M.~A.,  {Schneider}, D.~P., {Weinstein}, M.~A., {Reichard}, T.~A., {York}, D.~G., {et al.} 2003, \aj, 126, 1131

\bibitem[{Richards}, {Nichol}, {Gray},  {Brunner}, {Lupton}, {Vanden Berk}, {Chong}, {Weinstein}, {Schneider},  {Anderson}, {Munn}, {Harris}, {Strauss}, {Fan}, {Gunn}, {Ivezi{\' c}},  {York}, {Brinkmann}, \& {Moore} 2004a]{rng+04}
{Richards}, G.~T., {Nichol}, R.~C., {Gray}, A.~G., {Brunner}, R.~J., {Lupton},  R.~H., {Vanden Berk}, D.~E., {Chong}, S.~S., {Weinstein}, M.~A., {et al.} 2004a, \apjs, 155, 257

\bibitem[{Richards}, {Strauss},  {Pindor}, {Haiman}, {Fan}, {Eisenstein}, {Schneider}, {Bahcall}, {Brinkmann},  \& {Brunner} 2004b]{rsp+04}
{Richards}, G.~T., {Strauss}, M.~A., {Pindor}, B., {Haiman}, Z., {Fan}, X.,  {Eisenstein}, D., {Schneider}, D.~P., {Bahcall}, N.~A., {et al.} 2004b, \aj, 127, 1305

\bibitem[{Scannapieco} \& {Oh} 2004]{so04}
{Scannapieco}, E. \& {Oh}, S.~P. 2004, \apj, 608, 62

\bibitem[{Schlegel}, {Finkbeiner}, \&  {Davis} 1998]{sfd98}
{Schlegel}, D.~J., {Finkbeiner}, D.~P., \& {Davis}, M. 1998, \apj, 500, 525

\bibitem[{Schmidt} 1963]{sch63}
{Schmidt}, M. 1963, \nat, 197, 1040

\bibitem[{Schmidt} 1968]{sch68}
---. 1968, \apj, 151, 393

\bibitem[{Schmidt} \& {Green} 1983]{sg83}
{Schmidt}, M. \& {Green}, R.~F. 1983, \apj, 269, 352

\bibitem[{Schmidt}, {Schneider}, \& {Gunn} 1995]{ssg95}
{Schmidt}, M., {Schneider}, D.~P., \& {Gunn}, J.~E. 1995, \aj, 110, 68

\bibitem[{Schneider}, {Fan}, {Hall}, {Jester},  {Richards}, {Stoughton}, {Strauss}, {SubbaRao}, {Vanden Berk}, {Anderson},  {Brandt}, {Gunn}, {Gray}, {Trump}, {Voges}, {Yanny}, {Bahcall}, {Blanton},  {Boroski}, {Brinkmann}, {Brunner}, {Burles}, {Castander}, {Doi},  {Eisenstein}, {Frieman}, {Fukugita}, {Heckman}, {Hennessy}, {Ivezi{\' c}},  {Kent}, {Knapp}, {Lamb}, {Lee}, {Loveday}, {Lupton}, {Margon}, {Meiksin},  {Munn}, {Newberg}, {Nichol}, {Niederste-Ostholt}, {Pier}, {Richmond},  {Rockosi}, {Saxe}, {Schlegel}, {Szalay}, {Thakar}, {Uomoto}, \&  {York} 2003]{sfh+03}
{Schneider}, D.~P., {Fan}, X., {Hall}, P.~B., {Jester}, S., {Richards}, G.~T.,  {Stoughton}, C., {Strauss}, M.~A., {SubbaRao}, M., {et al.} 2003, \aj, 126, 2579

\bibitem[{Schneider}, {Hall}, {Richards}, {Vanden  Berk}, {Anderson}, {Fan}, {Jester}, {Stoughton}, {Strauss}, {SubbaRao},  {Brandt}, {Gunn}, {Yanny}, {Bahcall}, {Barentine}, {Blanton}, {Boroski},  {Brewington}, {Brinkmann}, {Brunner}, {Csabai}, {Doi}, {Eisenstein},  {Frieman}, {Fukugita}, {Gray}, {Harvanek}, {Heckman}, {Ivezi{\' c}}, {Kent},  {Kleinman}, {Knapp}, {Kron}, {Krzesinski}, {Long}, {Loveday}, {Lupton},  {Margon}, {Munn}, {Neilsen}, {Newberg}, {Newman}, {Nichol}, {Nitta}, {Pier},  {Rockosi}, {Saxe}, {Schlegel}, {Snedden}, {Szalay}, {Thakar}, {Uomoto},  {Voges}, \& {York} 2005]{shr+05}
{Schneider}, D.~P., {Hall}, P.~B., {Richards}, G.~T., {Vanden Berk}, D.~E.,  {Anderson}, S.~F., {Fan}, X., {Jester}, S., {Stoughton}, C., {et al.} 2005, \aj, 130, 367

\bibitem[{Schneider}, {Ehlers}, \& {Falco} 1992]{sef92}
{Schneider}, P., {Ehlers}, J., \& {Falco}, E.~E. 1992, {Gravitational Lenses}  (Springer-Verlag Berlin)

\bibitem[{Scott}, {Kriss}, {Brotherton}, {Green},  {Hutchings}, {Shull}, \& {Zheng} 2004]{skb+04}
{Scott}, J.~E., {Kriss}, G.~A., {Brotherton}, M., {Green}, R.~F., {Hutchings},  J., {Shull}, J.~M., \& {Zheng}, W. 2004, \apj, 615, 135

\bibitem[{Scranton}, {Johnston}, {Dodelson},  {Frieman}, {Connolly}, {Eisenstein}, {Gunn}, {Hui}, {Jain}, {Kent},  {Loveday}, {Narayanan}, {Nichol}, {O'Connell}, {Scoccimarro}, {Sheth},  {Stebbins}, {Strauss}, {Szalay}, {Szapudi}, {Tegmark}, {Vogeley}, {Zehavi},  {Annis}, {Bahcall}, {Brinkman}, {Csabai}, {Hindsley}, {Ivezi{\' c}}, {Kim},  {Knapp}, {Lamb}, {Lee}, {Lupton}, {McKay}, {Munn}, {Peoples}, {Pier},  {Richards}, {Rockosi}, {Schlegel}, {Schneider}, {Stoughton}, {Tucker},  {Yanny}, \& {York} 2002]{sjd+02}
{Scranton}, R., {Johnston}, D., {Dodelson}, S., {Frieman}, J.~A., {Connolly},  A., {Eisenstein}, D.~J., {Gunn}, J.~E., {Hui}, L., {et al.} 2002,  \apj, 579, 48

\bibitem[{Seyfert} 1943]{sey43}
{Seyfert}, C.~K. 1943, \apj, 97, 28

\bibitem[{Silk} \& {Rees} 1998]{sr98}
{Silk}, J. \& {Rees}, M.~J. 1998, \aap, 331, L1

\bibitem[{Smith}, {Tucker}, {Kent}, {Richmond},  {Fukugita}, {Ichikawa}, {Ichikawa}, {Jorgensen}, {Uomoto}, {Gunn}, {Hamabe},  {Watanabe}, {Tolea}, {Henden}, {Annis}, {Pier}, {McKay}, {Brinkmann}, {Chen},  {Holtzman}, {Shimasaku}, \& {York} 2002]{stk+02}
{Smith}, J.~A., {Tucker}, D.~L., {Kent}, S., {Richmond}, M.~W., {Fukugita}, M.,  {Ichikawa}, T., {Ichikawa}, S., {Jorgensen}, A.~M., {et al.} 2002, \aj, 123, 2121

\bibitem[{Spergel}, {Verde}, {Peiris}, {Komatsu},  {Nolta}, {Bennett}, {Halpern}, {Hinshaw}, {Jarosik}, {Kogut}, {Limon},  {Meyer}, {Page}, {Tucker}, {Weiland}, {Wollack}, \& {Wright} 2003]{svp+03}
{Spergel}, D.~N., {Verde}, L., {Peiris}, H.~V., {Komatsu}, E., {Nolta}, M.~R.,  {Bennett}, C.~L., {Halpern}, M., {Hinshaw}, G., {et al.} 2003, \apjs, 148, 175

\bibitem[{Stoughton}, {Lupton}, {Bernardi},  {Blanton}, {Burles}, {Castander}, {Connolly}, {Eisenstein}, {Frieman},  {Hennessy}, {Hindsley}, {Ivezi{\' c}}, {Kent}, {Kunszt}, {Lee}, {Meiksin},  {Munn}, {Newberg}, {Nichol}, {Nicinski}, {Pier}, {Richards}, {Richmond},  {Schlegel}, {Smith}, {Strauss}, {SubbaRao}, {Szalay}, {Thakar}, {Tucker},  {Vanden Berk}, {Yanny}, {Adelman}, {Anderson}, {Anderson}, {Annis},  {Bahcall}, {Bakken}, {Bartelmann}, {Bastian}, {Bauer}, {Berman}, {B{\"  o}hringer}, {Boroski}, {Bracker}, {Briegel}, {Briggs}, {Brinkmann},  {Brunner}, {Carey}, {Carr}, {Chen}, {Christian}, {Colestock}, {Crocker},  {Csabai}, {Czarapata}, {Dalcanton}, {Davidsen}, {Davis}, {Dehnen},  {Dodelson}, {Doi}, {Dombeck}, {Donahue}, {Ellman}, {Elms}, {Evans}, {Eyer},  {Fan}, {Federwitz}, {Friedman}, {Fukugita}, {Gal}, {Gillespie}, {Glazebrook},  {Gray}, {Grebel}, {Greenawalt}, {Greene}, {Gunn}, {de Haas}, {Haiman},  {Haldeman}, {Hall}, {Hamabe}, {Hansen}, {Harris}, {Harris}, {Harvanek},  {Hawley}, {Hayes}, {Heckman}, {Helmi}, {Henden}, {Hogan}, {Hogg}, {Holmgren},  {Holtzman}, {Huang}, {Hull}, {Ichikawa}, {Ichikawa}, {Johnston}, {Kauffmann},  {Kim}, {Kimball}, {Kinney}, {Klaene}, {Kleinman}, {Klypin}, {Knapp},  {Korienek}, {Krolik}, {Kron}, {Krzesi{\' n}ski}, {Lamb}, {Leger},  {Limmongkol}, {Lindenmeyer}, {Long}, {Loomis}, {Loveday}, {MacKinnon},  {Mannery}, {Mantsch}, {Margon}, {McGehee}, {McKay}, {McLean}, {Menou},  {Merelli}, {Mo}, {Monet}, {Nakamura}, {Narayanan}, {Nash}, {Neilsen},  {Newman}, {Nitta}, {Odenkirchen}, {Okada}, {Okamura}, {Ostriker}, {Owen},  {Pauls}, {Peoples}, {Peterson}, {Petravick}, {Pope}, {Pordes}, {Postman},  {Prosapio}, {Quinn}, {Rechenmacher}, {Rivetta}, {Rix}, {Rockosi}, {Rosner},  {Ruthmansdorfer}, {Sandford}, {Schneider}, {Scranton}, {Sekiguchi}, {Sergey},  {Sheth}, {Shimasaku}, {Smee}, {Snedden}, {Stebbins}, {Stubbs}, {Szapudi},  {Szkody}, {Szokoly}, {Tabachnik}, {Tsvetanov}, {Uomoto}, {Vogeley}, {Voges},  {Waddell}, {Walterbos}, {Wang}, {Watanabe}, {Weinberg}, {White}, {White},  {Wilhite}, {Wolfe}, {Yasuda}, {York}, {Zehavi}, \& {Zheng} 2002]{slb+02}
{Stoughton}, C., {Lupton}, R.~H., {Bernardi}, M., {Blanton}, M.~R., {Burles},  S., {Castander}, F.~J., {Connolly}, A.~J., {Eisenstein}, D.~J., {et al.} 2002, \aj, 123, 485

\bibitem[{Strateva}, {Brandt}, {Schneider}, {Vanden  Berk}, \& {Vignali} 2005]{sbs+05}
{Strateva}, I.~V., {Brandt}, W.~N., {Schneider}, D.~P., {Vanden Berk}, D.~G.,  \& {Vignali}, C. 2005, \aj, 130, 387

\bibitem[{Tananbaum}, {Avni}, {Branduardi}, {Elvis},  {Fabbiano}, {Feigelson}, {Giacconi}, {Henry}, {Pye}, {Soltan}, \&  {Zamorani} 1979]{tab+79}
{Tananbaum}, H., {Avni}, Y., {Branduardi}, G., {Elvis}, M., {Fabbiano}, G.,  {Feigelson}, E., {Giacconi}, R., {Henry}, J.~P., {et al.} 1979, \apjl, 234, L9

\bibitem[{Telfer}, {Zheng}, {Kriss}, \&  {Davidsen} 2002]{tzk+02}
{Telfer}, R.~C., {Zheng}, W., {Kriss}, G.~A., \& {Davidsen}, A.~F. 2002, \apj,  565, 773

\bibitem[{Treister} \& {Urry} 2005]{tu05}
{Treister}, E. \& {Urry}, C.~M. 2005, \apj, 630, 115

\bibitem[{Treister}, {Urry}, {Chatzichristou},  {Bauer}, {Alexander}, {Koekemoer}, {Van Duyne}, {Brandt}, {Bergeron},  {Stern}, {Moustakas}, {Chary}, {Conselice}, {Cristiani}, \&  {Grogin} 2004]{tuc+04}
{Treister}, E., {Urry}, C.~M., {Chatzichristou}, E., {Bauer}, F., {Alexander},  D.~M., {Koekemoer}, A., {Van Duyne}, J., {Brandt}, W.~N., {et al.} 2004, \apj, 616, 123

\bibitem[{Tremaine}, {Gebhardt}, {Bender}, {Bower},  {Dressler}, {Faber}, {Filippenko}, {Green}, {Grillmair}, {Ho}, {Kormendy},  {Lauer}, {Magorrian}, {Pinkney}, \& {Richstone} 2002]{tgb+02}
{Tremaine}, S., {Gebhardt}, K., {Bender}, R., {Bower}, G., {Dressler}, A.,  {Faber}, S.~M., {Filippenko}, A.~V., {Green}, R., {et al.} 2002, \apj, 574, 740

\bibitem[{Tucker et al.} 2005]{tucker+05}
{Tucker et al.} 2005, AJ, submitted

\bibitem[{Ueda}, {Akiyama}, {Ohta}, \& {Miyaji} 2003]{Ueda03}
{Ueda}, Y., {Akiyama}, M., {Ohta}, K., \& {Miyaji}, T. 2003, \apj, 598, 886

\bibitem[{Vanden Berk}, {Richards}, {Bauer},  {Strauss}, {Schneider}, {Heckman}, {York}, {Hall}, {Fan}, {Knapp},  {Anderson}, {Annis}, {Bahcall}, {Bernardi}, {Briggs}, {Brinkmann}, {Brunner},  {Burles}, {Carey}, {Castander}, {Connolly}, {Crocker}, {Csabai}, {Doi},  {Finkbeiner}, {Friedman}, {Frieman}, {Fukugita}, {Gunn}, {Hennessy},  {Ivezi{\' c}}, {Kent}, {Kunszt}, {Lamb}, {Leger}, {Long}, {Loveday},  {Lupton}, {Meiksin}, {Merelli}, {Munn}, {Newberg}, {Newcomb}, {Nichol},  {Owen}, {Pier}, {Pope}, {Rockosi}, {Schlegel}, {Siegmund}, {Smee}, {Snir},  {Stoughton}, {Stubbs}, {SubbaRao}, {Szalay}, {Szokoly}, {Tremonti}, {Uomoto},  {Waddell}, {Yanny}, \& {Zheng} 2001]{vrb+01}
{Vanden Berk}, D.~E., {Richards}, G.~T., {Bauer}, A., {Strauss}, M.~A.,  {Schneider}, D.~P., {Heckman}, T.~M., {York}, D.~G., {Hall}, P.~B., {et al.} 2001, \aj, 122, 549

\bibitem[{Vanden Berk}, {Schneider}, {Richards},  {Hall}, {Strauss}, {Brunner}, {Fan}, {Baldry}, {York}, {Gunn}, {Nichol},  {Meiksin}, \& {Brinkmann} 2005]{vsr+05}
{Vanden Berk}, D.~E., {Schneider}, D.~P., {Richards}, G.~T., {Hall}, P.~B.,  {Strauss}, M.~A., {Brunner}, R., {Fan}, X., {Baldry}, I.~K., {et al.} 2005, \aj,  129, 2047

\bibitem[{Wall}, {Jackson}, {Shaver}, {Hook}, \&  {Kellermann} 2005]{wjs+05}
{Wall}, J.~V., {Jackson}, C.~A., {Shaver}, P.~A., {Hook}, I.~M., \&  {Kellermann}, K.~I. 2005, \aap, 434, 133

\bibitem[{Wampler} \& {Ponz} 1985]{wp85}
{Wampler}, E.~J. \& {Ponz}, D. 1985, \apj, 298, 448

\bibitem[{Warren}, {Hewett}, \& {Osmer} 1994]{who94}
{Warren}, S.~J., {Hewett}, P.~C., \& {Osmer}, P.~S. 1994, \apj, 421, 412

\bibitem[{Weinstein}, {Richards}, {Schneider},  {Younger}, {Strauss}, {Hall}, {Budav{\' a}ri}, {Gunn}, {York}, \&  {Brinkmann} 2004]{wrs+04}
{Weinstein}, M.~A., {Richards}, G.~T., {Schneider}, D.~P., {Younger}, J.~D.,  {Strauss}, M.~A., {Hall}, P.~B., {Budav{\' a}ri}, T., {Gunn}, J.~E., {et al.} 2004, \apjs, 155, 243

\bibitem[{Wisotzki} 2000]{wis00}
{Wisotzki}, L. 2000, \aap, 353, 861

\bibitem[{Wisotzki}, {Christlieb}, {Bade},  {Beckmann}, {K{\" o}hler}, {Vanelle}, \& {Reimers} 2000]{wcb+00}
{Wisotzki}, L., {Christlieb}, N., {Bade}, N., {Beckmann}, V., {K{\" o}hler},  T., {Vanelle}, C., \& {Reimers}, D. 2000, \aap, 358, 77

\bibitem[{Wolf}, {Wisotzki}, {Borch}, {Dye},  {Kleinheinrich}, \& {Meisenheimer} 2003]{wwb+03}
{Wolf}, C., {Wisotzki}, L., {Borch}, A., {Dye}, S., {Kleinheinrich}, M., \&  {Meisenheimer}, K. 2003, \aap, 408, 499

\bibitem[{Wyithe} \& {Loeb} 2003]{wl03}
{Wyithe}, J.~S.~B. \& {Loeb}, A. 2003, \apj, 595, 614

\bibitem[{York}, {Adelman}, {Anderson}, {Anderson},  {Annis}, {Bahcall}, {Bakken}, {Barkhouser}, {Bastian}, {Berman}, {Boroski},  {Bracker}, {Briegel}, {Briggs}, {Brinkmann}, {Brunner}, {Burles}, {Carey},  {Carr}, {Castander}, {Chen}, {Colestock}, {Connolly}, {Crocker}, {Csabai},  {Czarapata}, {Davis}, {Doi}, {Dombeck}, {Eisenstein}, {Ellman}, {Elms},  {Evans}, {Fan}, {Federwitz}, {Fiscelli}, {Friedman}, {Frieman}, {Fukugita},  {Gillespie}, {Gunn}, {Gurbani}, {de Haas}, {Haldeman}, {Harris}, {Hayes},  {Heckman}, {Hennessy}, {Hindsley}, {Holm}, {Holmgren}, {Huang}, {Hull},  {Husby}, {Ichikawa}, {Ichikawa}, {Ivezi{\' c}}, {Kent}, {Kim}, {Kinney},  {Klaene}, {Kleinman}, {Kleinman}, {Knapp}, {Korienek}, {Kron}, {Kunszt},  {Lamb}, {Lee}, {Leger}, {Limmongkol}, {Lindenmeyer}, {Long}, {Loomis},  {Loveday}, {Lucinio}, {Lupton}, {MacKinnon}, {Mannery}, {Mantsch}, {Margon},  {McGehee}, {McKay}, {Meiksin}, {Merelli}, {Monet}, {Munn}, {Narayanan},  {Nash}, {Neilsen}, {Neswold}, {Newberg}, {Nichol}, {Nicinski}, {Nonino},  {Okada}, {Okamura}, {Ostriker}, {Owen}, {Pauls}, {Peoples}, {Peterson},  {Petravick}, {Pier}, {Pope}, {Pordes}, {Prosapio}, {Rechenmacher}, {Quinn},  {Richards}, {Richmond}, {Rivetta}, {Rockosi}, {Ruthmansdorfer}, {Sandford},  {Schlegel}, {Schneider}, {Sekiguchi}, {Sergey}, {Shimasaku}, {Siegmund},  {Smee}, {Smith}, {Snedden}, {Stone}, {Stoughton}, {Strauss}, {Stubbs},  {SubbaRao}, {Szalay}, {Szapudi}, {Szokoly}, {Thakar}, {Tremonti}, {Tucker},  {Uomoto}, {Vanden Berk}, {Vogeley}, {Waddell}, {Wang}, {Watanabe},  {Weinberg}, {Yanny}, \& {Yasuda} 2000]{yaa+00}
{York}, D.~G., {Adelman}, J., {Anderson}, J.~E., {Anderson}, S.~F., {Annis},  J., {Bahcall}, N.~A., {Bakken}, J.~A., {Barkhouser}, R., {et al.} 2000, \aj, 120, 1579

\bibitem[{Yu} \& {Tremaine} 2002]{yt02}
{Yu}, Q. \& {Tremaine}, S. 2002, \mnras, 335, 965

\bibitem[{Zakamska}, {Strauss}, {Krolik}, {Collinge},  {Hall}, {Hao}, {Heckman}, {Ivezi{\' c}}, {Richards}, {Schlegel}, {Schneider},  {Strateva}, {Vanden Berk}, {Anderson}, \& {Brinkmann} 2003]{zsk+03}
{Zakamska}, N.~L., {Strauss}, M.~A., {Krolik}, J.~H., {Collinge}, M.~J.,  {Hall}, P.~B., {Hao}, L., {Heckman}, T.~M., {Ivezi{\' c}}, {\v Z}., {et al.} 2003, \aj, 126,  2125

\end{thebibliography}
\end{document}